\documentclass[twocolumn,twocolappendix]{aastex631}

\usepackage{enumerate}
\usepackage{natbib,ifthen}
\usepackage{graphicx}
\usepackage{fancyhdr}
\usepackage{amssymb,amsmath}
\usepackage{multirow,bigdelim}
\usepackage{longtable}
\usepackage{xcolor}
\usepackage{booktabs}

\shorttitle{Hot Cores in the LMC}
\shortauthors{Broadmeadow et al.}

\graphicspath{{./}{}}

\begin{document}

\title{ALMA Observations of Molecular Complexity in the Large Magellanic Cloud: Probing the Star-forming Region N\,160}

\correspondingauthor{Amanda Broadmeadow}
\email{abroadme@umd.edu}

\author[0000-0002-1048-9618]{Amanda Broadmeadow}
\affiliation{Department of Astronomy, University of Maryland, College Park, MD 20742, USA}

\author[0000-0003-2248-6032]{Marta Sewi{\l}o}
\affiliation{Exoplanets and Stellar Astrophysics Laboratory, NASA Goddard Space Flight Center, Greenbelt, MD 20771, USA}
\affiliation{Department of Astronomy, University of Maryland, College Park, MD 20742, USA}
\affiliation{Center for Research and Exploration in Space Science and Technology, NASA Goddard Space Flight Center, Greenbelt, MD 20771} 

\author[0000-0002-8876-0690]{Lee Mundy}
\affiliation{Department of Astronomy, University of Maryland, College Park, MD 20742, USA}

\author[0000-0003-1993-2302]{Roya Hamedani Golshan} 
\affiliation{I. Physikalisches Institut der Universit{\"a}t zu K{\"o}ln, Z{\"u}lpicher Str. 77, 50937, K{\"o}ln, Germany}

\author[0000-0002-2062-1600]{Kazuki Tokuda}
\affiliation{Department of Earth and Planetary Sciences, Faculty of Sciences, Kyushu University, Nishi-ku, Fukuoka 819-0395, Japan}
\affiliation{National Astronomical Observatory of Japan, National Institutes of Natural Sciences, 2-21-1 Osawa, Mitaka, Tokyo 181-8588, Japan}
\affiliation{Faculty of Education, Kagawa University, 1-1 Saiwaicho, Takamatsu, Kagawa 760-0016, Japan}

\author[0000-0002-9277-8025]{Thomas M{\"o}ller}
\affiliation{I. Physikalisches Institut der Universit{\"a}t zu K{\"o}ln, Z{\"u}lpicher Str. 77, 50937, K{\"o}ln, Germany}

\author[0000-0002-4663-6827]{Remy Indebetouw}
\affiliation{Department of Astronomy, University of Virginia, PO Box 400325, Charlottesville, VA 22904, USA}
\affiliation{National Radio Astronomy Observatory, 520 Edgemont Rd, Charlottesville, VA 22903, USA}

\author[0000-0002-0861-7094]{Joana M. Oliveira}
\affiliation{Lennard-Jones Laboratories, Keele University, ST5 5BG, UK}

\author[0000-0001-6752-5109]{Steven B. Charnley}
\affiliation{Astrochemistry Laboratory, NASA Goddard Space Flight Center, Greenbelt, MD 20771, USA}

\author[0000-0002-1143-6710]{Jennifer Wiseman}
\affiliation{Exoplanets and Stellar Astrophysics Laboratory, NASA Goddard Space Flight Center, Greenbelt, MD 20771, USA}

\author[0000-0002-8217-7509]{Naoto Harada}
\affiliation{Department of Astronomy, Graduate School of Science, The University of Tokyo, 7-3-1 Hongo, Bunkyo-ku, Tokyo 113-0033, Japan}

\author[0000-0003-2141-5689]{Peter Schilke}
\affiliation{I. Physikalisches Institut der Universit{\"a}t zu K{\"o}ln, Z{\"u}lpicher Str. 77, 50937, K{\"o}ln, Germany}

\begin{abstract}
Hot cores are small ($\lesssim$0.1~pc), dense ($\geq$10$^6$ cm$^{-3}$), and hot ($>$100 K) regions around massive protostars and are one of the main production sites of complex organic molecules (COMs, $\geq6$ atoms, including carbon). The Large Magellanic Cloud (LMC) is an ideal place to study hot core and COM formation in an environment that is different from our Galaxy, though prior to this study there have only been nine detections of extragalactic hot cores (seven in the LMC and two in the Small Magellanic Cloud, SMC). Here, we report 1.2 mm continuum and molecular line observations with the Atacama Large Millimeter/submillimeter Array (ALMA) in the star-forming region N\,160 that we named N\,160A--mm. We identify six 1.2 mm continuum sources, four of which are associated with methanol (CH$_3$OH) emission. Another COM, methyl cyanide (CH$_3$CN) is associated with the brightest source, N\,160A--mm\,A, the most chemically rich source in the field. Using the XCLASS software, we perform spectral modeling to estimate rotational temperatures and total column densities of detected molecular species for four sources. Based on the temperature exceeding 100 K, small size, and high H$_2$ number density, we identify N\,160A--mm\,A as a hot core. We compare the molecular abundances of this newly detected hot core with those previously detected in the LMC and SMC, as well as with a sample of Galactic hot cores, and discuss the complex nature of N\,160A--mm\,A. 

\end{abstract}

\section{Introduction} \label{s:intro}
High-mass stars are particularly important sources of chemical enrichment in the interstellar medium, as well as sources of general feedback in star-forming regions. Their formation however, is less understood compared to that of low mass stars. One of the earliest stages of high-mass star formation is the hot core, defined as compact ($\lesssim$0.1~pc) and hot ($>$100 K) with high densities ($\geq$10$^6$ cm$^{-3}$) forming around a massive protostar \citep{kurtz2000, vanderTak2004}. In the Galaxy, hot cores show a rich chemical makeup with many molecular lines, particularly those from complex organic molecules (COMs, $\geq$6 atoms, contain carbon, \citealt{herbst2009}) which are believed to be a precursor of life on Earth \citep{ehrenfreund2000, mumma2011}. COMs are thought to primarily form on the icy grain mantles and are released into the gas near the protostar by thermal evaporation and sputtering due to shocks. These processes lead to the chemical complexity of hot cores \citep{Palau2011, oberg2016, jorgensen2020}.

By studying hot cores in a variety of environments, we can learn about the impact of changing physical conditions on the formation and destruction of COMs. The Large Magellanic Cloud (LMC), a satellite galaxy of the Milky Way, is an ideal laboratory for carrying out such studies. The low-metallicity environment of the LMC (0.3--0.5 Z$_\odot$, \citealt{russell1992, rolleston2002}) is a good analog for galaxies at the peak of cosmic star formation ($z\sim2$, \citealt{pei1999, mehlert2002, madau2014}). In addition, the LMC has larger gas-to-dust ratios than those seen in the Milky Way \citep{Koornneef1984,Roman-Duval2014} as well as higher UV radiation fields \citep{browning2003, welty2006}. These two properties together lead to overall higher dust temperatures in the LMC \citep{vanLoon2010}. The LMC's close proximity ($49.59\pm0.09$ (statistical) $\pm0.54$ (systematic) kpc; \citealt{pietrzynski2019}) allows us to resolve individual sources.

\begin{figure*}[t!]
    \centering
    \includegraphics[width=1\textwidth]{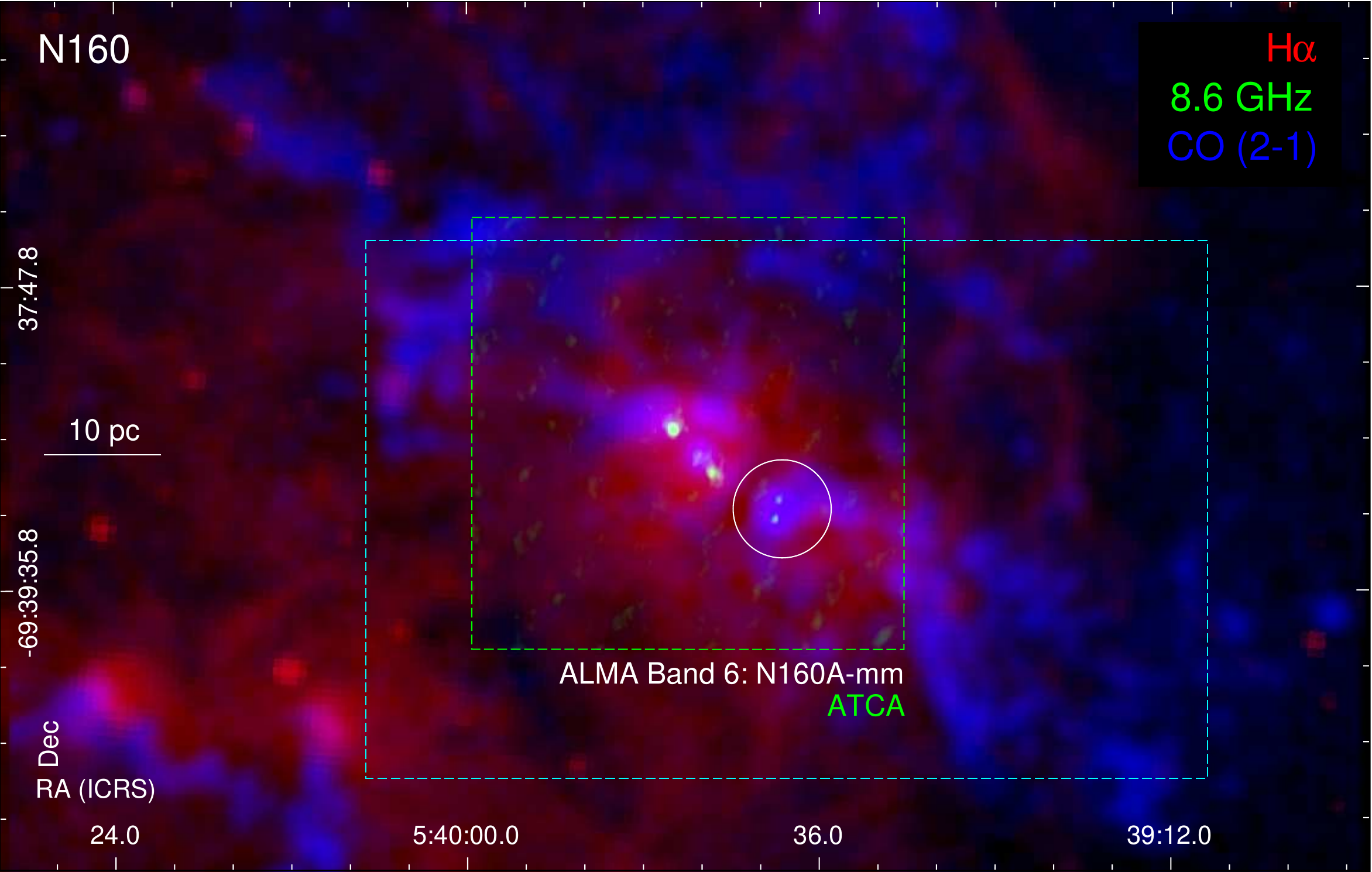} \\
     \includegraphics[width=1\textwidth]{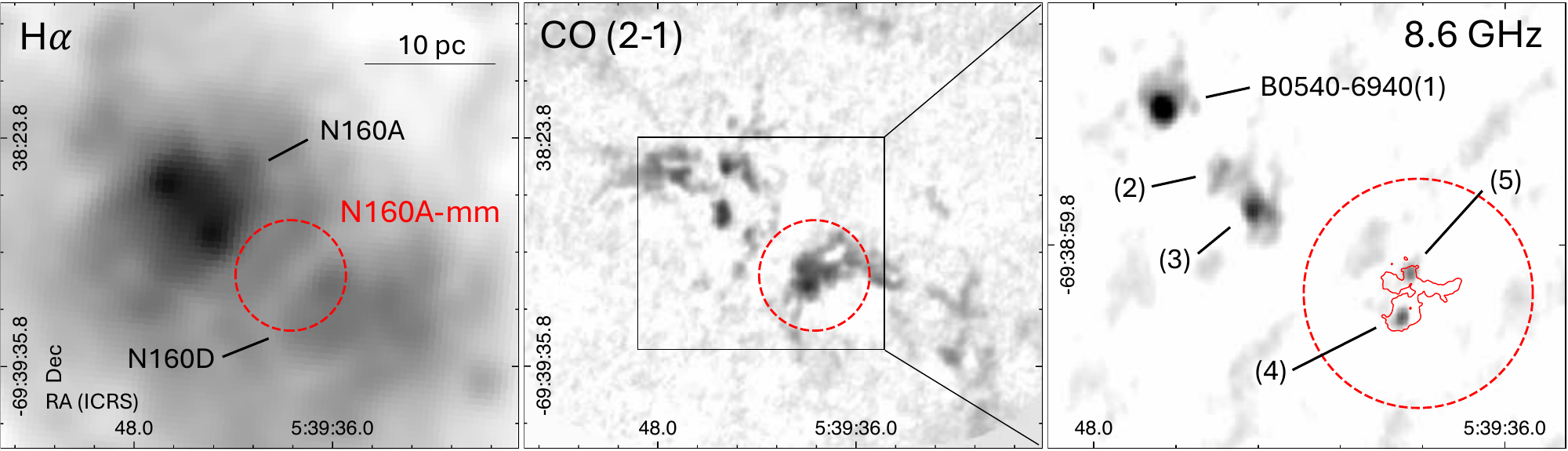}
    \caption{{\it Upper panel}: Three-color mosaic of the star-forming region N\,160 combining the H$\alpha$ image from the MCELS survey (red; $\sim$2$''$ resolution, \citealt{smith1998}), the Australia Telescope Compact Array (ATCA) 8.6 GHz / 3 cm image (green; a half-power beam width, HPBW, of $\sim$1$\rlap.{''}$5, \citealt{indebetouw2004}), and the Atacama Compact Array (ACA) $^{12}$CO (2--1) image (blue; HPBW=7$''$, Tarantino et al., in prep.). The dashed green box indicates the field-of-view of the ATCA image, while the white circle shows the field-of-view of the ALMA Band 6 observations discussed in this paper (N\,160A--mm). The dashed cyan box indicates the area shown in Fig.~\ref{fig:N160RGB}. {\it Lower panel}: Individual images used in the three-color mosaic shown in the upper panel (from left to right: H$\alpha$, CO (2--1), and 8.6 GHz) zoomed-in on the central part of N\,160. However, for CO (2--1), we show the higher resolution ALMA archival image (HPBW$\sim$2$''$ or 0.5 pc; based on the combined 12m-Array, ACA, and Total Power data). The ALMA N\,160A--mm field is indicated with the dashed red circle in the lower three panels. The 3 cm / 6 cm radio sources from \citet{indebetouw2004} are labeled in the right panel. The red contour in the right panel corresponds to 1\% of the 1.2 mm continuum peak (see e.g., Fig.~\ref{fig:Mom0COMs}).
    \label{fig:Ha_8.6_CO}}
\end{figure*}

The sample of extragalactic hot cores is very limited. To date, there have only been 9 hot cores detected outside the Milky Way. In the LMC: ST11 \citep{shimonishi2016b}, N\,113 A1 and B3 \citep{sewilo2018}, ST16 \citep{shimonishi2020}, N\,105--2\,A and 2\,B \citep{sewilo2022}, and N\,132-14A \citep{Golshan2024}. In the Small Magellanic Cloud (SMC): S07 and S09 \citep{Shimonishi2023}. Though the sample size is still small, it reveals large variations (over an order of magnitude) in the CH$_3$OH abundance with respect to H$_2$ while the SO$_2$ and SO abundances do not change significantly from source to source. This could be a result of their different formation mechanisms, with CH$_3$OH forming earlier on dust grains and SO$_2$ forming later in the gas phase.

This project is the second in a series of papers studying molecular complexity in star-forming regions of the LMC based on the Atacama Large Millimeter/submillimeter Array (ALMA) observations covering four $\sim$2 GHz-wide spectral windows between $\sim$241.5 GHz and $\sim$260.6 GHz (in ALMA Band 6). One of the goals of the project is to identify hot cores, expanding a very small sample of known hot cores in the LMC. The program includes seven fields in the LMC that have common characteristics with regions in N\,113 hosting bona fide hot cores, at the time of the target selection, the only known extragalactic sources with hot core complex chemistry. Specifically, they are associated with high-mass young stellar objects (YSOs), masers (H$_2$O, OH, or CH$_3$OH), and SO emission. \citet{sewilo2022} presented the results for three out of seven ALMA fields, all located in the star-forming region N\,105. Two bona fide hot cores (N\,105--2\,A and 2\,B) and a handful of hot core candidates were identified in N\,105. In the present paper, we report the results of the observations of an ALMA field located in the star-forming region LHA 120-N\,160 (hereafter N\,160; \citealt{henize1956}) that we refer to as N\,160A--mm. 

\begin{figure*}[t!]
    \centering
    \includegraphics[width=\textwidth]{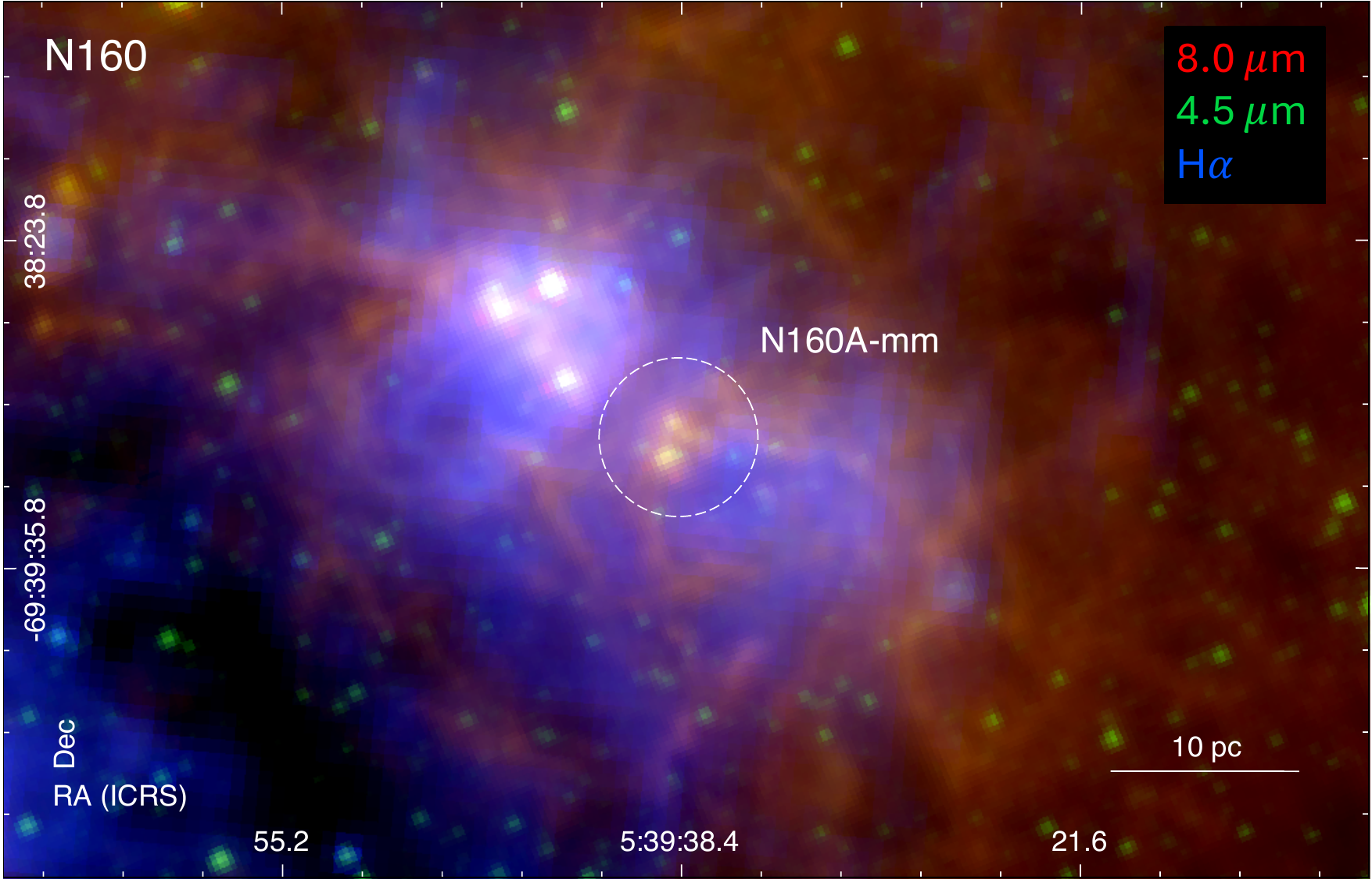}
    \caption{Three-color mosaic of the star-forming region N\,160, combining the Spitzer/IRAC 8.0 $\mu$m (red) and 4.5 $\mu$m (green) images from the SAGE survey \citep{meixner2006}, and the MCELS H$\alpha$ image (blue; \citealt{smith1998}). The ALMA field N\,160A--mm is shown as the white circle and labeled. The 8.0 $\mu$m emission traces hot dust and Polycyclic Aromatic Hydrocarbons (PAHs), while the 4.5 $\mu$m and H$\alpha$ emission traces stars and the ionized gas, respectively.} \label{fig:N160RGB}
\end{figure*}

N\,160 is part of the prominent molecular cloud complex in the LMC referred to as ``the molecular ridge,'' stretching from 30 Doradus (the most massive star-forming region in the Local Group) south for $\sim$1.6 kpc (e.g., \citealt{cohen1988}; \citealt{fukui1999}; \citealt{mizuno2001}). N\,160 lies $\sim$33$'$/0.48 kpc south from the R136 cluster in the center of the 30 Dor nebula, in the northern half of the molecular ridge characterized by vigorous star formation (N\,158, N\,160, N\,159E, and N\,159W), in contrast to its quiescent southern part. N\,160 shows numerous signs of ongoing star formation including YSOs, H$_2$O/OH/CH$_3$OH maser emission, and compact radio sources \citep{epchtein1984,jones1986,caswell1981,gardner1985,caswell1995,indebetouw2004,ellingsen2010}. Additionally, there are a number of more evolved high-mass stars associated with N\,160, including 41 members of the LH\,103 OB association \citep{LuckeHodge1970,lucke1974}.

The star-forming region N\,160 is composed of a number of smaller H$\alpha$ nebulae (N\,160A--F; \citealt{henize1956}). The brightest of these nebulae is N\,160A, seen in the center of the three-color mosaic incorporating the H$\alpha$ image shown in Figure~\ref{fig:Ha_8.6_CO}, with the dimmer nebula N\,160D toward southwest. Other H$\alpha$ nebulae are located toward southeast from N\,160A/N\,160D with N\,160B and N\,160C seen in the lower left corner of the three-color mosaic. Three 3 cm/ 6 cm compact radio sources, B0540$-$6940(1)--(3), are associated with the main optical nebula N\,160A \citep{indebetouw2004}. An additional two, B0540$-$6940(4) and B0540$-$6940(5), lie in the H$\alpha$-dim region between N\,160A and N\,160D, within our ALMA field N\,160A--mm (see Figure~\ref{fig:Ha_8.6_CO}). 

N\,160A--mm is embedded in the brightest $^{12}$CO, $^{13}$CO, and C$^{18}$O (2--1) clump in N\,160 as traced by the Atacama Compact Array (ACA) observations probing $\sim$1.7 pc scales (Tarantino et al., in prep.; see Figure~\ref{fig:Ha_8.6_CO}). Several fainter CO clumps are associated with N\,160A. The bulk of the molecular gas has a filamentary structure and extends to the northeast and southwest from N\,160A/N\,160D and northwest from the northeast--southwest line going through these two H$\alpha$ nebulae. 

The {\it Spitzer} Space Telescope images shown in Figure \ref{fig:N160RGB} reveal multiple 8.0 $\mu$m bubbles in N\,160, produced by strong winds and radiation from young massive stars (e.g, \citealt{churchwell2006}; \citealt{deharveng2010}). The bubbles are filled with the ionized gas traced by the H$\alpha$ emission. N\,160A--mm is located on a rim of one of the bubbles, indicating that high-mass star formation in N\,160A--mm may have been triggered by its expansion.

This paper is organized as follows. In Section~\ref{s:observations}, we describe our ALMA observations of N\,160A--mm and the ALMA archival data utilized in our analysis. In Section~\ref{s:Analysis}, we present the analysis of the 1.2 mm continuum and spectral line data. We provide a discussion on our results in Section~\ref{s:Discussion} and in Section~\ref{s:Conclusions} we summarize our results and conclusions.

\section{The Data} 
\label{s:observations}

\subsection{ALMA Observations}

The ALMA field N\,160A--mm in the star-forming region N\,160 was observed with the 12m-Array as part of the Cycle 7 project 2019.1.01720.S (PI M. Sewi{\l}o). A detailed description of the observations can be found in \cite{sewilo2022}. The spectral setup included four 1875 MHz spectral windows centered on frequencies of $\sim$242.4 GHz, $\sim$244.8 GHz, $\sim$257.85 GHz, and $\sim$259.7 GHz, each with 3840 channels of 488.3 kHz (or $\sim$0.6 km s$^{-1}$) width. The maximum recoverable angular scale calculated from the 5$^{th}$ percentile baseline length for the final data set varied between 5$\rlap.{''}$6 and 5$\rlap.{''}$2 for a sky frequency range covered by our observations ($\sim$241.3--260.4 GHz).

The data were calibrated with version 5.6.1-8 of the ALMA pipeline \citep{Hunter2023} in Common Astronomy Software Applications (CASA; \citealt{casa2022}) and imaged with CASA version 6.5. The CASA task \texttt{tclean} was used for imaging with the Hogbom deconvolver, standard gridder, Briggs weighting with a robust parameter of 0.5, and the 'auto-multithresh' masking. The spectral cubes have a cell size of $0\rlap.{''}092\times0\rlap.{''}092\times0.6$ km s$^{-1}$. All the images were corrected for primary beam attenuation.

The position of the ALMA pointing in N\,160, as well as the 242.4 GHz (1.2 mm) continuum image and spectral cube parameters are listed in Table \ref{t:contcubedata}. The typical synthetic beam size is $\sim$$0\rlap.{''}5$ which corresponds to $\sim$$0.1$ pc. The rms noise per channel for the spectral cubes was estimated based on line-free channels.

\begin{deluxetable*}{lcccccc}
\centering
\tablecaption{ALMA Pointing and 1.2 mm Continuum Image and Spectral Cube Parameters\label{t:contcubedata}}
\tablewidth{0pt}
\tablehead{
\colhead{Field} &
\colhead{RA} &
\colhead{Decl.} &
\colhead{Spectral } &
\colhead{Frequency Range} &
\colhead{Synth. Beam: ($\Theta_B$, PA)} &
\colhead{Image rms ($\sigma$)} \\
\colhead{} & 
\colhead{($^{\rm h}$ $^{\rm m}$ $^{\rm s}$)} &
\colhead{($^{\rm \circ}$ $'$ $''$)} &
\colhead{Window} &
\colhead{(GHz)} &
\colhead{($'' \times ''$, $^{\circ}$)} &
\colhead{(10$^{-4}$ Jy beam$^{-1}$, mK)} 
}
\startdata
N\,160A--mm & 05:39:38.51 & $-$69:39:07.0 & 242 GHz & 241.27607 -- 243.14925 & $0.534\times0.497$, 19.6 & 19.0, 150 \\
\nodata & \nodata & \nodata & 245 GHz & 243.66722 -- 245.54040 & $0.530\times0.491$, 16.1 & 18.1, 142 \\
\nodata & \nodata & \nodata & 258 GHz & 256.71446 -- 258.58764 & $0.508\times0.469$, 19.1 & 19.5, 151 \\
\nodata & \nodata & \nodata & 260 GHz & 258.54756 -- 260.42074 & $0.504\times0.466$, 17.6 & 23.4, 181 \\
\nodata & \nodata & \nodata & \multicolumn{2}{c}{Continuum (242.4 GHz)} & $0.503\times0.468$, 16.4 & 1.01, 8.29 \\
\enddata
\end{deluxetable*}

\subsection{ALMA Archival Data}

We retrieved the Band 7 data from the ALMA archive for target Lh08 corresponding to our ALMA field in N\,160. The observations were part of the Cycle 7 project 2019.1.01770.S (PI K. Tanaka). The data were calibrated and imaged using CASA version 6.1.2.7. 

The 345.798 GHz ($\sim$870 $\mu$m) continuum image was constructed from line-free channels in the Band 7 data. The CASA task \texttt{tclean} was used for imaging with the multi-scale deconvolver and the Briggs weighting with a robust parameter of 0.5. The sensitivity of $2.8\times10^{-4}$ Jy beam$^{-1}$ was achieved in the continuum; the beam size is $0\rlap.{''}46\times0\rlap.{''}$39. The image was corrected for primary beam attenuation.

\section{The Analysis and Results} \label{s:Analysis}

\subsection{1.2 mm Continuum Emission and Source Identification} \label{ss:Continuum identification}

\begin{figure*}
    \centering
    \includegraphics[width=1\textwidth]{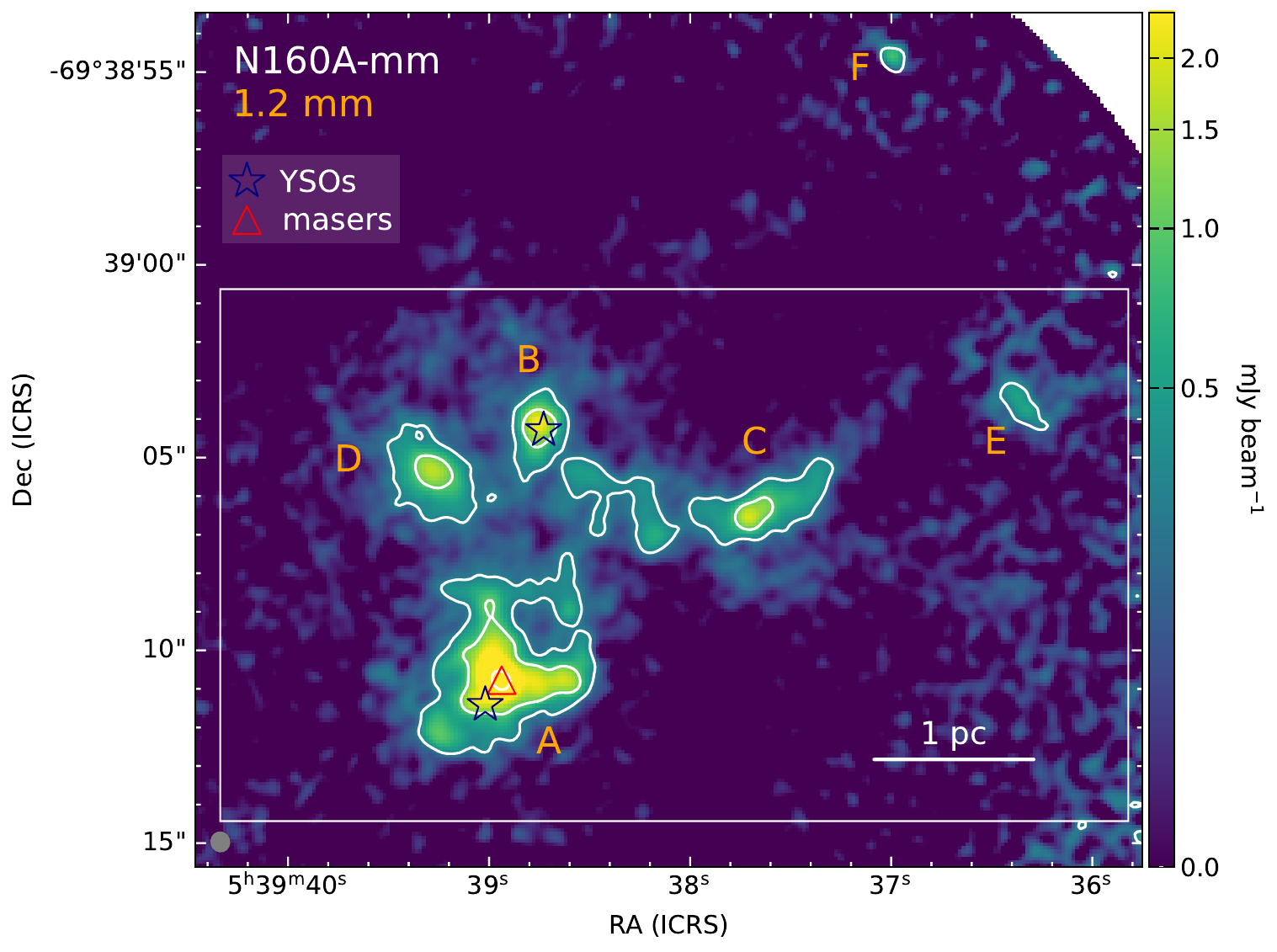}
    \caption{The 1.2 mm continuum image of the ALMA field N\,160A--mm. No continuum or molecular line emission was detected outside the shown field of view. Contours represent 3$\sigma$, 10$\sigma$, and 100$\sigma$ where $\sigma$ is the continuum image rms of $1.01 \times 10^{-4}$ Jy beam$^{-1}$. The white box shows the field of view for the moment maps in Figures \ref{fig:Mom0COMs}-\ref{fig:Mom0Maps}. The blue stars show the locations of Spitzer-identified massive YSOs while the red triangle shows the location of the water, methanol, and OH masers. The six continuum sources identified are labeled A through F and the ALMA synthesized beam size ($0.503''\times0.468''$) is shown as the gray ellipse in the lower left (Table \ref{t:contcubedata}).}
    \label{continuum}
\end{figure*}

The 1.2 mm continuum image of N\,160A--mm shown in Figure~\ref{continuum} reveals several compact continuum sources surrounded by an extended emission. Individual sources are identified based on the 10$\sigma$ contours, where $\sigma$ is the continuum image rms of $1.01 \times 10^{-4}$ Jy beam$^{-1}$. We assign names to all the continuum sources that will be discussed in this paper. These include four sources with the maximum intensity of at least $10\sigma$: N\,160A--mm\,A, N\,160A--mm\,B, N\,160A--mm\,C, and N\,160A--mm\,D, in order of decreasing continuum peak intensity (see Figure~\ref{continuum}). 

Two spectroscopically confirmed YSOs are present in the ALMA field: 053938.73-693904.3 and 053939.02-693911.4 (e.g., \citealt{seale2009}). YSO 053938.73-693904.3 is associated with N\,160A--mm\,B, and 053939.02-693911.4 is N\,160A--mm\,A's nearest neighbor in the Spitzer images. N\,160A--mm\,A is coincident with the position of the H$_2$O, OH, and CH$_3$OH masers (e.g., \citealt{ellingsen2010}; see Figure~\ref{continuum} and Section~\ref{sss: hot core}). The presence of YSOs and masers in the targeted fields was one of the selection criteria for our observations (\citealt{sewilo2022}).

In the ALMA field N\,160A--mm, we also identify two fainter continuum sources that are associated with the molecular line emission peaks. We assign them names N\,160A--mm\,E (a $\sim$6$\sigma$ detection in the continuum) and N\,160A--mm\,F ($\sim$8$\sigma$). These two sources lie close to the edge of the 1.2 mm continuum image where the noise is significantly higher than in the central part of the image hosting sources N\,160A--mm\,A--D. For simplicity, we refer to sources (N\,160A--mm\,A, N\,160A--mm\,B, N\,160A--mm\,C, ...) as (A, B, C, ...) throughout the paper.

\subsection{Spectral Analysis} \label{ss:Spectral Analysis}
\begin{figure*}
    \centering
    \includegraphics[width=\textwidth]{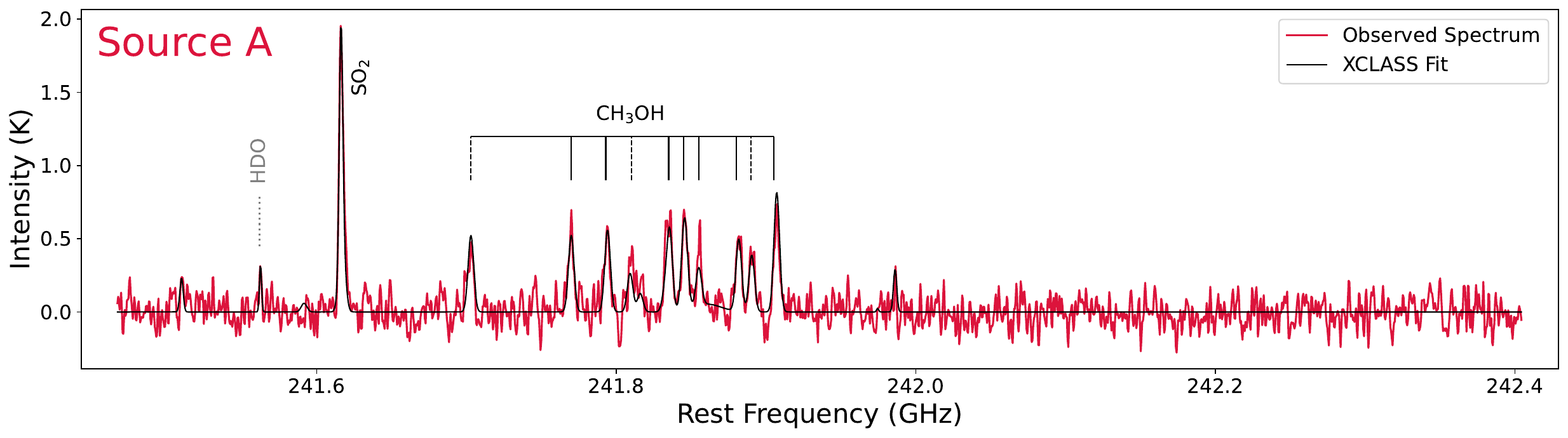} \\
    \includegraphics[width=\textwidth]{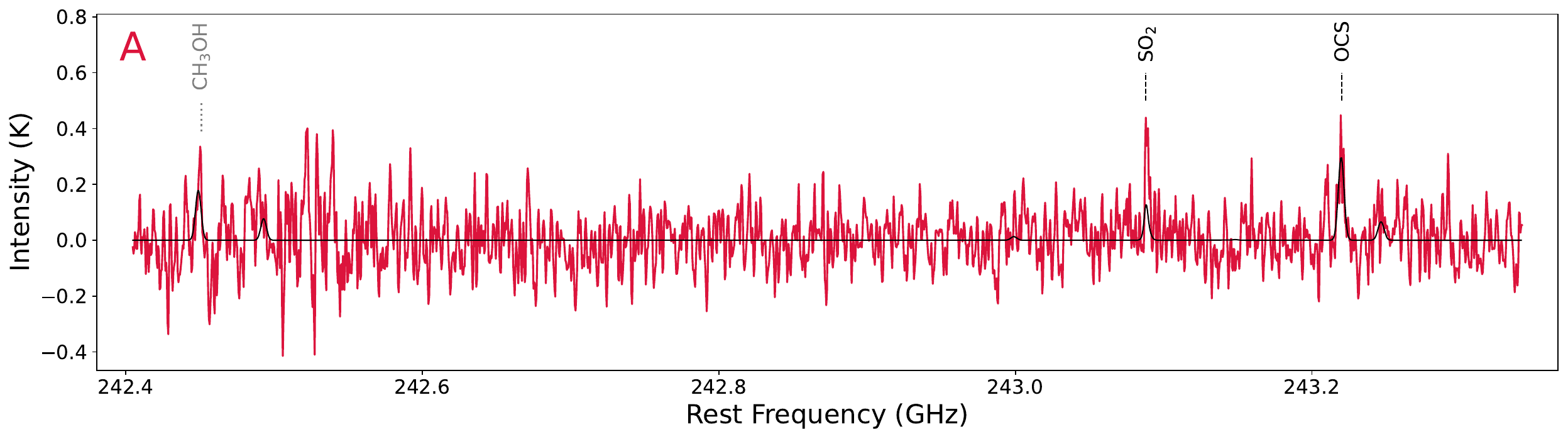} \\
    \includegraphics[width=\textwidth]{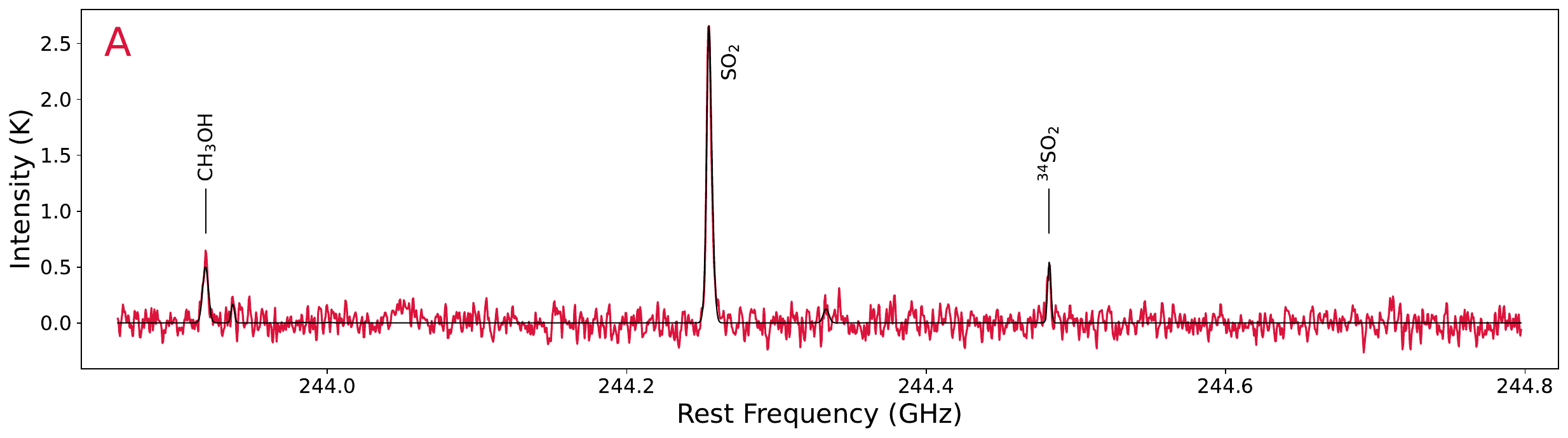} \\
    \includegraphics[width=\textwidth]{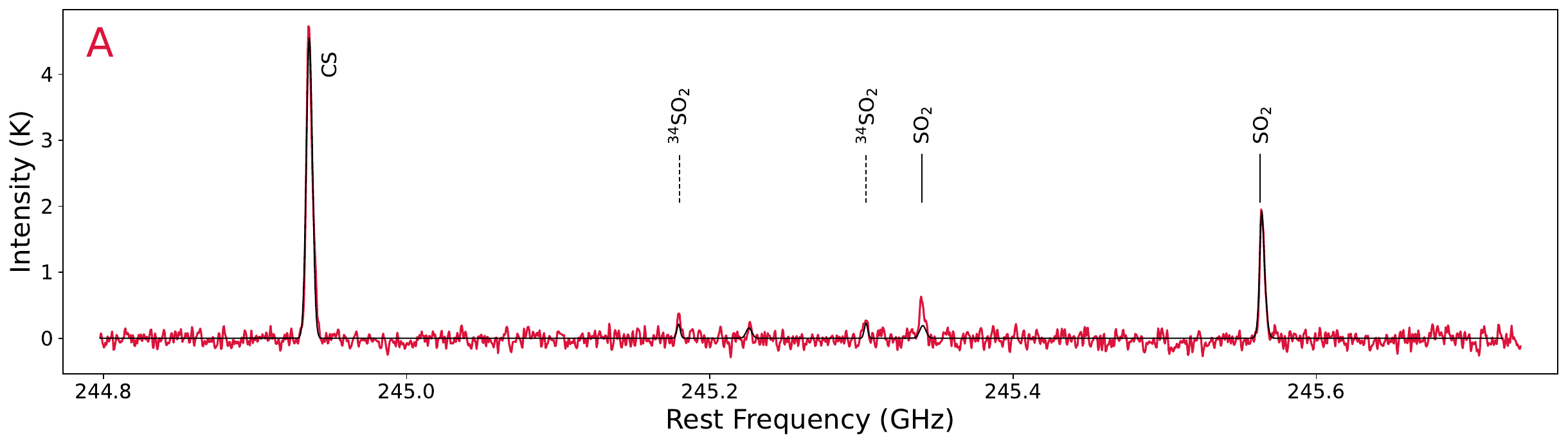} \\
    \caption{ALMA spectra of N\,160A--mm\,A for spectral windows 242 GHz and 245 GHz. The spectra are extracted as the mean intensity of the area enclosed by the 50\% contour of the peak continuum emission of source A. The red solid line shows the observed spectrum and the solid black line shows the XCLASS best fit (see Section \ref{xclass}). Detected lines above $5\sigma_{\rm s,i}$ are labeled with black solid lines, detected lines between $3\sigma_{\rm s,i}$ and $5\sigma_{\rm s,i}$ are labeled with black dashed lines, and tentative detections are labeled with gray dotted lines.}
    \label{fig:SpectraA1}
\end{figure*}

\begin{figure*}
    \centering
    \includegraphics[width=\textwidth]{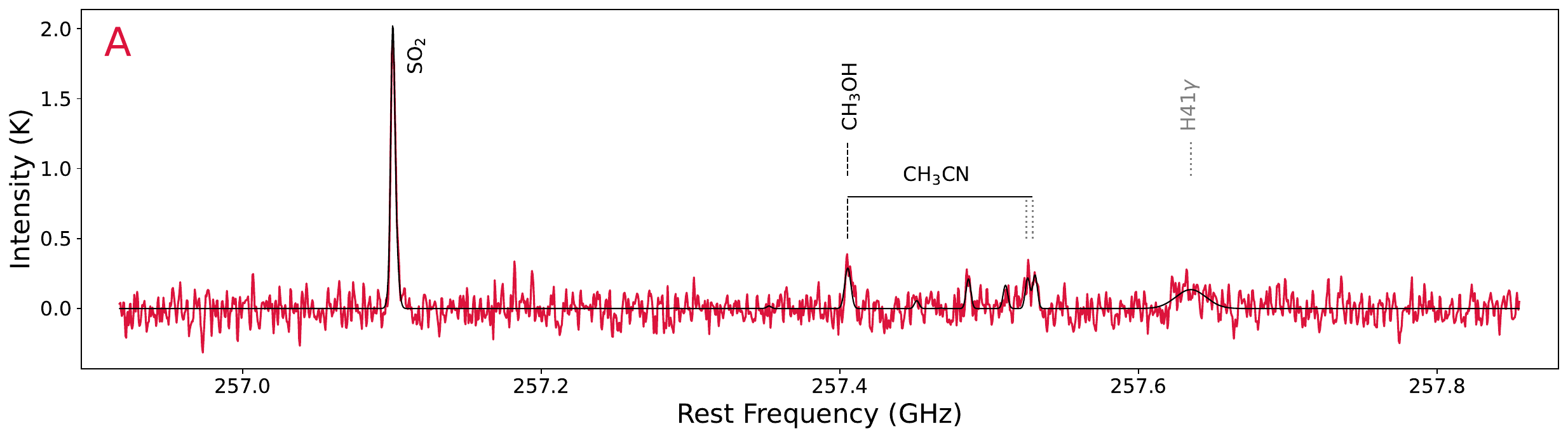} \\
    \includegraphics[width=\textwidth]{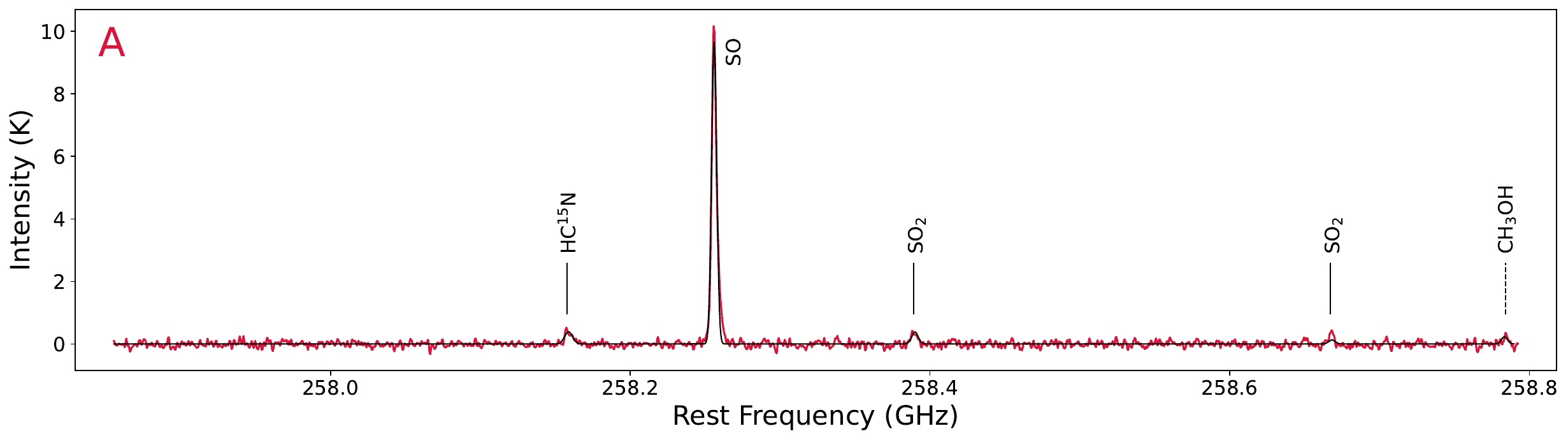} \\
    \includegraphics[width=\textwidth]{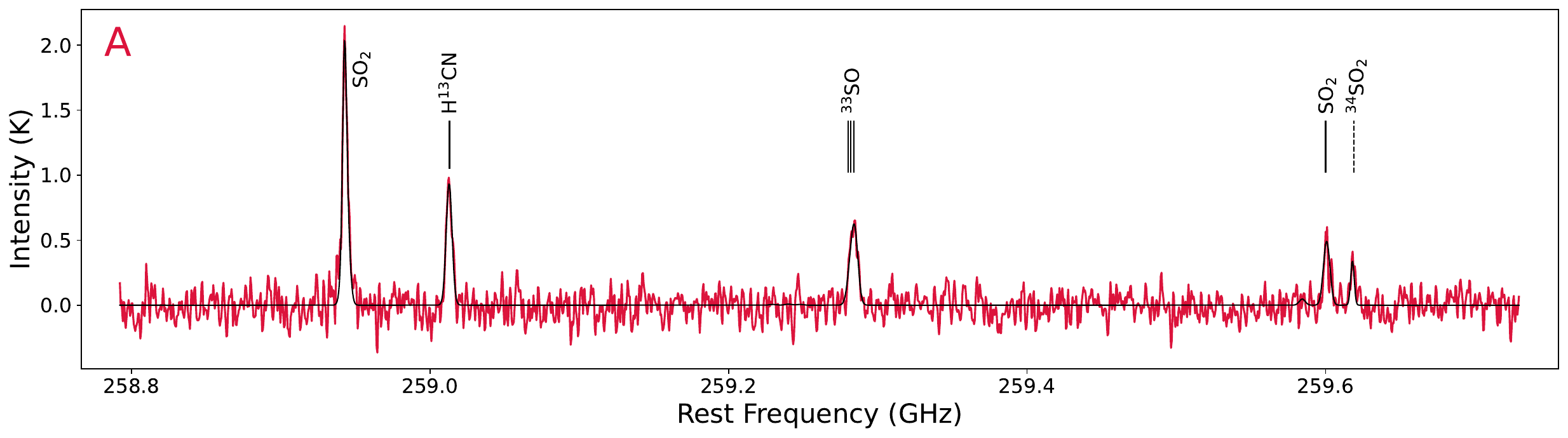} \\
    \includegraphics[width=\textwidth]{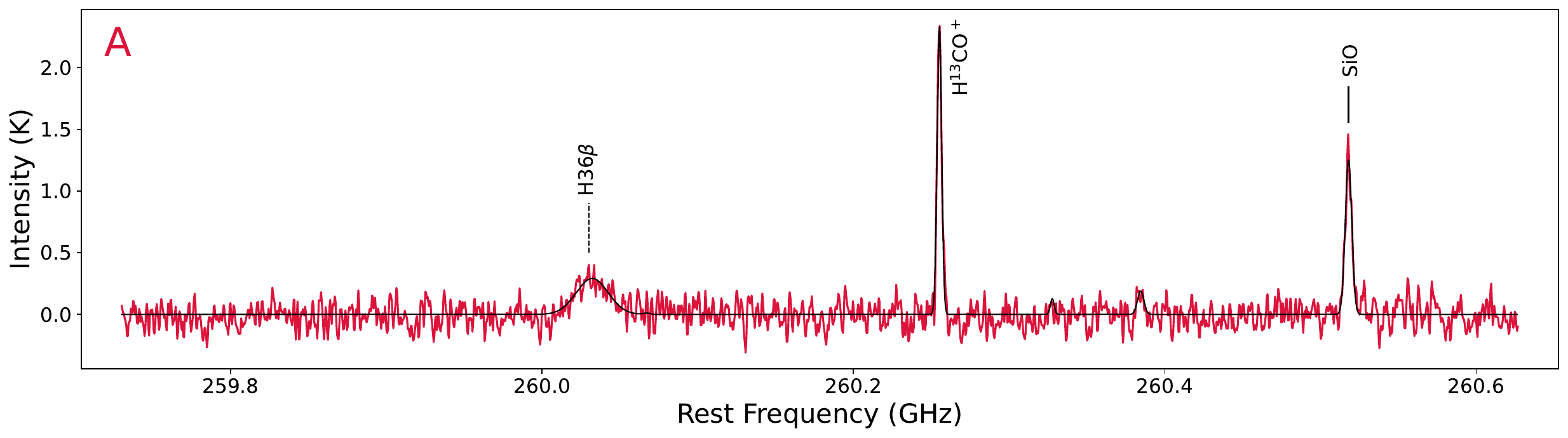} \\
    \caption{ALMA spectra of N\,160A--mm\,A for spectral windows 258 GHz and 260 GHz.}
    \label{fig:SpectraA2}
\end{figure*}

For each source, the spectra were extracted from the data cubes as the mean of the emission enclosed by the half-maximum 1.2 mm continuum emission contour. Figures \ref{fig:SpectraA1} and \ref{fig:SpectraA2} show the spectrum for source A, the brightest continuum source with the most chemically-rich spectrum. The remaining spectra can be found in Appendix \ref{a:spectra}.


\subsubsection{Line Identification} \label{s:Line identification}
We performed the initial spectral line identification using the CASA task \texttt{imview} which uses the NRAO spectral line database Splatalogue.\footnote{http://www.cv.nrao.edu/php/splat} To consider a line detected, three frequency channels in a row must be at least three times the noise level ($3\sigma_{\rm s,i}$). This noise level is calculated for each source "s" and spectral window "i" using a sigma clipping analysis. Lines that do not satisfy this criteria but have four channels above $2\sigma_{\rm s,i}$ with at least one above $3\sigma_{\rm s,i}$, are considered tentative detections. We detected two COMs, CH$_3$OH and CH$_3$CN, a variety of N-bearing (HC$^{15}$N, H$^{13}$CN) and S-bearing (SO, OCS, H$_2$CS, CS, C$^{33}$S, SO$_2$, $^{34}$SO$_2$, $^{33}$SO) species, as well as SiO, H$^{13}$CO$^+$, and tentatively HDO. We also detect a hydrogen recombination line, H36$\beta$. Source A is the most chemically rich source. Source B is the least chemically rich. The full list of detected and tentatively detected spectral line transitions with rest frequencies and upper state energies can be found in Table \ref{t:detections}. We use the molecular data from the Cologne Database for Molecular Spectroscopy (CDMS\footnote{https://cdms.astro.uni-koeln.de}, \citealt{muller2005}) for all species except for HDO where we use the JPL database\footnote{http://spec.jpl.nasa.gov/} \citep{pickett1998}.

\clearpage
\startlongtable
\begin{deluxetable*}{lccccccccc}
\centering
\tablecaption{Spectral Lines Detected Toward Continuum Sources in N\,160A--mm\tablenotemark{\footnotesize a} \label{t:detections}}
\tablewidth{0pt}
\tablehead{
\multicolumn{1}{c}{Species} &
\multicolumn{1}{c}{Transition} &
\colhead{Frequency} &
\colhead{$E_{\rm u}$}&
\colhead{A} & 
\colhead{B} & 
\colhead{C} & 
\colhead{D} & 
\colhead{E} &
\colhead{F} \\
\colhead{} &
\colhead{} & 
\colhead{(MHz)} &
\colhead{(K)} &
\colhead{} &
\colhead{} &
\colhead{} &
\colhead{} &
\colhead{} &
}
\startdata
\multicolumn{10}{c}{COMs} \\
\hline
CH$_3$OH & 5$_{-0, 5}$--4$_{-0, 4}$ E, v$_t$=0 & 241700.159 & 47.94 & \checkmark & - & \checkmark\checkmark & \checkmark\checkmark & \checkmark\checkmark & - \\
CH$_3$OH & 5$_{1, 5}$--4$_{1, 4}$ E, v$_t$=0 & 241767.234 & 40.39 & \checkmark\checkmark & - & \checkmark\checkmark & \checkmark\checkmark & \checkmark\checkmark & - \\
CH$_3$OH & 5$_{0, 5}$--4$_{0, 4}$ A, v$_t$=0 & 241791.352 & 34.82 & \checkmark\checkmark & - & \checkmark\checkmark & \checkmark\checkmark & \checkmark\checkmark & - \\
CH$_3$OH & 5$_{4, 2}$--4$_{4, 1}$ A, v$_t$=0 & 241806.524 & 115.17 & \checkmark & - & \checkmark? & - & - & - \\
CH$_3$OH & 5$_{4, 1}$--4$_{4, 0}$ A, v$_t$=0 & 241806.525 & 115.17 & \checkmark & - & \checkmark? & - & - & - \\
CH$_3$OH & 5$_{-4, 1}$--4$_{-4, 0}$ E, v$_t$=0 & 241829.629 & 130.82 & \checkmark\checkmark & - & - & - & - & - \\
CH$_3$OH & 5$_{3, 3}$--4$_{3, 2}$ A, v$_t$=0 & 241832.718 & 84.62 & \checkmark\checkmark & - & \checkmark\checkmark & \checkmark & - & - \\
CH$_3$OH & 5$_{3, 2}$--4$_{3, 1}$ A, v$_t$=0 & 241833.106 & 84.62 & \checkmark\checkmark & - & \checkmark\checkmark & \checkmark & - & - \\
CH$_3$OH & 5$_{2, 4}$--4$_{2, 3}$ A, v$_t$=0 & 241842.284 & 72.53 & \checkmark\checkmark & - & \checkmark? & - & - & - \\
CH$_3$OH & 5$_{-3, 3}$--4$_{-3, 2}$ E, v$_t$=0 & 241843.604 & 82.53 & \checkmark\checkmark & - & \checkmark? & - & - & - \\
CH$_3$OH & 5$_{3, 2}$--4$_{3, 1}$ E, v$_t$=0 & 241852.299 & 97.53 & \checkmark\checkmark & - & - & - & - & - \\
CH$_3$OH & 5$_{-1, 4}$--4$_{-1, 3}$ E, v$_t$=0 & 241879.025 & 55.87 & \checkmark\checkmark & - & \checkmark\checkmark & \checkmark\checkmark & \checkmark & - \\
CH$_3$OH & 5$_{2, 3}$--4$_{2, 2}$ A, v$_t$=0 & 241887.674 & 72.54 & \checkmark & - & \checkmark & - & - & - \\
CH$_3$OH & 5$_{2, 3}$--4$_{2, 2}$ E, v$_t$=0 & 241904.147 & 60.73 & \checkmark\checkmark & - & \checkmark\checkmark & \checkmark\checkmark & \checkmark\checkmark & - \\
CH$_3$OH & 5$_{-2, 4}$--4$_{-2, 3}$ E, v$_t$=0 & 241904.643 & 57.07 & \checkmark\checkmark & - & \checkmark\checkmark & \checkmark\checkmark & \checkmark\checkmark & - \\
CH$_3$OH & 14$_{1, 14}$--13$_{2, 11}$ E, v$_t$=0 & 242446.084 & 248.94 & \checkmark? & - & - & - & - & - \\
CH$_3$OH & 5$_{1, 4}$--4$_{1, 3}$ A, v$_t$=0 & 243915.788 & 49.66 & \checkmark\checkmark & - & \checkmark\checkmark & \checkmark\checkmark & \checkmark\checkmark & - \\
CH$_3$OH & 18$_{3, 16}$--18$_{2, 17}$ A, v$_t$=0 & 257402.086 & 446.55 &\checkmark \tablenotemark{\footnotesize b} & - & - & - & - & - \\
CH$_3$OH & 19$_{3, 17}$--19$_{2, 18}$ A, v$_t$=0 & 258780.248 & 490.58 & \checkmark & - & - & - & - & - \\
CH$_3$CN & 14$_5$--13$_5$ & 257403.585 & 271.23 & \checkmark \tablenotemark{\footnotesize b} & - & - & - & - & - \\
CH$_3$CN & 14$_3$--13$_3$ & 257482.792 & 156.99 & \checkmark? & - & - & - & - & - \\
CH$_3$CN & 14$_1$--13$_1$ & 257522.428 & 99.85 & \checkmark? & - & - & - & - & - \\
CH$_3$CN & 14$_0$--13$_0$ & 257527.384 & 92.71 & \checkmark? & - & - & - & - & - \\
\hline
\multicolumn{10}{c}{Other Molecules} \\
\hline
HC$^{15}$N & 3--2 & 258156.996 & 24.78 & \checkmark\checkmark & - & \checkmark\checkmark & \checkmark\checkmark & - & - \\
H$^{13}$CN & 3--2 & 259011.798 & 24.86 & \checkmark\checkmark & - & \checkmark\checkmark & \checkmark\checkmark & - & - \\
H$^{13}$CO$^{+}$ & 3--2 & 260255.339 & 24.98 & \checkmark\checkmark & \checkmark? & \checkmark\checkmark & \checkmark\checkmark & \checkmark\checkmark & - \\
SO $^{3}\Sigma$ & 6$_6$--5$_5$ & 258255.826 & 56.50 & \checkmark\checkmark & - & \checkmark\checkmark & \checkmark\checkmark & \checkmark\checkmark & \checkmark? \\
OCS & 20--19 & 243218.036 & 122.58 & \checkmark & - & - & - & - & - \\
H$_2$CS & 7$_{1, 6}$--6$_{1, 5}$ & 244048.504 & 60.03 & - & - & \checkmark\checkmark & \checkmark\checkmark & \checkmark\checkmark & - \\
CS & 5--4 & 244935.557 & 35.27 & \checkmark\checkmark & \checkmark\checkmark & \checkmark\checkmark & \checkmark\checkmark & \checkmark\checkmark & \checkmark\checkmark \\
C$^{33}$S & 5--4 & 242913.610 & 34.98 & - & - & \checkmark\checkmark & \checkmark\checkmark & \checkmark\checkmark & - \\
SO$_2$ & 5$_{2, 4}$--4$_{1, 3}$ & 241615.797 & 23.59 & \checkmark\checkmark & - & \checkmark\checkmark & \checkmark\checkmark & - & - \\
SO$_2$ & 5$_{4, 2}$--6$_{3, 3}$ & 243087.647 & 53.07 & \checkmark & - & - & - & - & - \\
SO$_2$ & 14$_{0, 14}$--13$_{1, 13}$ & 244254.218 & 93.90 & \checkmark\checkmark & - & \checkmark & \checkmark\checkmark & - & - \\ 
SO$_2$ & 26$_{3, 23}$--25$_{4, 22}$ & 245339.233 & 350.79 & \checkmark\checkmark & - & - & - & - & - \\
SO$_2$ & 10$_{3, 7}$--10$_{2, 8}$ & 245563.422 & 72.72 & \checkmark\checkmark & - & - & \checkmark? & - & - \\
SO$_2$ & 7$_{3, 5}$--7$_{2, 6}$ & 257099.966 & 47.84 & \checkmark\checkmark & - & \checkmark & \checkmark\checkmark & - & - \\
SO$_2$ & 32$_{4, 28}$--32$_{3, 29}$ & 258388.716 & 531.12 & \checkmark\checkmark & - & - & - & - & - \\
SO$_2$ & 20$_{7, 13}$--21$_{6, 16}$ & 258666.969 & 313.19 & \checkmark\checkmark & - & - & - & - & - \\
SO$_2$ & 9$_{3, 7}$--9$_{2, 8}$ & 258942.199 & 63.47 & \checkmark\checkmark & - & - & - & - & - \\
SO$_2$ & 30$_{4, 26}$--30$_{3, 27}$ & 259599.448 & 471.52 & \checkmark\checkmark & - & - & - & - & - \\
$^{34}$SO$_2$ & 14$_{0, 14}$--13$_{1, 13}$ & 244481.517 & 93.54 & \checkmark\checkmark & - & - & - & - & - \\
$^{34}$SO$_2$ & 15$_{2, 14}$--15$_{1, 15}$ & 245178.587 & 118.72 & \checkmark & - & - & - & - & - \\
$^{34}$SO$_2$ & 6$_{3, 3}$--6$_{2, 4}$ & 245302.239 & 40.66 & \checkmark & - & - & - & - & - \\
$^{34}$SO$_2$ & 13$_{3, 11}$--13$_{2, 12}$ & 259617.203 & 104.91 & \checkmark & - & - & - & - & - \\
$^{33}$SO & 6$_{7, 6}$--5$_{6, 5}$ & 259280.331 & 47.12 & \checkmark\checkmark & - & - & - & - & - \\
$^{33}$SO & 6$_{7, 7}$--5$_{6, 6}$ & 259282.276 & 47.12 & \checkmark\checkmark & - & - & - & - & - \\
$^{33}$SO & 6$_{7, 8}$--5$_{6, 7}$ & 259284.027 & 47.12 & \checkmark\checkmark & - & - & - & - & - \\
$^{33}$SO & 6$_{7, 9}$--5$_{6, 8}$ & 259284.027 & 47.12 & \checkmark\checkmark & - & - & - & - & - \\
SiO & 6--5 & 260518.009 & 43.76 & \checkmark\checkmark & - & \checkmark & - & - & - \\
HDO & 2$_{1, 1}$--2$_{1, 2}$ & 241561.550 & 95.22 & \checkmark? & - & - & - & - & - \\
\hline
\multicolumn{10}{c}{Hydrogen Recombination Lines} \\
\hline
H & H41$\gamma$ & 257635.49 & \nodata & \checkmark? & - & - & - & - & - \\
H & H36$\beta$ & 260032.78 & \nodata & \checkmark & - & - & - & - & - \\
\enddata
\tablenotetext{a}{Spectroscopic parameters were taken from the CDMS catalog for all species except HDO for which the data were taken from the JPL database. The symbols `\checkmark', `\checkmark?', and '-' indicate, respectively, a detection, a tentative detection, and a non-detection of a given transition. The symbol '\checkmark\checkmark' indicates the detection is above $5\sigma_{\rm s,i}$.}
\tablenotetext{b}{The CH$_3$OH and CH$_3$CN transitions at 257402.086 MHz and 257403.585 MHz, respectively, are blended.}
\end{deluxetable*}

\subsubsection{Spatial Distribution of the Molecular Line Emission} \label{s:MomentMap}
\begin{figure}
    \centering
    \includegraphics[width=\columnwidth]{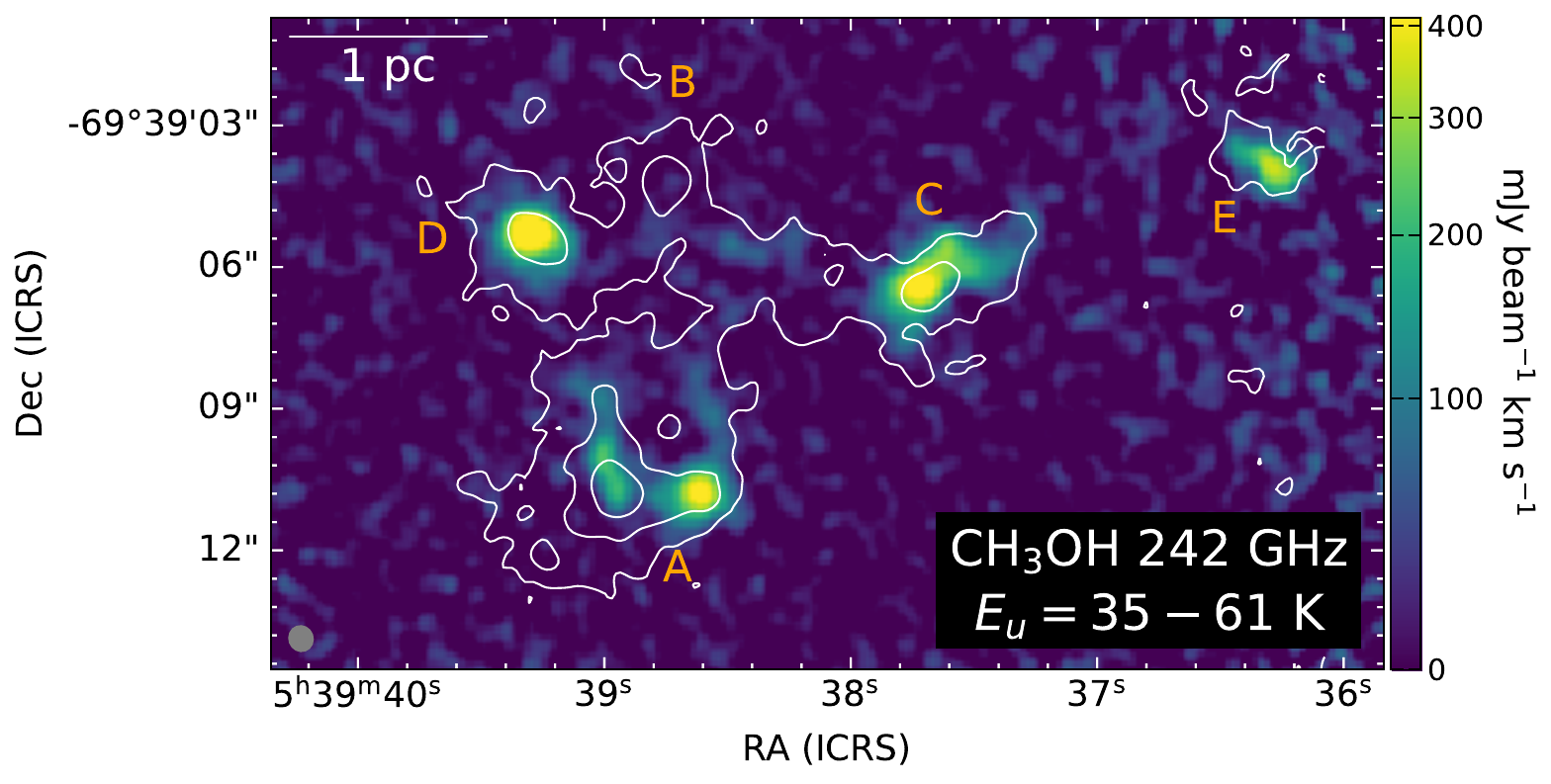} \\
    \includegraphics[width=\columnwidth]{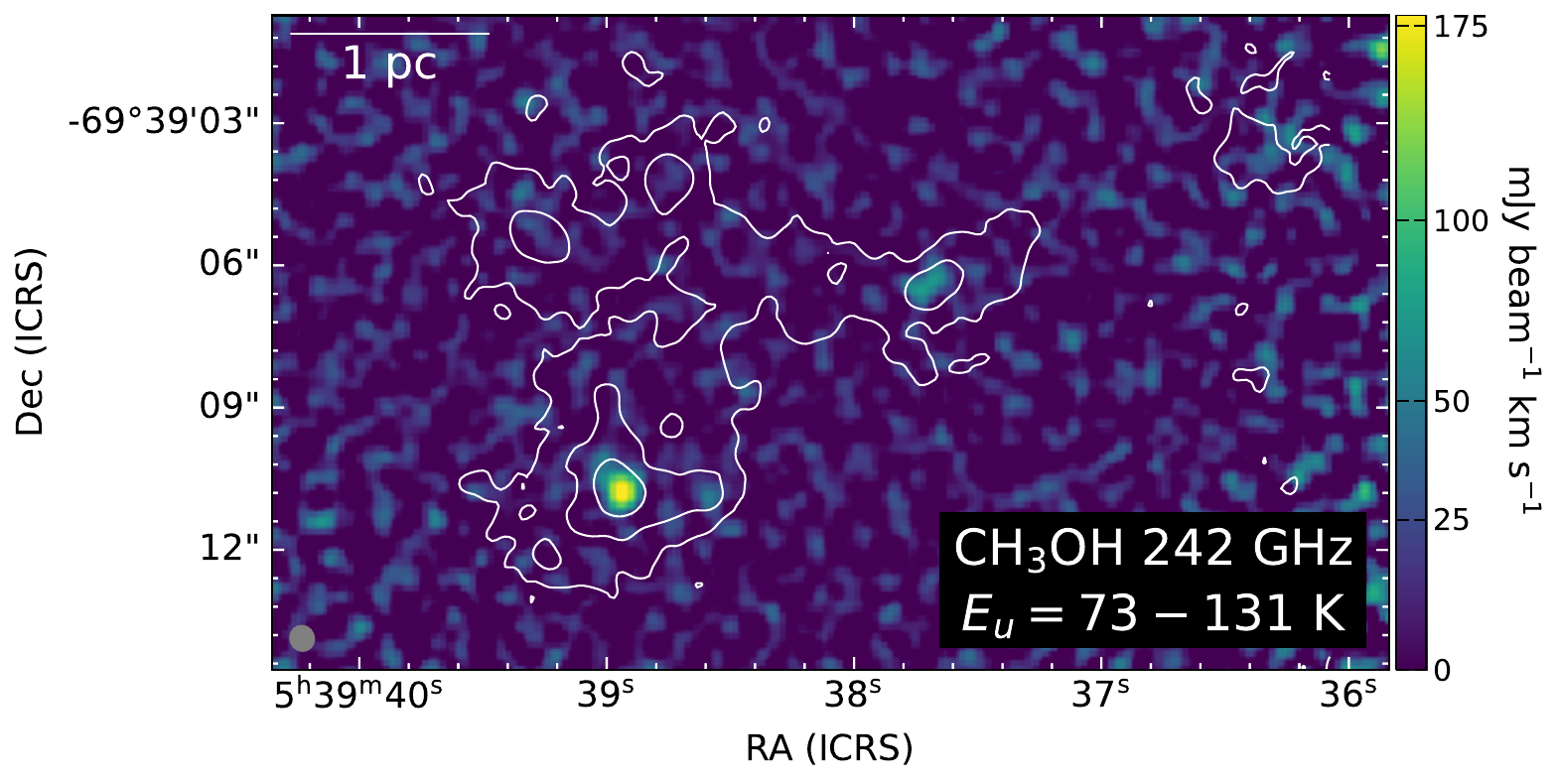} \\
    \includegraphics[width=\columnwidth]{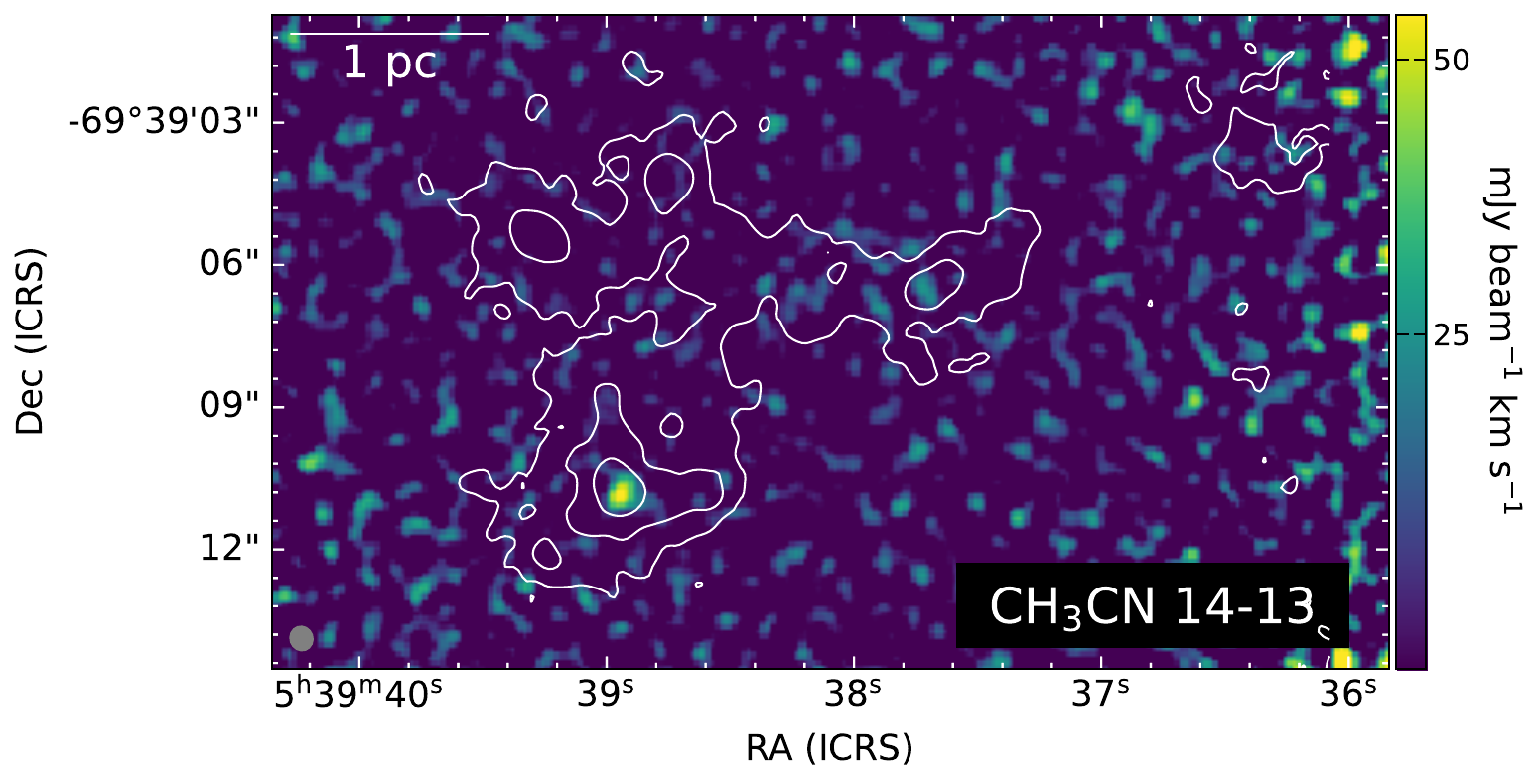}
    \caption{Integrated intensity maps of all COMs detected toward N\,160A--mm. This includes CH$_3$OH and CH$_3$CN emission. Contours represent 1\%, 5\%, and 20\% of 1.2mm continuum peak intensity. The methanol emission in the first panel was created using the 5$_{-0, 5}$--4$_{-0, 4}$ E, 5$_{1, 5}$--4$_{1, 4}$ E, 5$_{0, 5}$--4$_{0, 4}$ A, 5$_{2, 3}$--4$_{2, 2}$ E, and 5$_{-2, 4}$--4$_{-2, 3}$ E transitions. The second panel methanol map was created using the 5$_{-4,1}$--4$_{-4,0}$ E, $5_{3,3}$--$4_{3,2}$ A, $5_{3,2}$--$4_{3,1}$ A, $5_{2,4}$--$4_{2,3}$ A, $5_{-3,3}$--$4_{-3,2}$ E, and $5_{3,2}$--$4_{3,1}$ E transitions. The CH$_3$CN map was created using the $K=$ 0, 1, and 3 components of the $14_K-13_K$ ladder. The ALMA synthesized beam size is shown as the gray ellipse in the lower left (Table \ref{t:contcubedata}).}
    \label{fig:Mom0COMs}
\end{figure}

\begin{figure*}
\includegraphics[width=.48\textwidth]{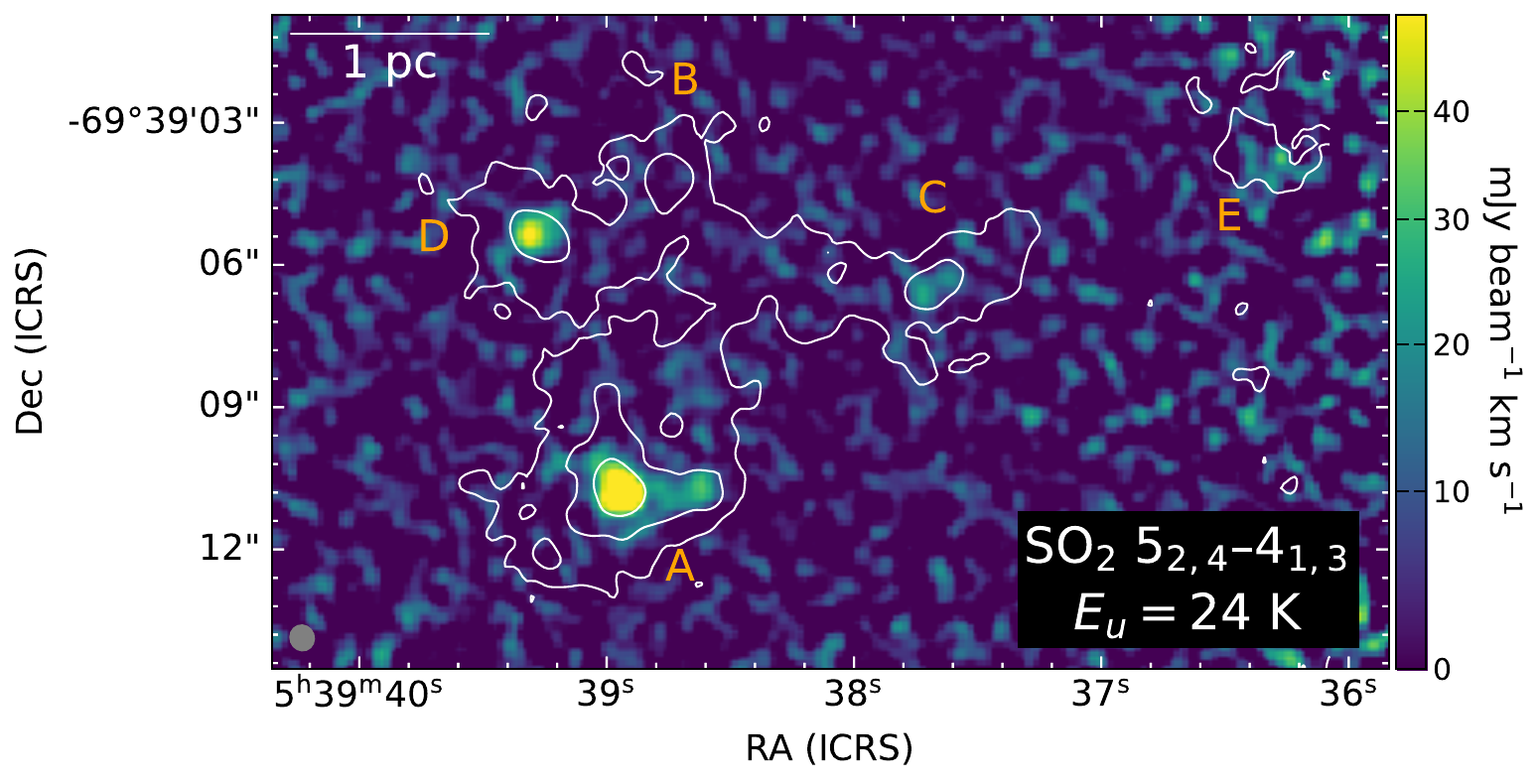} \hfill 
\includegraphics[width=.48\textwidth]{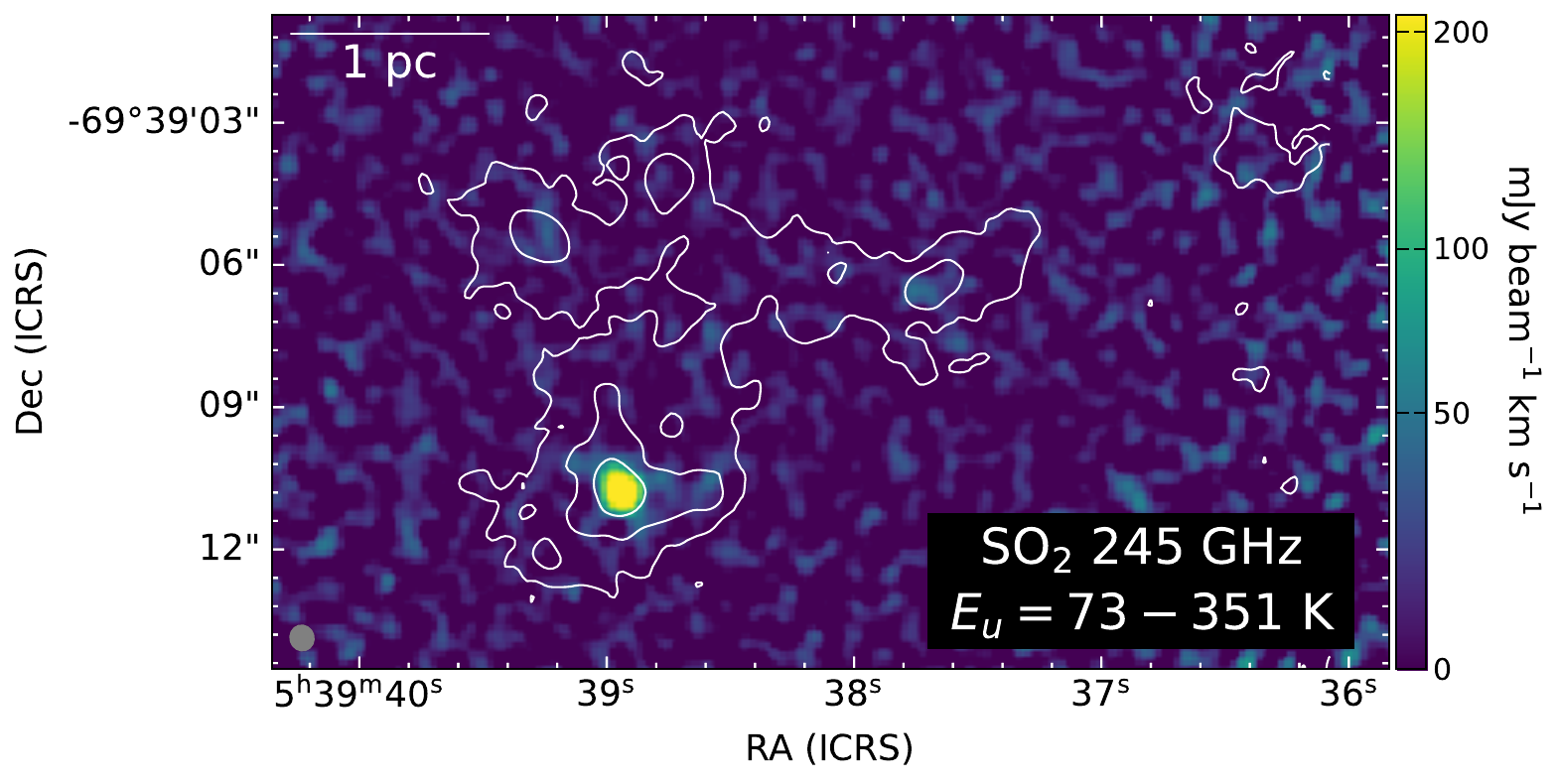} \hfill\\
\includegraphics[width=.48\textwidth]{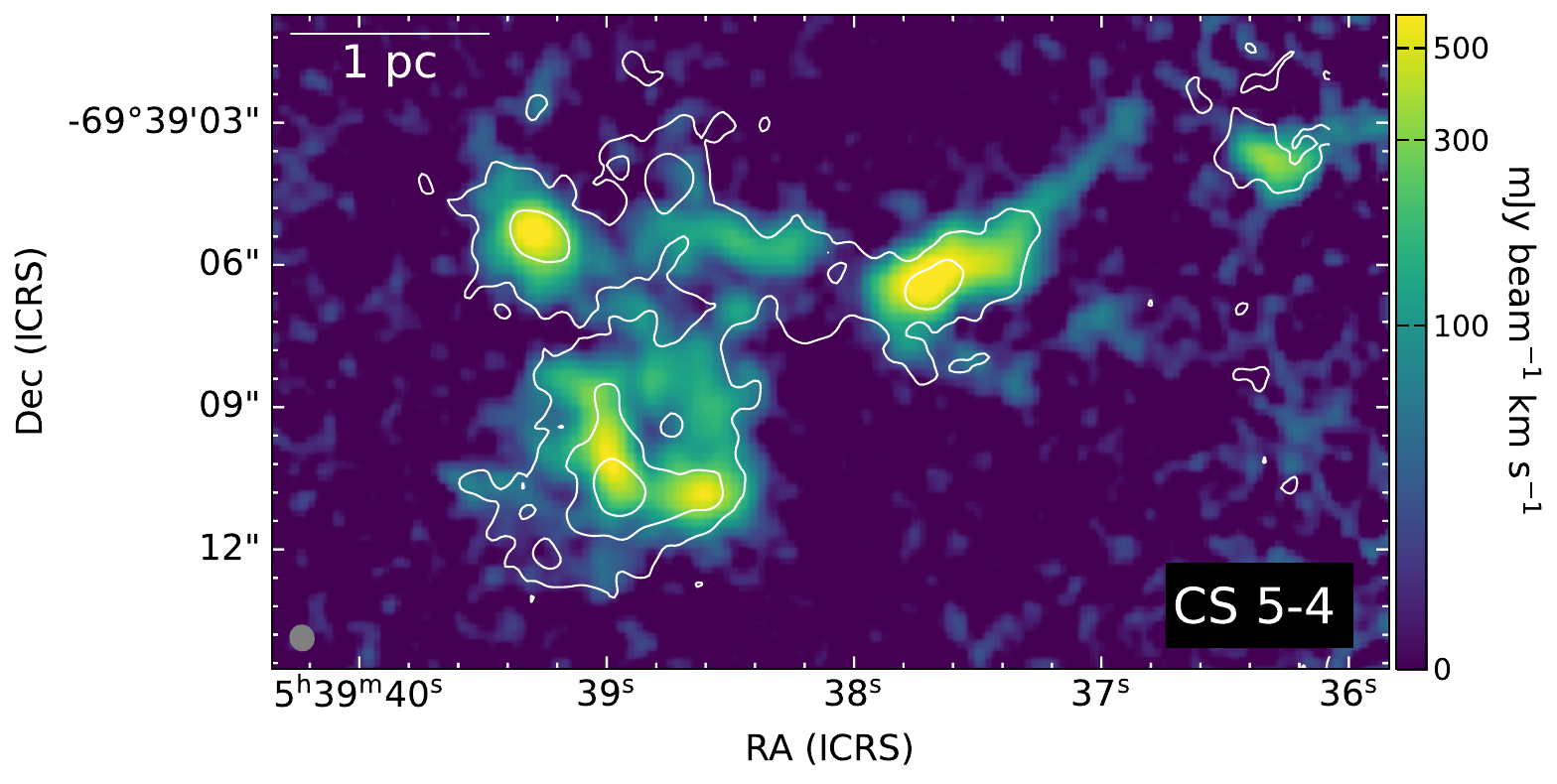} \hfill 
\includegraphics[width=.48\textwidth]{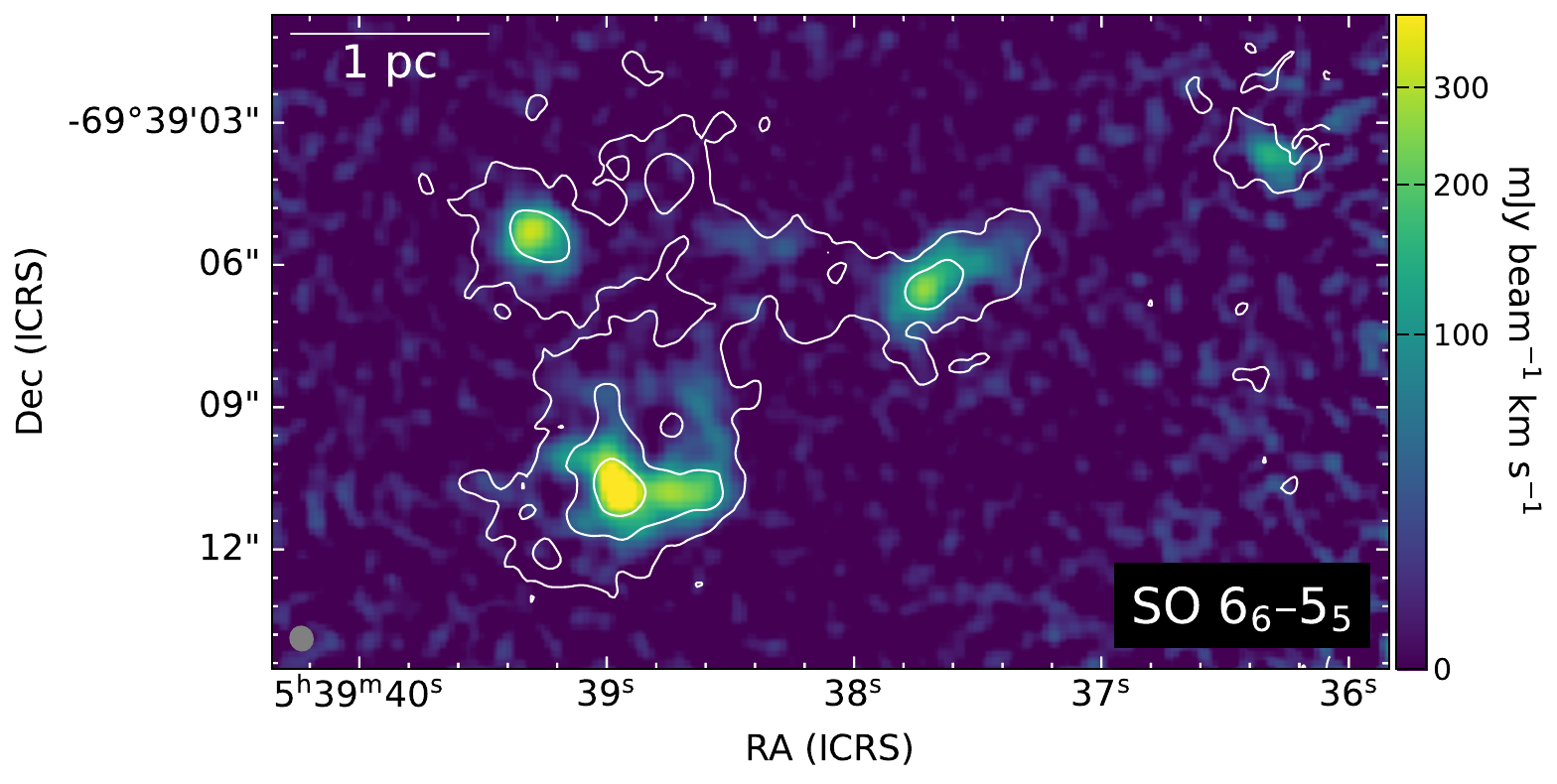} \hfill \\
\includegraphics[width=.48\textwidth]{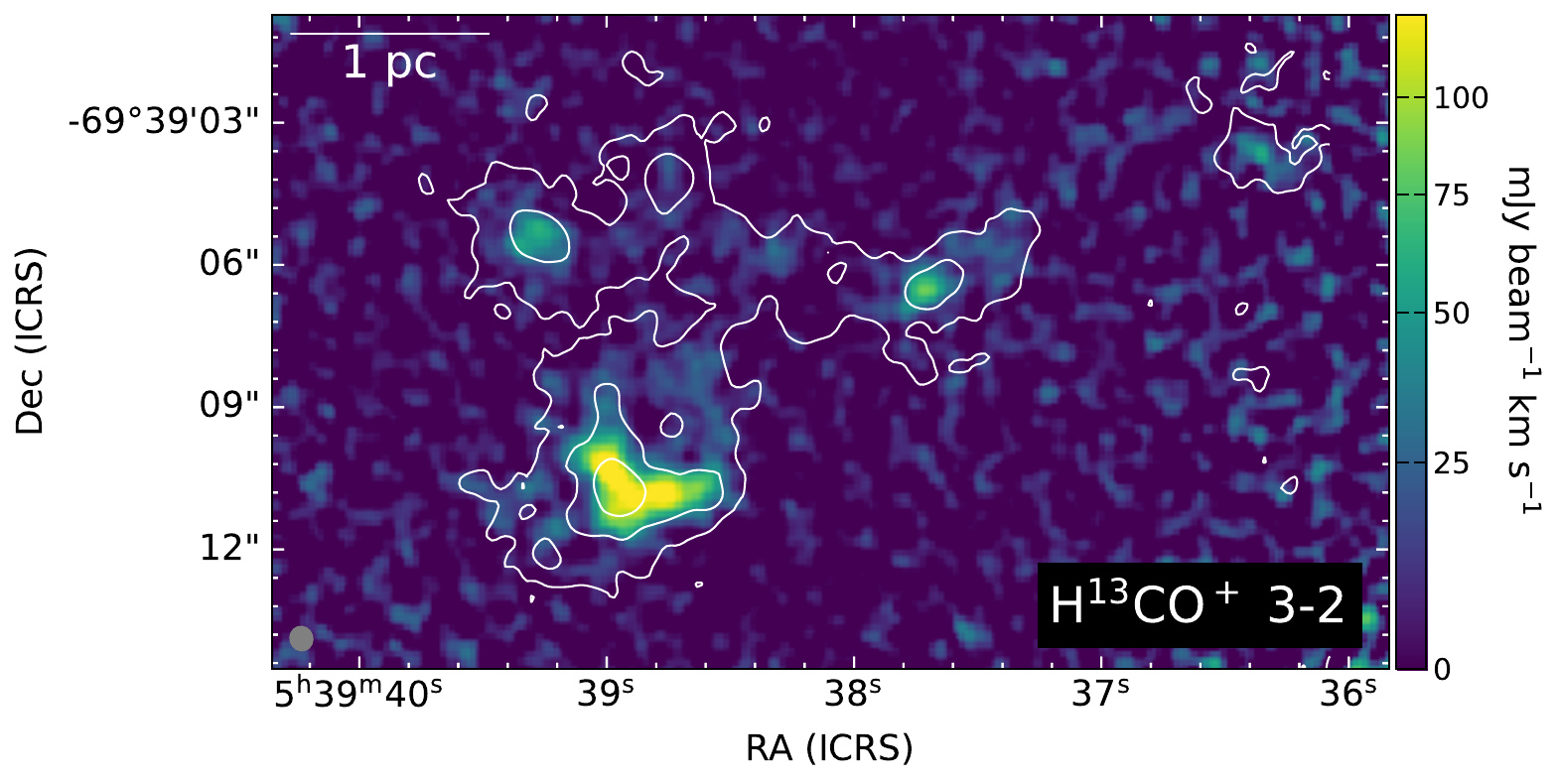} \hfill 
\includegraphics[width=.48\textwidth]{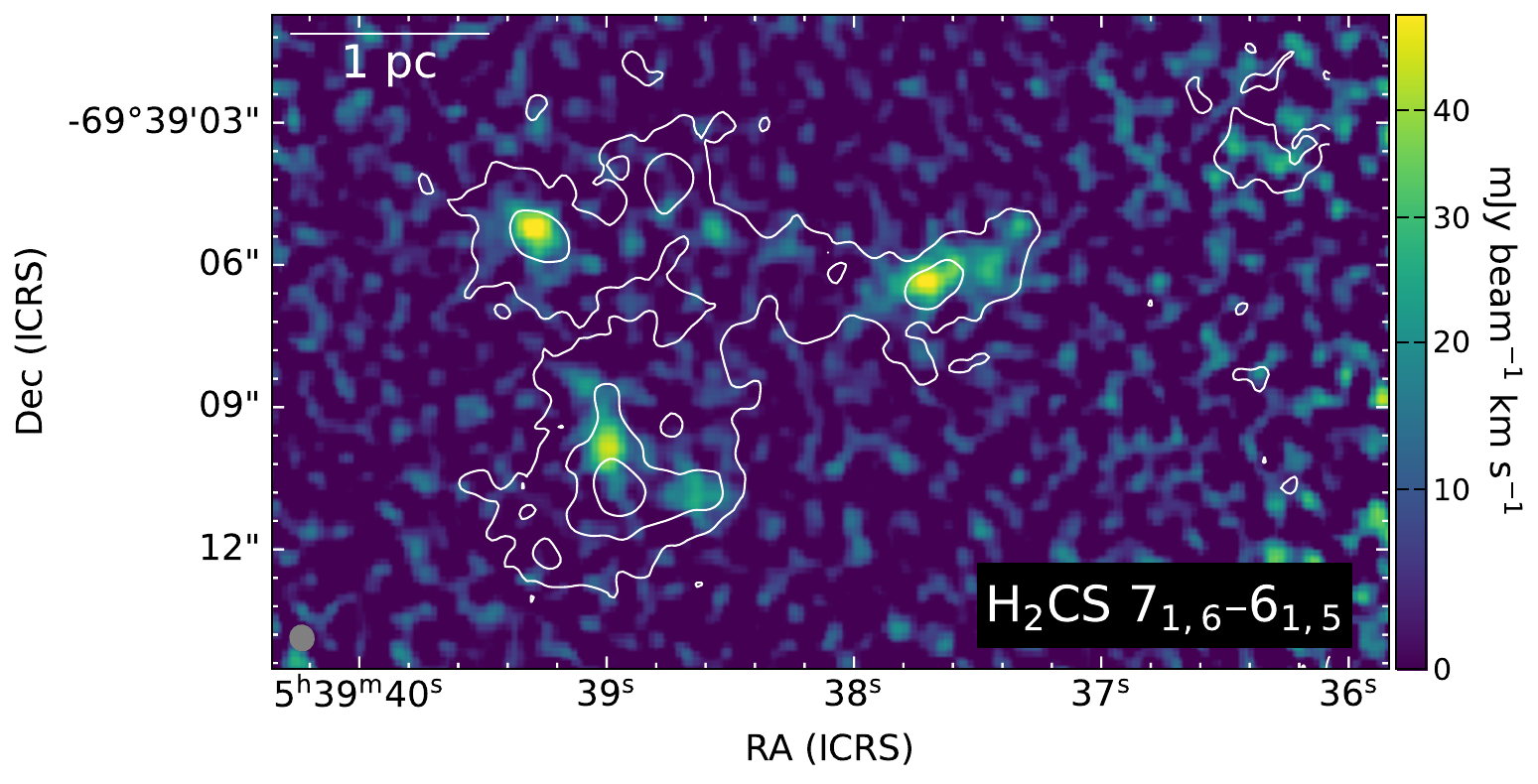} \hfill \\
\includegraphics[width=.48\textwidth]{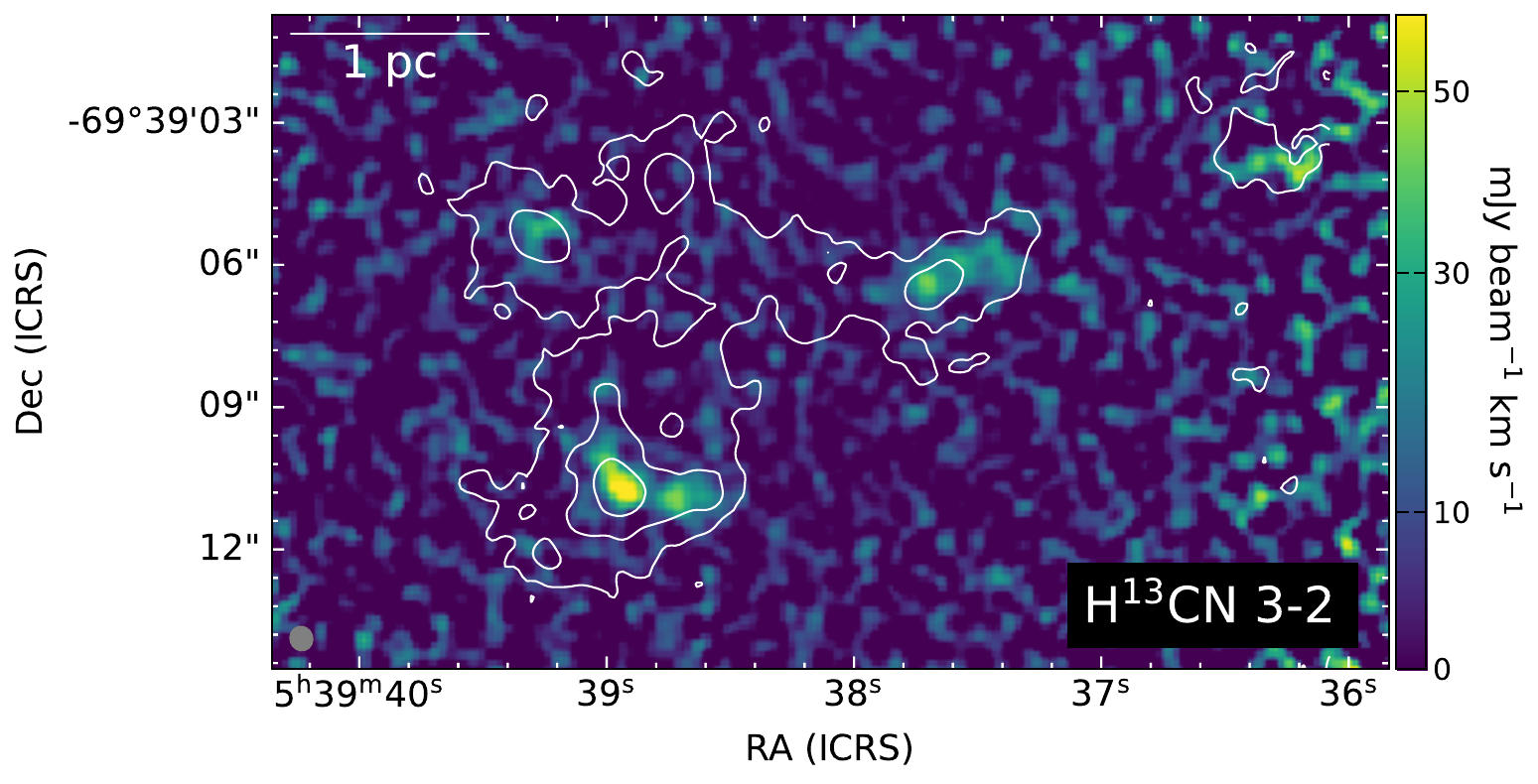} \hfill
\includegraphics[width=.48\textwidth]{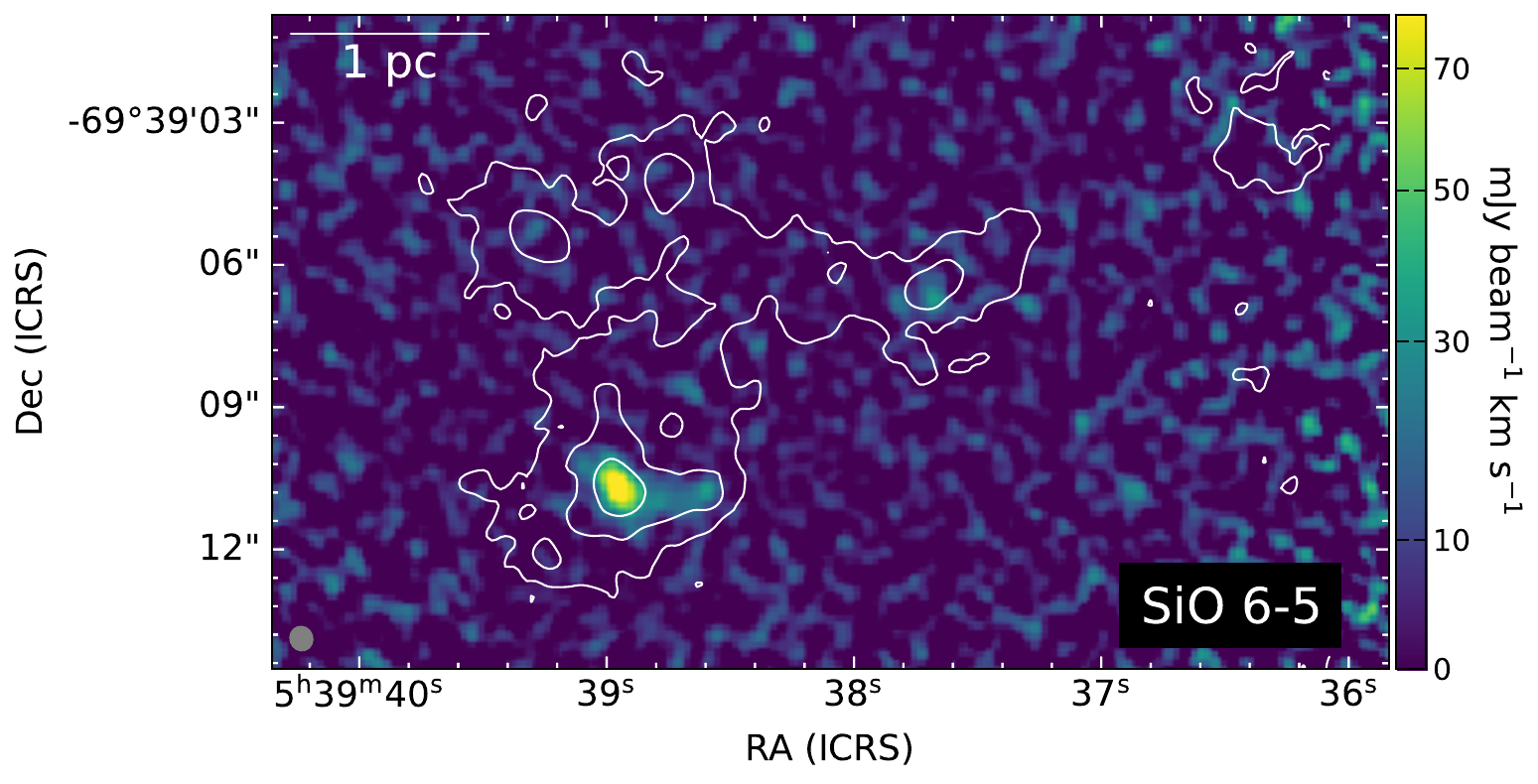} \hfill \\
\includegraphics[width=.48\textwidth]{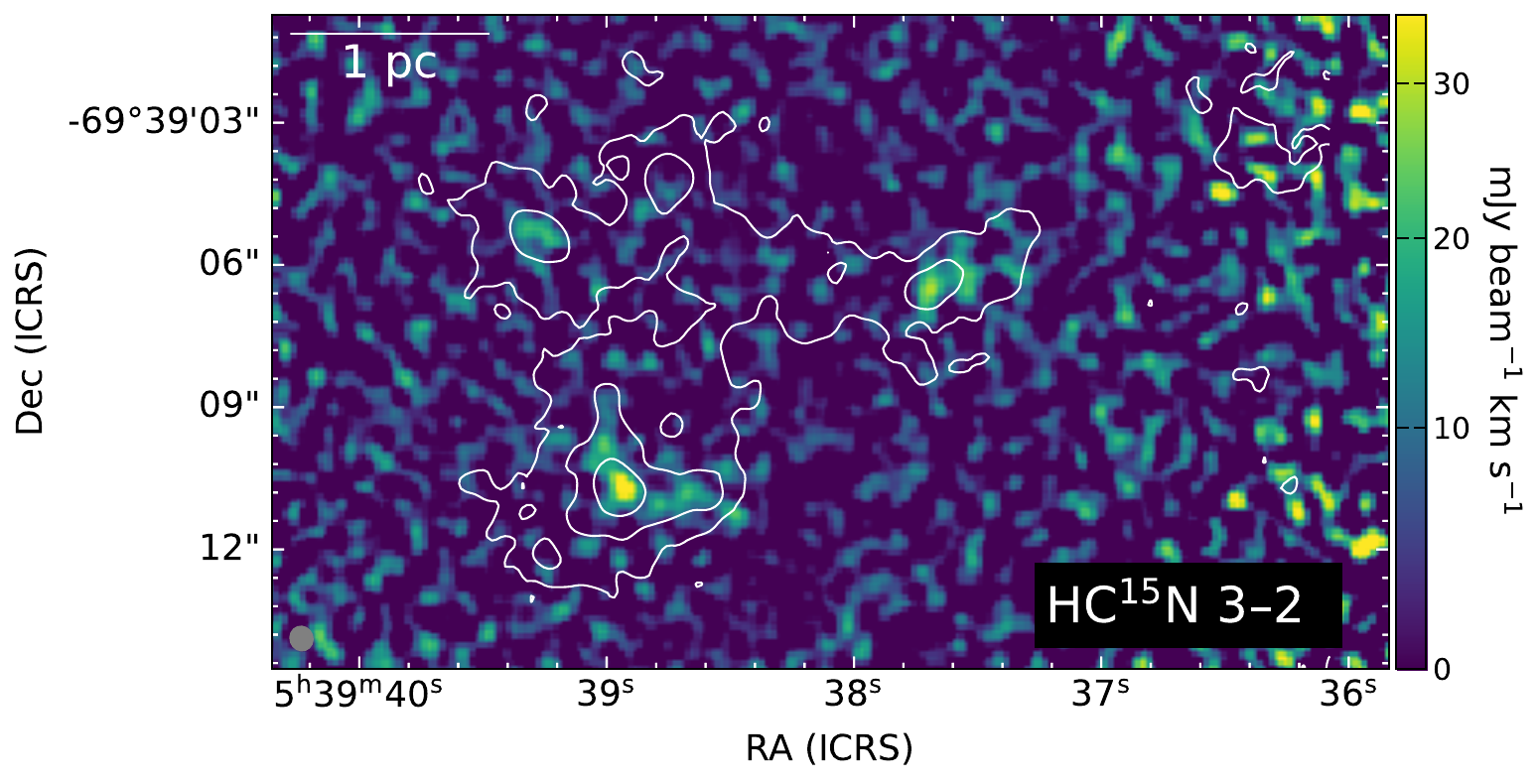}\hfill
\includegraphics[width=.48\textwidth]{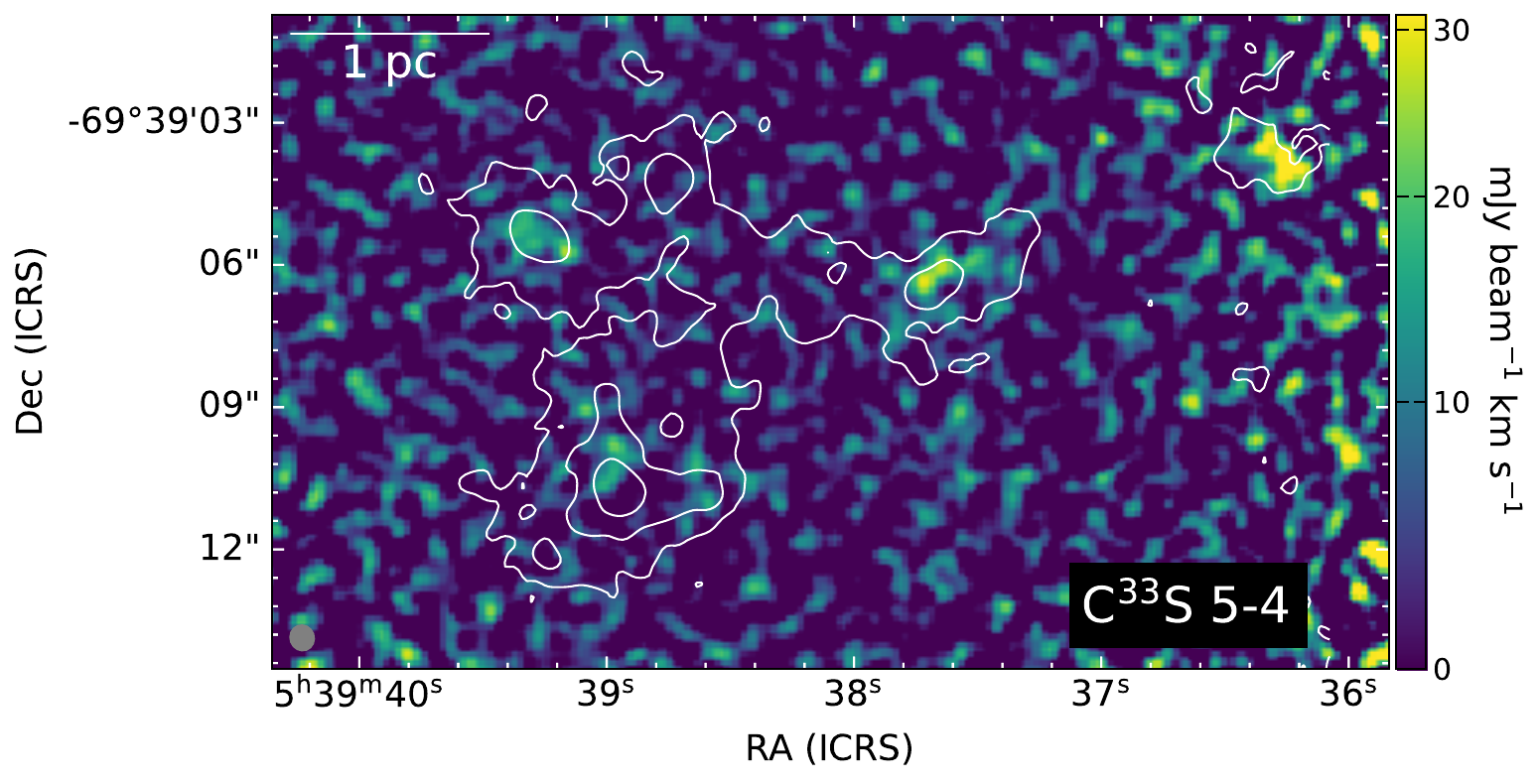}\hfill
\caption{Integrated intensity maps of N\,160A--mm for brightest lines detected toward multiple sources (SO$_2$, CS, SO, H$^{13}$CO$^+$, H$_2$CS, H$^{13}$CN, SiO, HC$^{15}$N, and C$^{33}$S). The SO$_2$ map in the first panel was created using the single transition $5_{2,4}$--$4_{1,3}$. The SO$_2$ map in the second panel was created using all transitions detected in the 245 GHz window ($14_{0,14}$--$13_{1,13}$, 26$_{3, 23}$--25$_{4, 22}$, 10$_{3, 7}$--10$_{2, 8}$). Contours represent 1\%, 5\%, and 20\% of 1.2mm continuum peak intensity. The ALMA synthesized beam size is shown as the gray ellipse in the lower left. 
\label{fig:Mom0Maps}}
\end{figure*}

\begin{figure*}
\centering
    \includegraphics[width=0.45\textwidth]{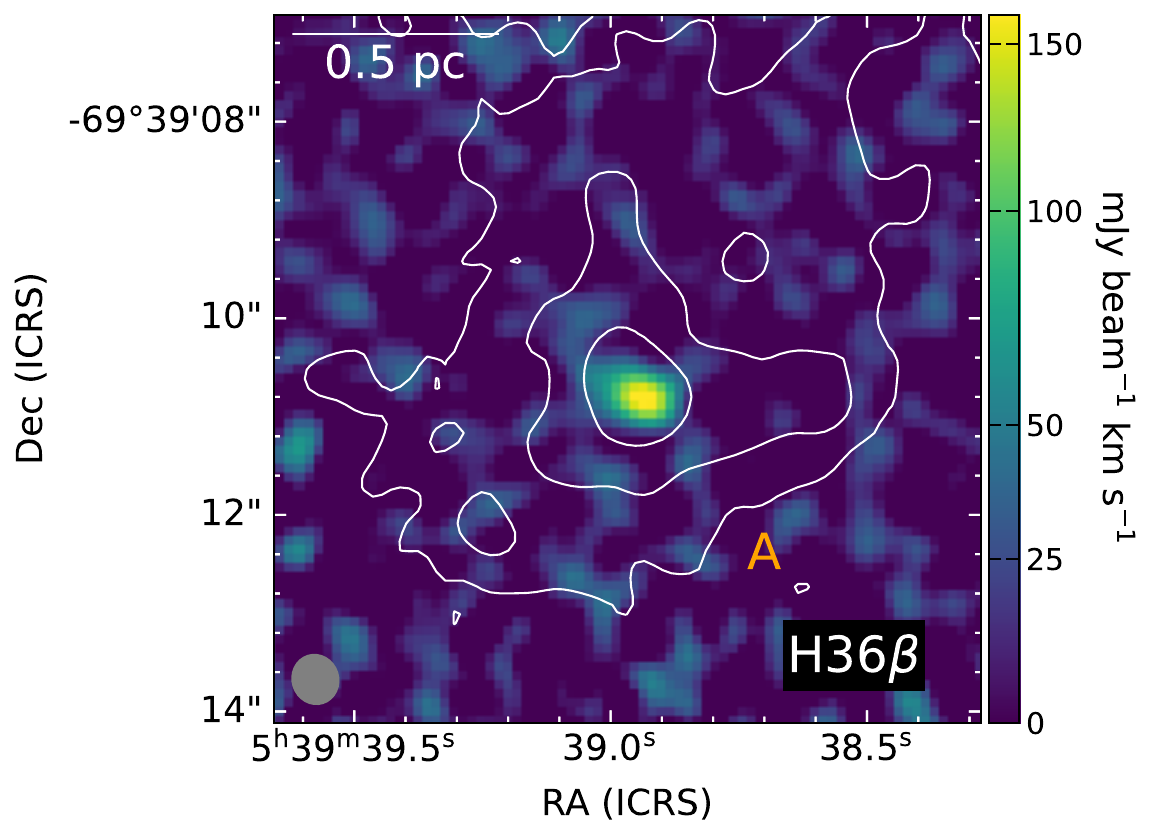}\hfill
    \includegraphics[width=0.45\textwidth]{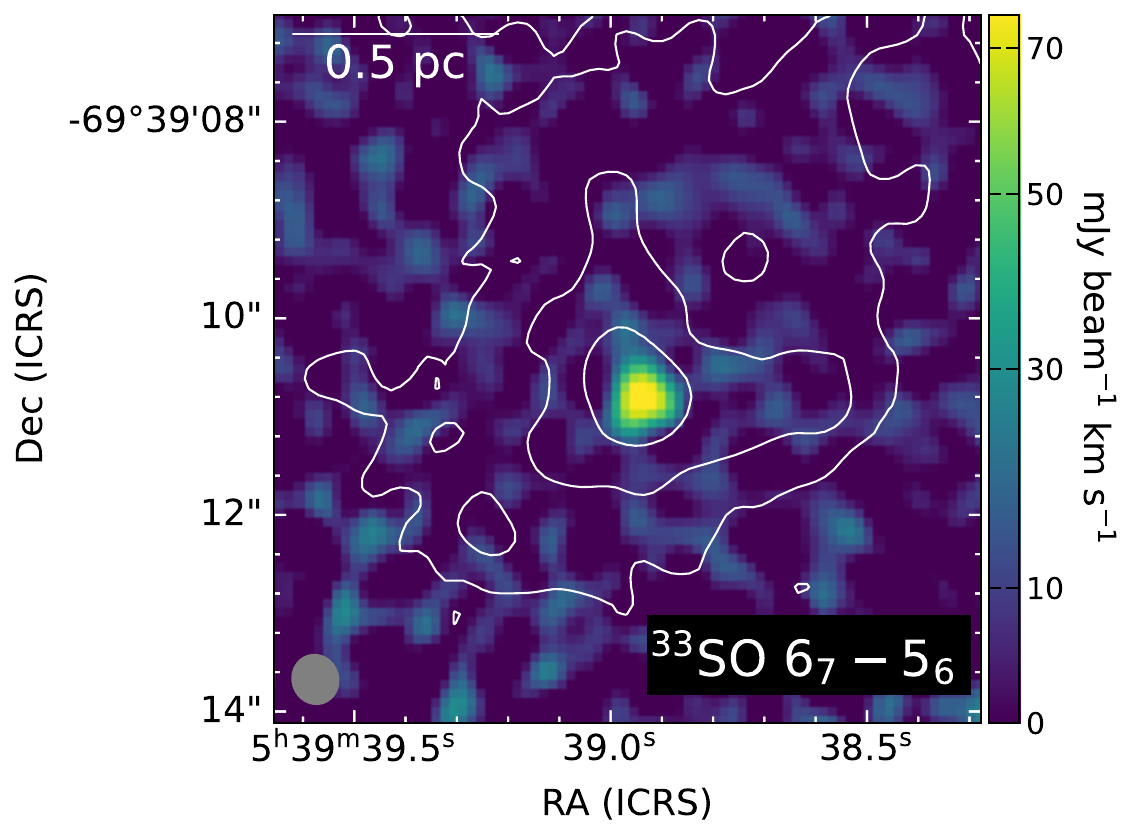}\hfill \\
    \includegraphics[width=0.45\textwidth]{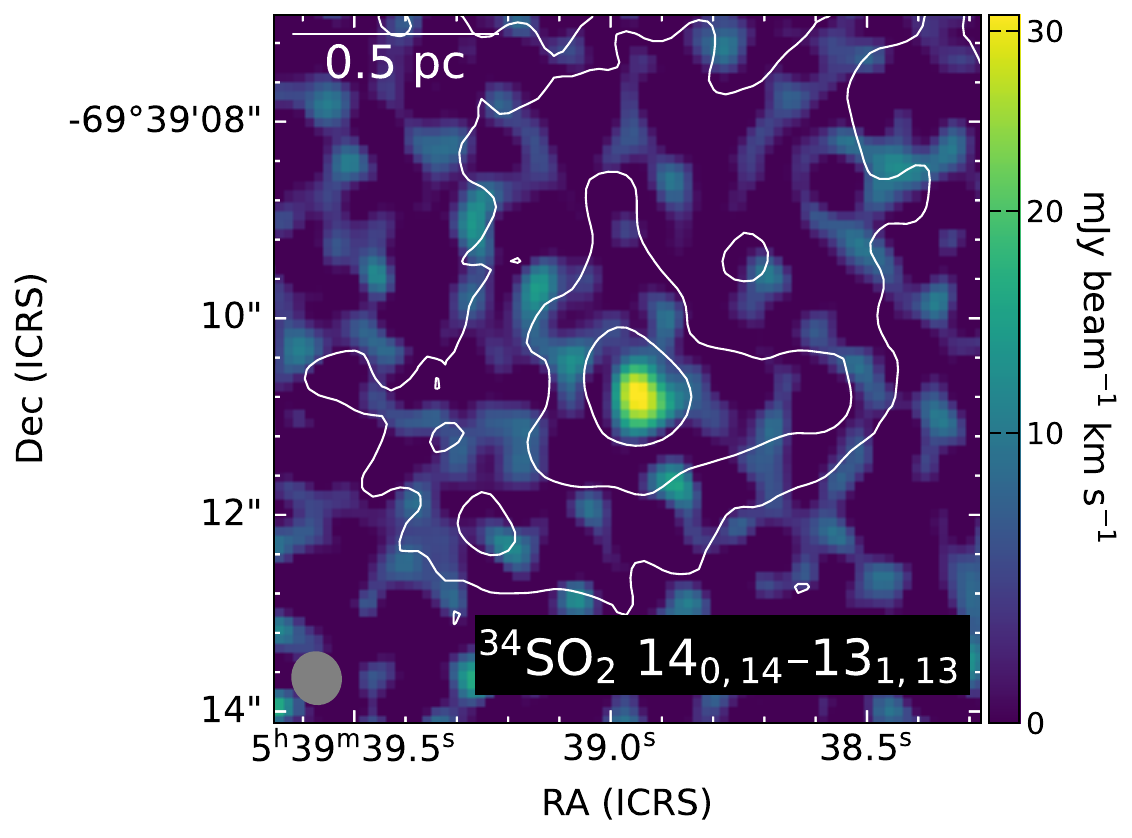}\hfill
    \includegraphics[width=0.45\textwidth]{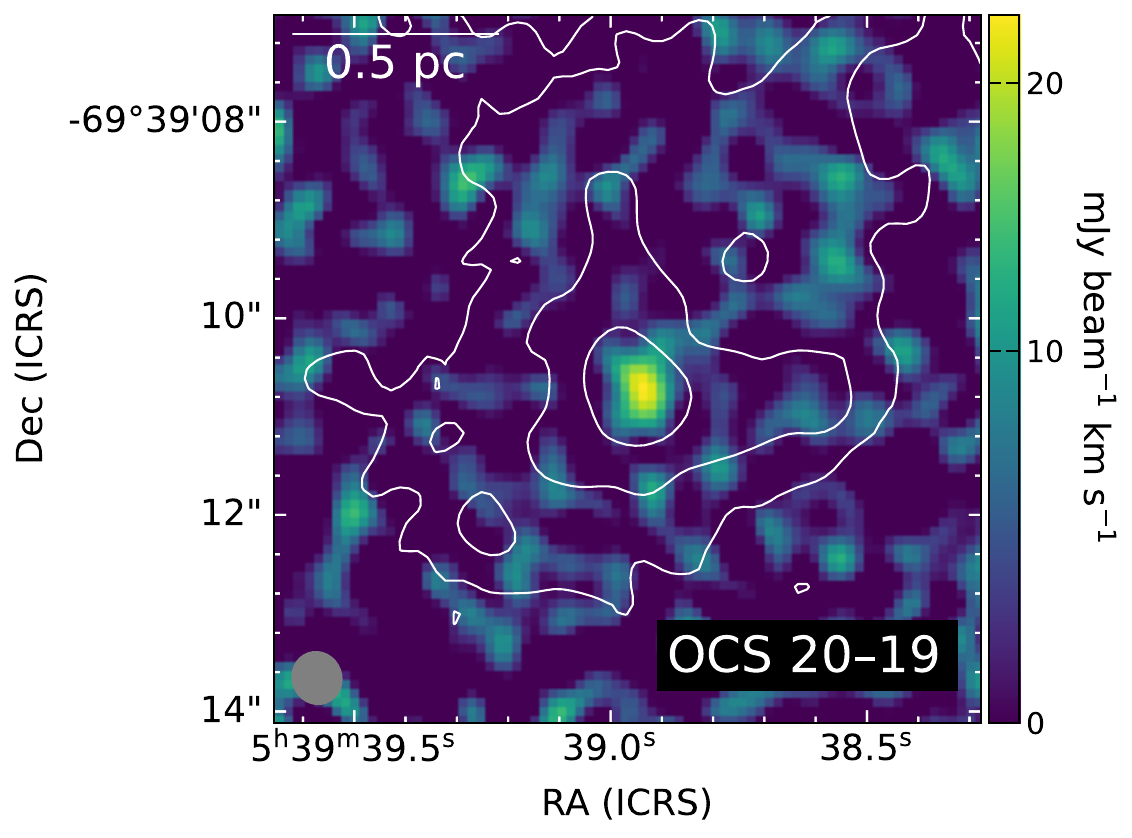}\hfill
    \caption{Integrated intensity maps for molecular species ($^{33}$SO, $^{34}$SO$_2$, OCS) and the H$36\beta$ recombination line detected toward source A only. The $^{33}$SO emission was created using the four blended transitions 6$_{7, 6}$--5$_{6, 5}$, 6$_{7, 7}$--5$_{6, 6}$, 6$_{7, 8}$--5$_{6, 7}$, and 6$_{7, 9}$--5$_{6, 8}$. The $^{34}$SO$_2$ map was created using the 14$_{0, 14}$--13$_{1, 13}$ transition. Contours represent 1\%, 5\%, and 20\% of the 1.2 mm continuum peak intensity. The ALMA synthesized beam size is shown as the gray ellipse in the lower left.}
    \label{fig:Mom0Amaps}
\end{figure*}

Figures \ref{fig:Mom0COMs} and \ref{fig:Mom0Maps} show the integrated intensity (moment 0) maps of all the species detected in the ALMA field N\,160A--mm toward at least one source (A--E). Figure \ref{fig:Mom0Amaps} shows the integrated intensity maps of molecules that were detected only toward source A. The maps were created by integrating over a single transition using channels where emission was present, except for CH$_3$OH, SO$_2$, CH$_3$CN, and $^{33}$SO. For CH$_3$OH, we made two integrated intensity maps using different sets of transitions. The first is a combination of the four brightest methanol lines in spectral window 242 GHz (5$_{-0, 5}$--4$_{-0, 4}$ E, 5$_{1, 5}$--4$_{1, 4}$ E, 5$_{0, 5}$--4$_{0, 4}$ A, and blended 5$_{2, 3}$--4$_{2, 2}$ E and 5$_{-2, 4}$--4$_{-2, 3}$ E) with the upper state energies ($E_u$) between 35 and 61 K. To make the second map, we used the higher $E_u$ transitions (73-131 K) from spectral window 242 GHz (blended 5$_{-4,1}$--4$_{-4,0}$ E, $5_{3,3}$--$4_{3,2}$ A, and $5_{3,2}$--$4_{3,1}$ A, blended $5_{2,4}$--$4_{2,3}$ A and $5_{-3,3}$--$4_{-3,2}$ E, and $5_{3,2}$--$4_{3,1}$ E). We also made two SO$_2$ integrated intensity maps. The first is based on a single transition ($5_{2,4}$--$4_{1,3}$) and the second utilizes all detected transitions in the 245 GHz window ($14_{0,14}$--$13_{1,13}$, 26$_{3, 23}$--25$_{4, 22}$, 10$_{3, 7}$--10$_{2, 8}$). The different CH$_3$OH and SO$_2$ maps highlight regions with detections of lower and higher excitation transitions. Additionally, the CH$_3$CN map was created using the $K=$ 0, 1, and 3 components of the $14_K-13_K$ ladder. The map for $^{33}$SO was integrated over the four blended detections in spectral window 260 GHz.

The distribution of the molecular line emission is most complex toward source A. It is associated with the CH$_3$OH, CH$_3$CN, SO$_2$, SO, H$^{13}$CN, SiO, HC$^{15}$N, H36$\beta$, $^{33}$SO, $^{34}$SO$_2$, and OCS emission, all with peaks coinciding with the 1.2 mm continuum emission. To the west, there are additional CH$_3$OH, SO$_2$, SO, and H$^{13}$CN emission peaks as well as CS and H$^{13}$CO$^+$ emission. These are offset from the 1.2 mm continuum emission peak of source A by approximately 1$\rlap.{''}$73/0.41 pc (CH$_3$OH, CS, SO$_2$) and 1$\rlap.{''}$04/0.25 pc (H$^{13}$CO$^+$, SO, H$^{13}$CN). To the north, there are CH$_3$OH, CS, and H$^{13}$CO$^+$ emission peaks offset from the 1.2 mm continuum emission peak of source A with offsets ranging from 0$\rlap.{''}$61/0.15 pc to 0$\rlap.{''}$80/0.19 pc. These regions are associated with source A through surrounding extended emission. In our analysis of the extracted spectra, H$_2$CS was not detected toward source A, although Figure \ref{fig:Mom0Maps} shows there is emission to the north and west corresponding with the northern and western peaks described above. There is no C$^{33}$S emission associated with source A.

No compact molecular line emission is associated with source B. The only detection in this region is extended CS emission, likely originating in the background.

Toward source C, the CH$_3$OH, SO$_2$, CS, SO, H$^{13}$CO$^+$, H$^2$CS, H$^{13}$CN, HC$^{15}$N, and C$^{33}$S emission peaks coincide with the 1.2 mm continuum. Toward source D, the CH$_3$OH, SO$_2$, CS, SO, H$^{13}$CO$^+$, and H$^2$CS emission are coincident with the 1.2 mm continuum. Extended emission is visible surrounding each source in CH$_3$OH, CS, and SO. There is a filamentary structure that appears to be stretching from source C to source E that is visible in both continuum and molecular emission, particularly in CS.

Sources E and F are located toward the edge of the field of view and are therefore in an area with higher noise. CH$_3$OH, CS, SO, and C$^{33}$S emission peaks are all detected toward E. Source F has only one detection, CS, but due to the high noise we do not include its integrated intensity map.

\subsubsection{Rotational Diagram Analysis} \label{s:rotational diagram}
We perform the rotational diagram analysis (e.g. \citealt{sutton1995, goldsmith1999}) for sources with the detection of multiple methanol or SO$_2$ transitions across a wide upper state energy range, to make the initial estimates of the rotational temperatures and total column densities for the more robust XCLASS analysis (see Section \ref{xclass}). Multiple methanol transitions are available for sources A, C, D, E; only source A has a sufficient number of the SO$_2$ transitions suitable for the analysis. 

The rotational diagram analysis assumes that the gas is in local thermodynamic equilibrium (LTE) and that the lines are optically thin (e.g., \citealt{goldsmith1999}). We select only lines that can be fit with a single Gaussian function to get the integrated line intensities; this excludes any lines that are blended with other nearby lines or lines with multiple velocity peaks. Using the integrated intensities we calculate the column density of molecules in the upper energy state ($N^{\rm{thin}}_u$) for each transition, 
\begin{equation}
    \frac{N_{\rm{u}}^{\rm thin}}{g_u} = \frac{3k\int T_{\rm mb}dv}{8\pi^3\nu S\mu^2} \, ,
\end{equation}
where $g_u$ is the degeneracy of the upper level, $k$ is the Boltzmann constant, $\int T_{\rm{mb}}dv$ is the integrated intensity, $\nu$ is the frequency of the transition, $S$ is the line strength, and $\mu$ is the dipole moment. The values for $S\mu^2$ were taken from the CDMS. Under the assumption of LTE, $\frac{N_{\rm{u}}^{\rm thin}}{g_u}$ is related to the total column density ($N^{\rm{total}}$) by the formula:
\begin{equation}
    \mathrm{ln} \left(\frac{N_u^{\rm{thin}}}{g_u} \right) = -\left( \frac{1}{T_{\rm{rot}}}\right)\left( \frac{E_{\rm u}}{k}\right) +\mathrm{ln}\left(\frac{N^{\rm{total}}}{Q(T_{\rm{rot}})} \right)
\end{equation}
where $T_{\rm{rot}}$ is the rotational temperature, $E_u$ is the upper state energy, and $Q(T_{\rm{rot}})$ is the partition function. The upper state energy and partition function values were also taken from CDMS. To get the partition function for the calculated temperature of the region, we interpolate the partition function values for the given temperatures in CDMS. In this form, we can fit a linear function to our data points and get the rotational temperature and column density from the slope and y-intercept, respectively.

\begin{figure*}
    \centering
    \includegraphics[width=0.48\textwidth]{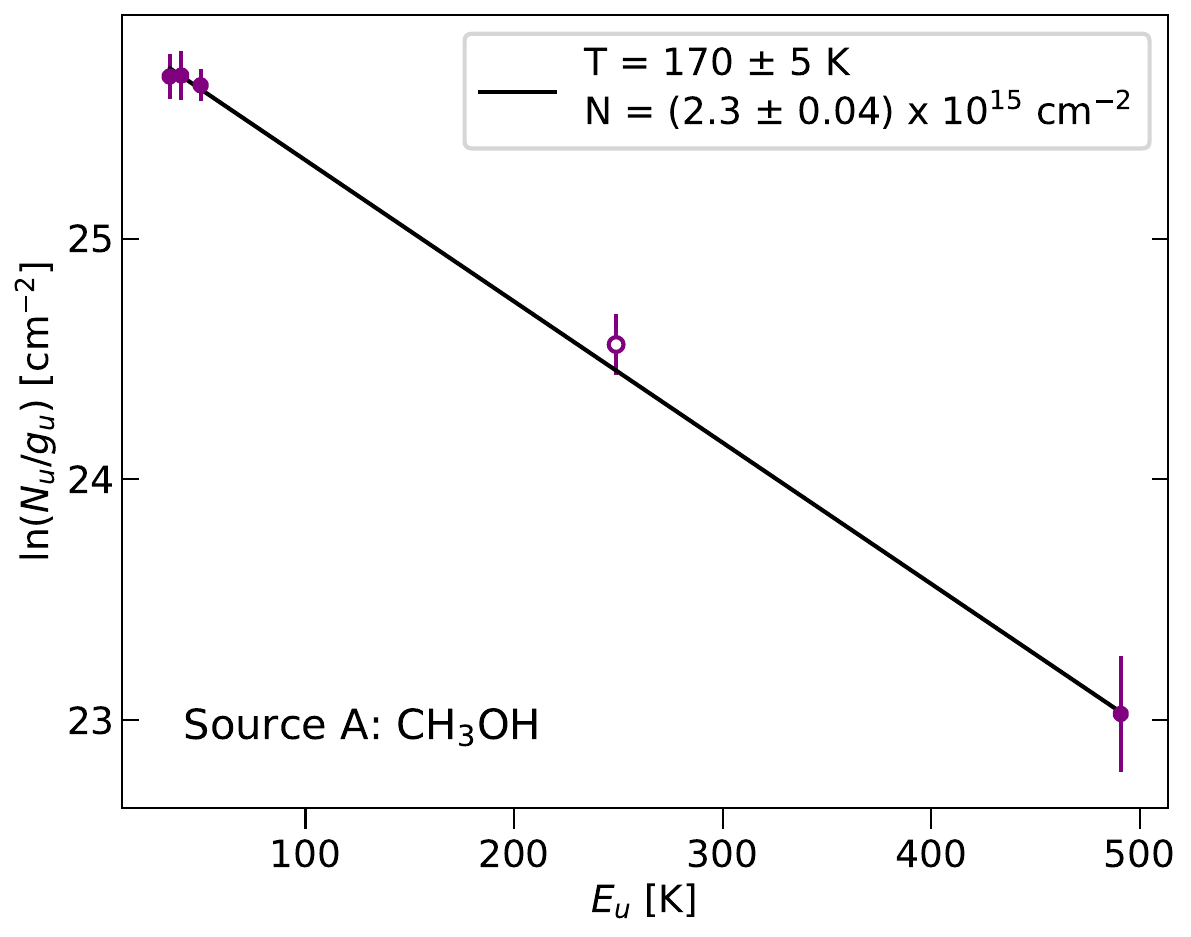} \includegraphics[width=0.48\textwidth]{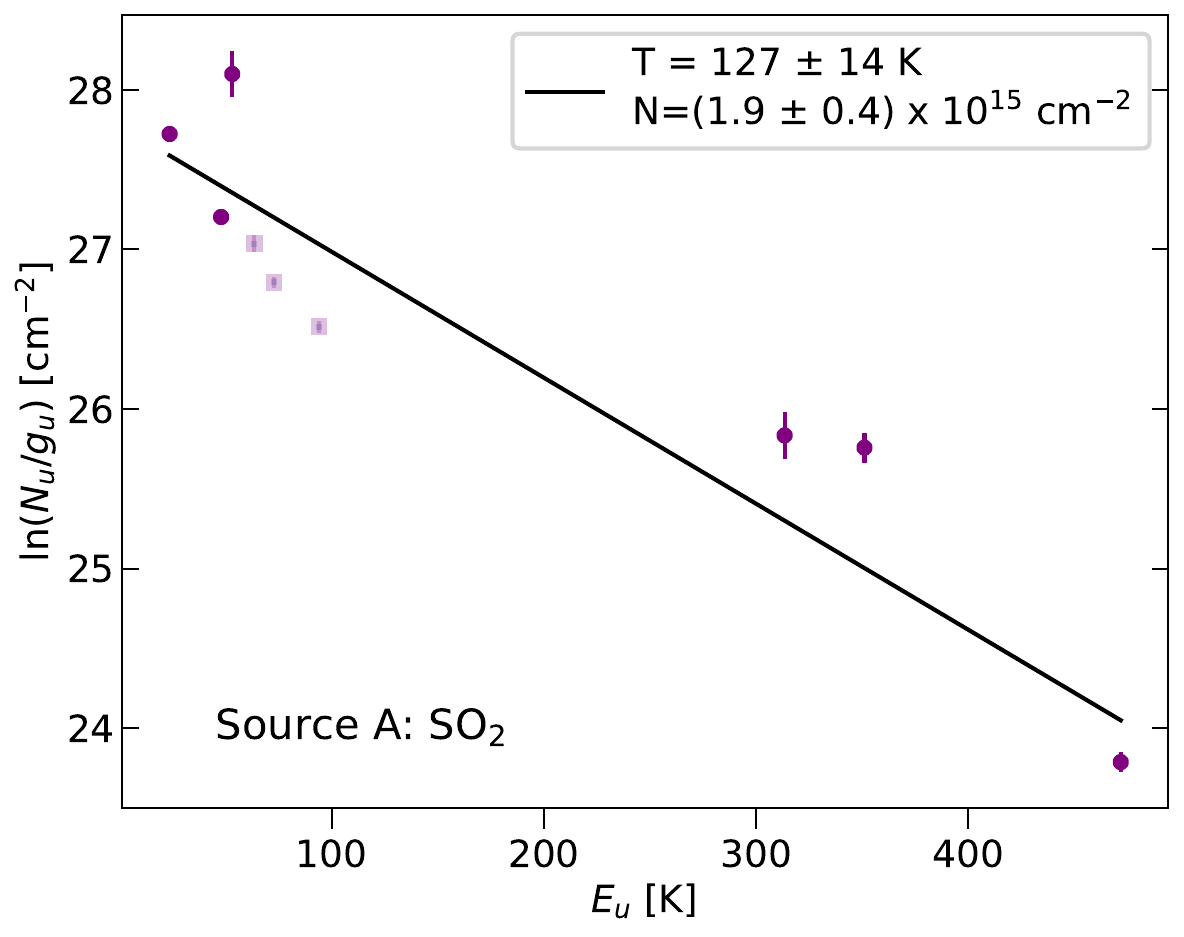} \\
    \includegraphics[width=0.3\textwidth]{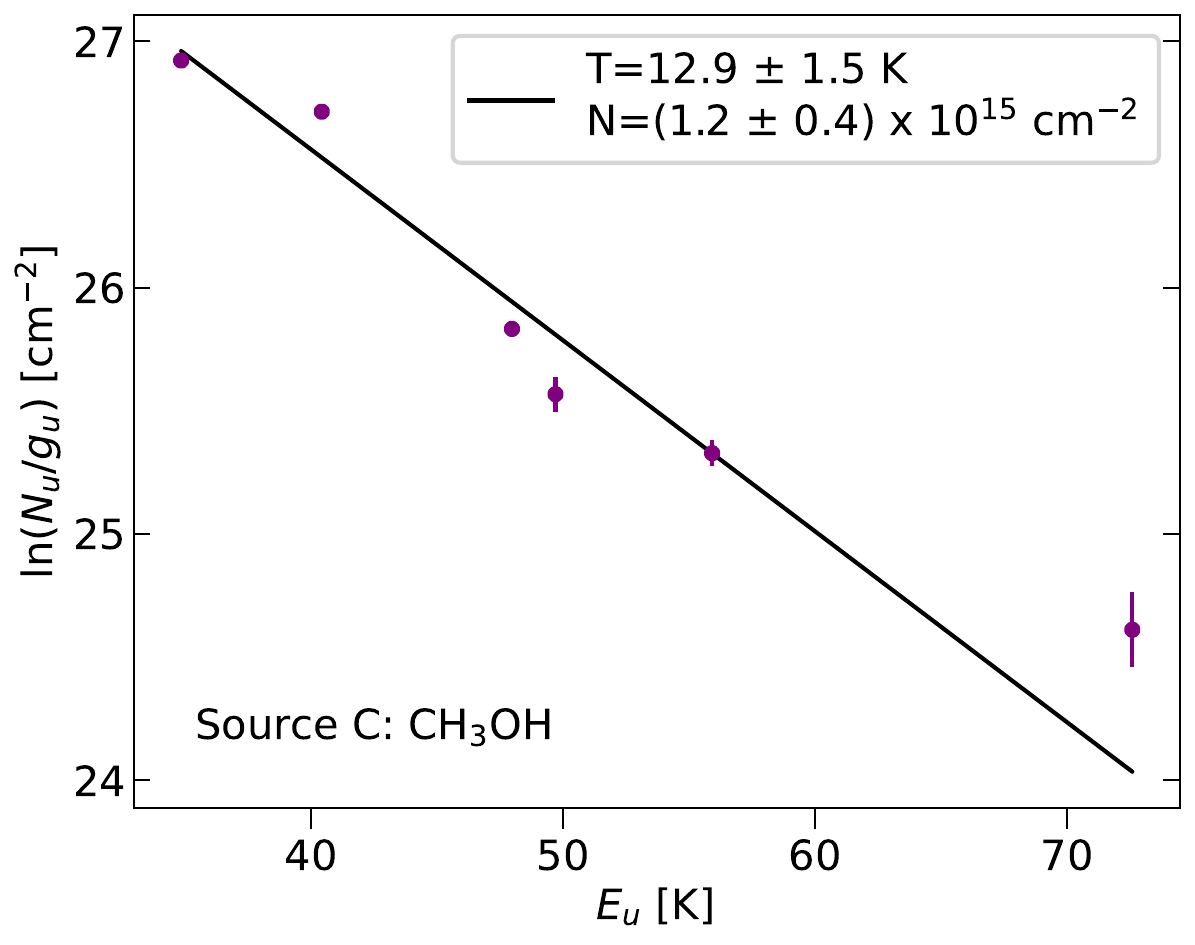} \includegraphics[width=0.3\textwidth]{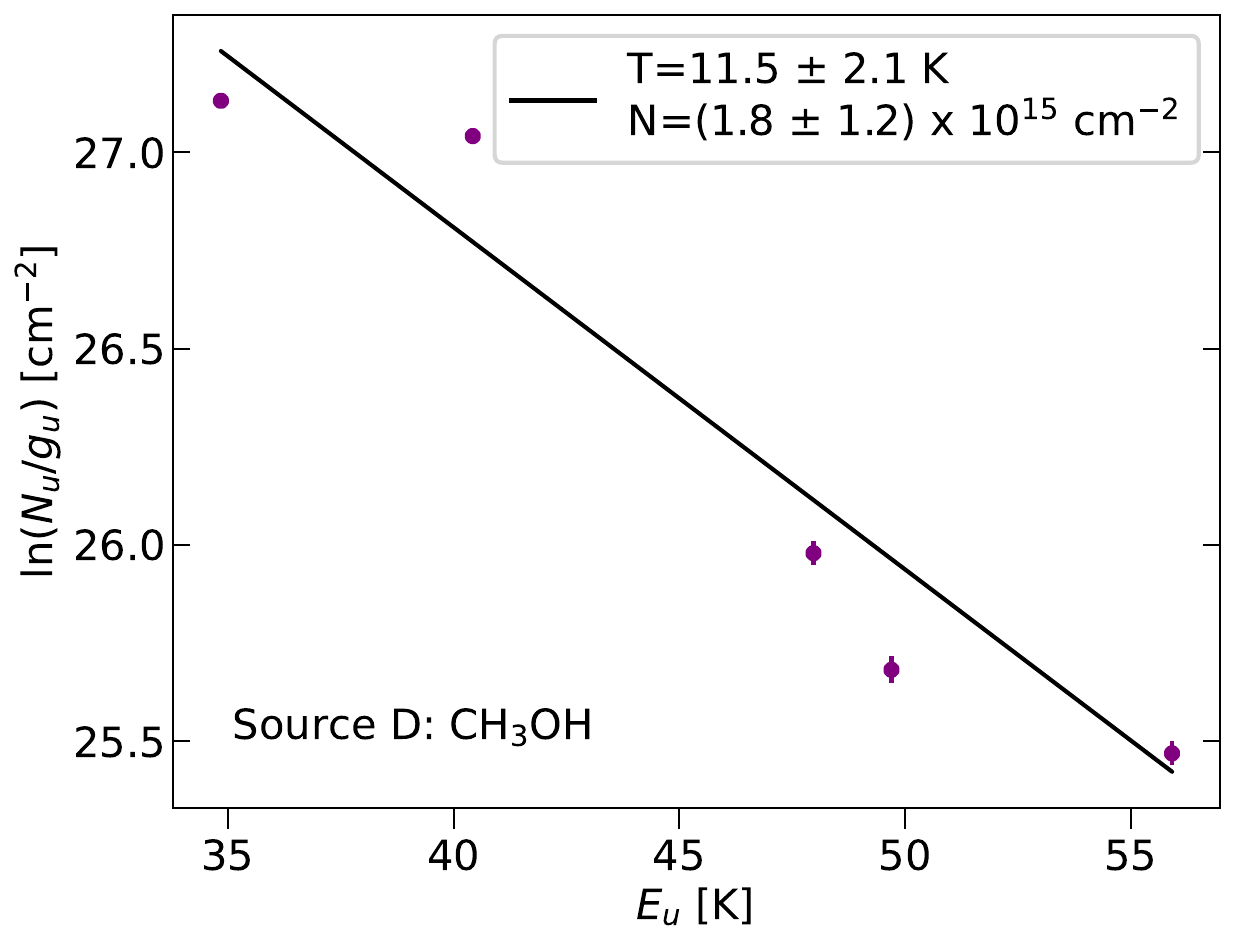} \includegraphics[width=0.3\textwidth]{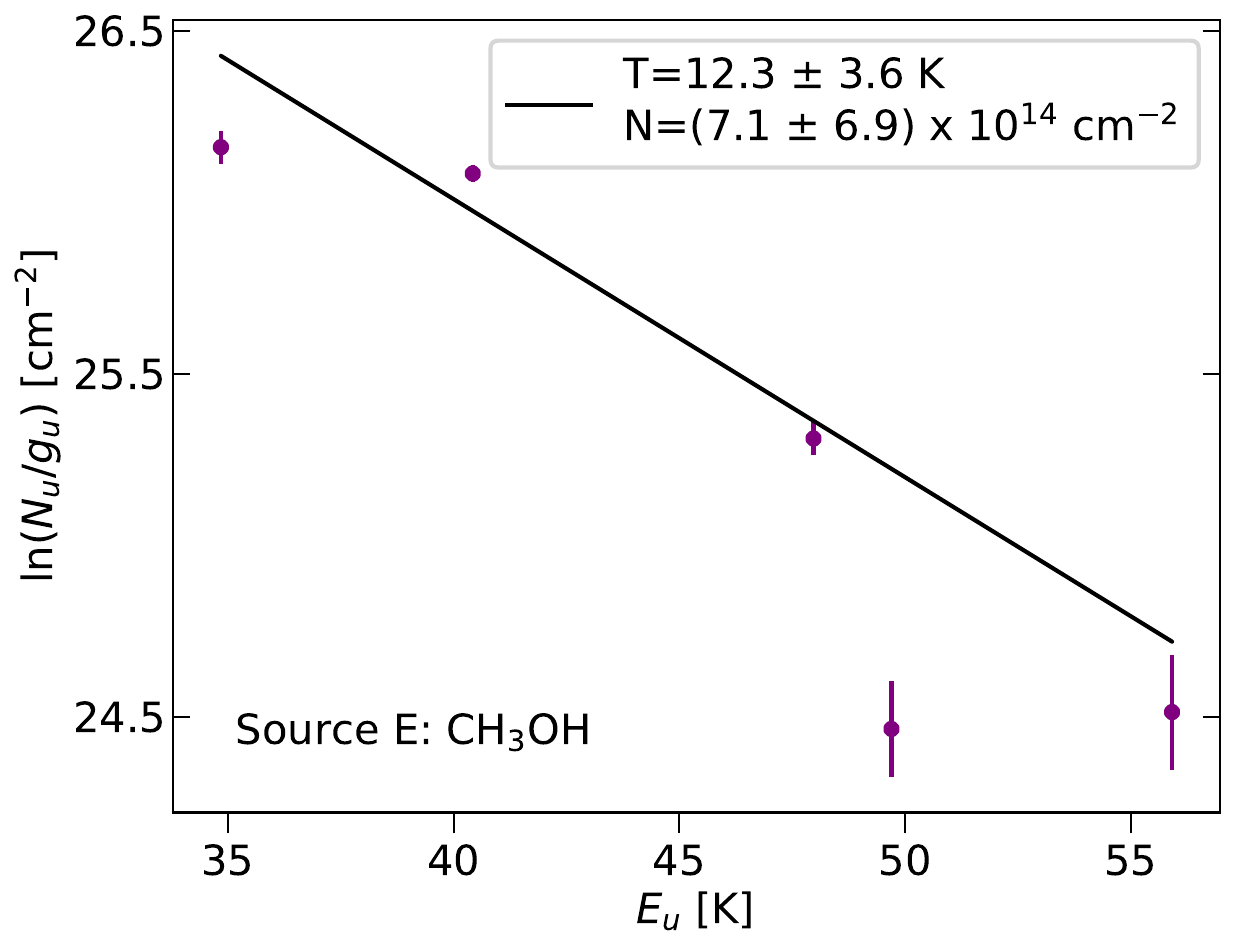}
    \caption{\textit{Upper panel}: Rotational diagrams for source A for methanol and SO$_2$. Rotational temperatures above 100 K indicate the presence of a hot core. The open circle in the CH$_3$OH plot is a tentative detection and was not included in the fit. The three pale squared markers in the SO$_2$ plot represent the three transitions that are likely impacted by opacity effects and are not included in the fit. \textit{Lower panel}: Rotational diagrams for sources C, D, and E for methanol. Sources C, D, and E are cold.}
    \label{fig:rot_diagrams}
\end{figure*}

Figure~\ref{fig:rot_diagrams} shows the CH$_3$OH rotational diagrams for sources A, C, D, and E, as well as the SO$_2$ rotational diagram for A. Three SO$_2$ transitions are excluded from the fit to avoid possible opacity effects. Specifically, these are SO$_2$ transitions with $E_{\rm u}<100$ K and log$(S\mu^2)>1$. The resulting methanol and SO$_2$ rotational temperatures and total column densities with uncertainties are provided in the plots. Our results indicate that sources C, D, and E are associated with cold methanol ($T_{\rm rot}<15$ K). For source A, both methanol and SO$_2$ are hot ($T_{\rm rot}>100$ K), indicating that the source is a hot core.

\subsubsection{Full Spectral Modeling} \label{xclass}

Using the rotational temperatures and column densities from the rotational diagram analysis as initial estimates, we perform a full spectral modeling analysis using the eXtended CASA Line Analysis Software Suite (XCLASS; \citealt{moller2017}, with additional extensions, T. M{\"o}ller, priv. comm., see also \citealt{Moller2023,sewilo2023}). For example, the extensions offer the possibility to treat molecules in non-LTE using the RADEX routines \citep{vandertak2007}.

XCLASS models a given spectrum by solving the radiative transfer equation for a one-dimensional object in either local thermodynamic equilibrium (LTE) or non-LTE based on user input. There can be any number of molecules with any number of components where each component is defined by being either a core component or a foreground component. Each molecule's component is defined by the following parameters: source size (in $''$), rotational temperature (K), column density (cm$^{-2}$), line width (km s$^{-1}$), and the velocity offset relative to the v$_{\rm LSR}$ (km s$^{-1}$). When fitting the spectra, these can either be fixed or allowed to vary as free parameters. Due to a degeneracy between source size and column density, we assume that the source is beam filling when a single component is present. For two components, one is beam filling and the other is smaller than the beam. 

Our fitting process has three steps. First, we fit only molecules that have multiple transitions detected, allowing the temperature to be determined. We assume that they can be well represented by a single core component in LTE. Next, if the fit is unsatisfactory, we either add a second component or use a single core component in non-LTE. Finally, once we are confident in our initial fit, we include all detected and tentatively detected molecules listed in Table \ref{t:detections}. Since only one transition is reliably detected for these species, it is not possible to determine the rotational temperature. We instead fix their rotational temperatures to the CH$_3$OH temperature from the initial fit or, for SO-bearing species, the SO$_2$ temperature if it was included in the initial run. We choose the CH$_3$OH temperature over the CH$_3$CN temperature for source A for the sake of consistency over the sample and because the CH$_3$CN detections are tentative and therefore the temperature determination is less reliable. For the molecules included in the initial fit, all the XCLASS input parameters are fixed to their best fit parameters for this final run. The spectra of the sources A, C, D, and E are analyzed using XCLASS; the spectral analysis was not possible for sources B and F as too few spectral lines were detected.

For each source, we estimate column density upper limits for molecules detected toward any other N\,160A--mm source. For source A, the only source associated with hot CH$_3$OH, we also estimate upper limits for COMs (HCOOCH$_3$, CH$_3$OCH$_3$, NH$_2$CHO, C$_2$H$_5$CN, CH$_3$CHO, C$_2$H$_5$OH) commonly detected toward Galactic and extragalactic (LMC and SMC) hot cores. To determine column density upper limits for a given molecule with XCLASS, we set its temperature, line width, and velocity offset to that of CH$_3$OH and their column density is allowed to vary. All other previously detected molecules are fixed to their best fit parameters.

We estimate the uncertainties of the parameters obtained in the XCLASS analysis using Monte Carlo noise resampling. We randomly injected noise into the final XCLASS model spectra for each source and passed that into XCLASS. The random noise is taken from a normal distribution multiplied by the rms in the spectrum ($\sigma_{\rm s,i}$) for a given source (s) and spectral window (i). We perform 300 iterations to obtain a sufficiently large number of parameter sets with the best fit parameters having a well-sampled Gaussian distribution. The final parameter errors are taken to be the standard deviation of the best fit results from these 300 runs. 

The resulting rotational temperatures ($T_{\rm rot}$), column densities ($N$), velocities ($v_{\rm LSR}$), and line widths ($\Delta v_{\rm FWHM}$) from the XCLASS fitting and error analysis can be found in Table \ref{t:XCLASS Params}, along with the estimated abundances with respect to H$_2$ (see Sections \ref{ss:h2columndensity} and \ref{ss:N160Abundances}). Below, we provide additional details on the XCLASS fitting for individual sources.

\noindent \textit{Source A:} The initial rotational diagram analysis indicated hot CH$_3$OH and SO$_2$ rotational temperatures. Using these as our temperature guesses, we performed our initial XCLASS fit with CH$_3$OH, CH$_3$CN, SO$_2$, and $^{34}$SO$_2$, and hydrogen radio recombination lines. For the hydrogen recombination lines, we fixed the electron temperature to 15000 K and fit the remaining three parameters resulting in an emission measure of $2.4\times10^7$ pc cm$^{-6}$, line width of 27.8 km s$^{-1}$, and a velocity offset of 237.63 km s$^{-1}$. The fit for SO$_2$ had several severely under and overestimated lines, so we tested two components as well as non-LTE SO$_2$. The best of these results includes two SO$_2$ core components one sub-beam sized (0$\rlap.{''}$4) and one that filled the beam uniformly, both in LTE. This second component might be required to describe a second velocity component or a temperature gradient. Though there are still a few lines that are underestimated, this result improves the majority of the poor line fits from the previous runs. 

\begin{figure*}
    \includegraphics[width=0.58\textwidth]{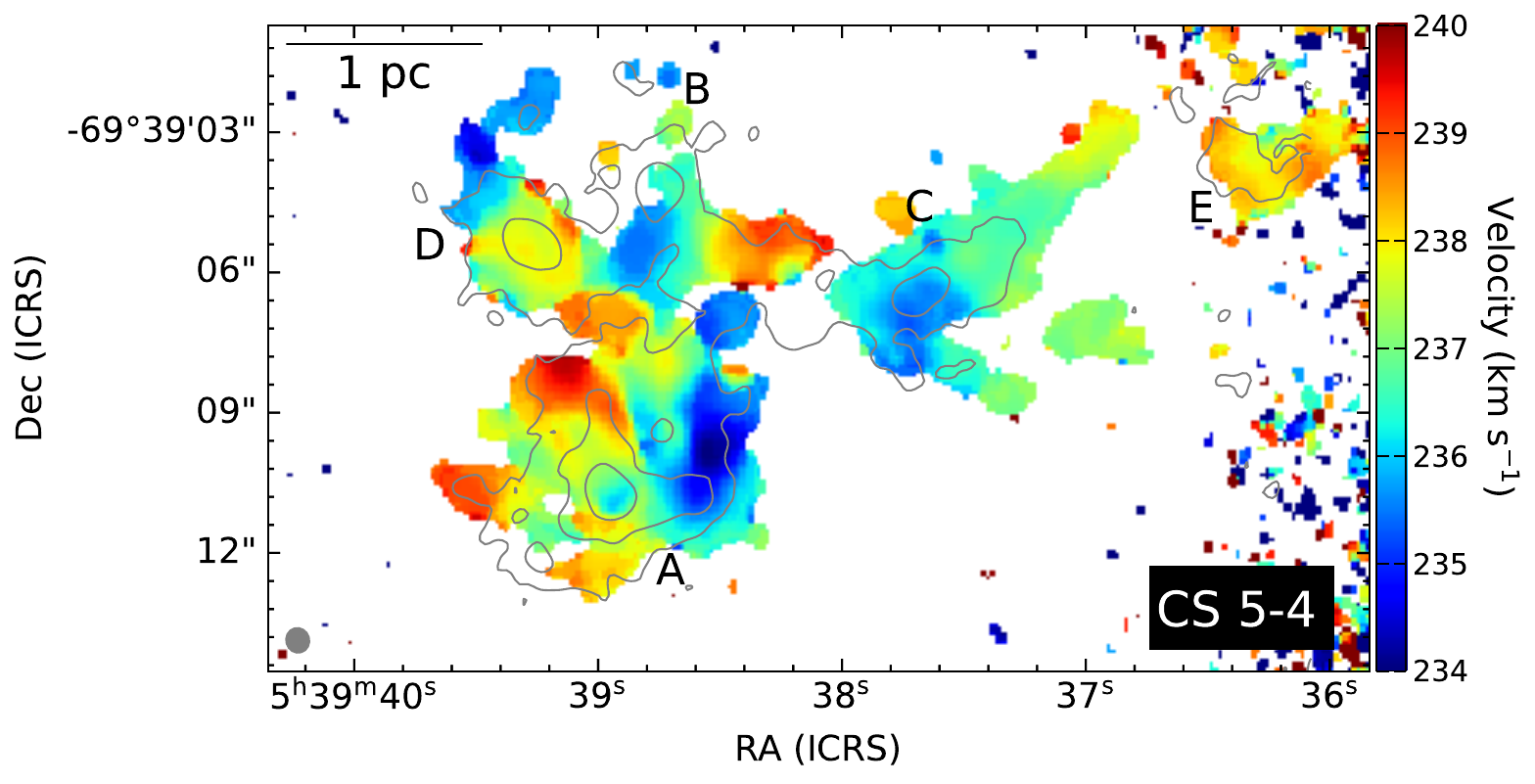} \includegraphics[width=0.4\textwidth]{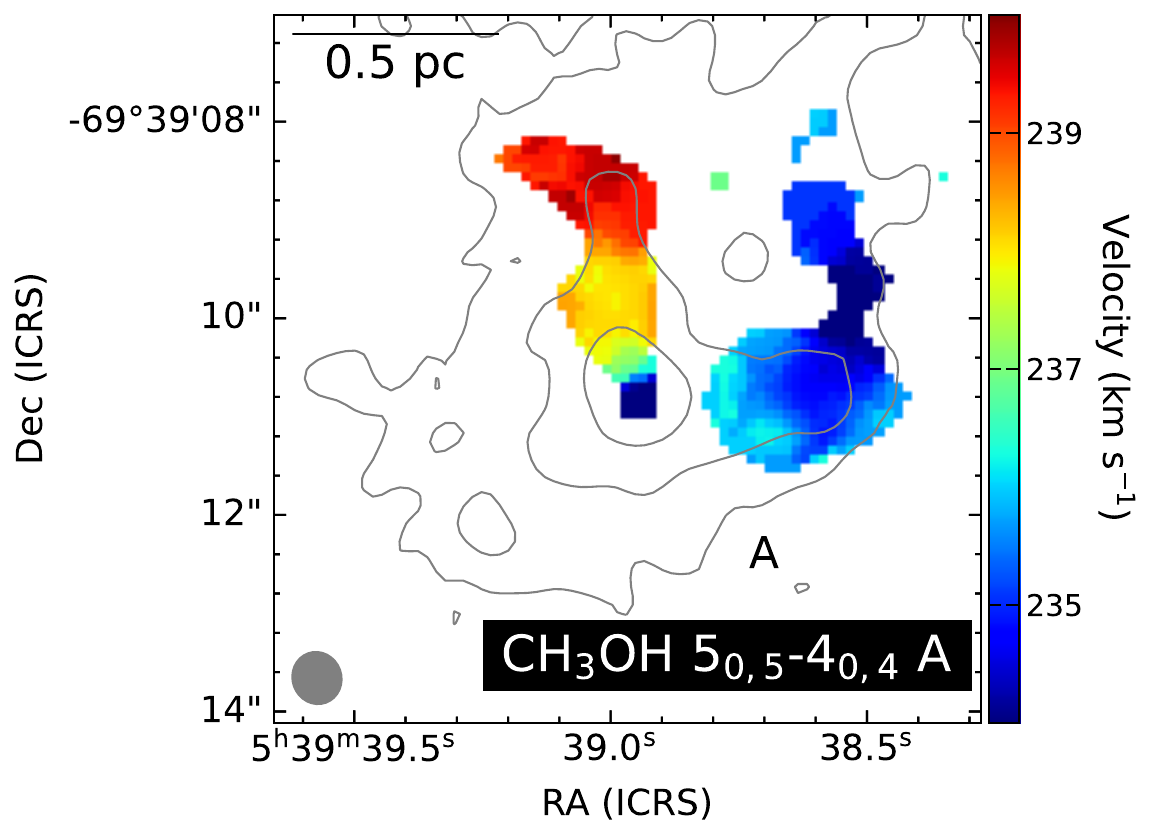} 
    \caption{The CS velocity (moment 1) map of N\,160A--mm A--E and the CH$_3$OH velocity map zoomed in on N\,160A--mm\,A. The 1.2 mm contours are overlaid with contour levels the same as in Figure~\ref{fig:Mom0COMs}. Any pixel with the intensity below $4\sigma$ in the corresponding integrated intensity map is excluded from the velocity map. The velocity maps reveal complex kinematics of the molecular gas in the ALMA field N\,160A--mm (see also Figure~\ref{fig:vmaps}). \label{fig:mom1maps}} 
\end{figure*}

Our initial results for CH$_3$OH and SO$_2$ were confirmed with the XCLASS analysis providing a CH$_3$OH temperature of 188 K and a SO$_2$ temperature of 178 K. Additionally, $^{34}$SO$_2$ and CH$_3$CN have temperatures around 100 K (see Table \ref{t:XCLASS Params}). The discrepancy between the CH$_3$OH and CH$_3$CN temperatures may be the result of CH$_3$CN being sub-thermally excited; however, some of the difference may be caused by the CH$_3$CN temperature determination being based on tentatively detected lines.

\noindent \textit{Source C:} For the initial XCLASS fit we include CH$_3$OH and SO$_2$. The best fit temperatures for CH$_3$OH and SO$_2$ are 29 K and 32 K respectively, indicating a cold core. The XCLASS analysis showed that a single non-LTE core component provides a good fit to the observed spectrum for CH$_3$OH. For the non-LTE case, the collision partner was H$_2$, the ratio of A and E CH$_3$OH was fixed at one, and the collision partner density was fit as a free parameter. The best fit n(H$_2$) was $5.8 \times 10^6$ cm$^{-3}$. For all other molecules, we assumed a single core component in LTE. The integrated intensity maps of CH$_3$OH reveal the presence of the faint emission from the high excitation CH$_3$OH transitions, though in our model these lines are underfit. We tested a two CH$_3$OH component run with XCLASS with a core in LTE and a foreground in non-LTE. The addition of a second component did not significantly improve the results indicating that a single core component model provides a good fit to the observed spectra.

\noindent \textit{Source D:} For the initial XCLASS fit we use only CH$_3$OH. The best fit XCLASS model included a single core component in non-LTE for CH$_3$OH with the same parameter setup as described above for source C. The best fit n(H$_2$) is $3.9 \times 10^6$ cm$^{-3}$. The best fit CH$_3$OH temperature is 25 K, indicating a cold core. 

\noindent \textit{Source E:} CH$_3$OH is the only molecule included in the initial XCLASS analysis. The XCLASS analysis indicates a low temperature of 9 K with a single non-LTE core component for CH$_3$OH being the best fit for the observations. The n(H$_2$) best fit value is $4.5 \times 10^8$ cm$^{-3}$.

\subsubsection{A Glimpse at the Molecular Gas Kinematics in N\,160A--mm} \label{s:Kinematics}

The asymmetric and multi-peaked spectral line profiles observed toward source A indicate a significant kinematic complexity toward this source. The intensity weighted velocity (moment 1) maps for different species confirm this initial assessment. Figure~\ref{fig:mom1maps} shows the CS velocity map of the field covering sources A--E, and the CH$_3$OH map zoomed in on source A. Additional maps are provided in Appendix \ref{a:kinematics} Figure~\ref{fig:vmaps} for SO and H$^{13}$CO$^{+}$. The velocity maps show that while elongated structures with velocity gradients are present near sources C and D, the region around source A exhibits the greatest kinematic complexity.

In Figure \ref{fig:vfwhm}, we show the distribution of line widths ($\Delta v_{\rm FWHM}$ in Table~\ref{t:XCLASS Params} for all species) for the four sources analyzed with XCLASS, i.e., A, C, D, and E. In general, the broadest molecular lines are observed toward source A, providing further evidence that this source is associated with more significant large-scale motions than other sources; these may include an outflow, rotation, accretion, or a combination thereof. The detection of SiO toward source A points to the presence of shocks. This is supported by the presence of the S-bearing species (SO, $^{33}$SO, SO$_2$, and $^{34}$SO$_2$) that can form in shocks (similarly to SiO); these species have higher abundances than those measured toward other sources (if available; see Figure~\ref{fig:N160Abundances}). The Si and S atoms are released to the gas in shocks by sputtering or destruction of refractory grain cores (e.g., \citealt{schilke1997,vandishoeck2018}). A detailed analysis of the kinematics in N\,160A--mm is out of scope of this paper.

\subsection{H$_2$ Column Density} \label{ss:h2columndensity}
The H$_2$ column density, N(H$_2$), can be calculated using the following formula (e.g \citealt{hildebrand1983, kauffmann2008}):
\begin{equation}
N({\rm H_2}) = \frac{S_{\rm \nu}^{\rm beam}\,R_{\rm gd}}{\Omega_{\rm A}\,\mu_{\rm H_2}\,m_{\rm H}\,\kappa_{{\rm \nu, d}}\,B_{\nu}(\rm T)},
\end{equation}
where $S_{\nu}^{\rm{beam}}$ is the flux per synthesized beam; $R_{\rm{gd}}$ is the gas-to-dust mass ratio; $\Omega_{\rm A}$ is the beam solid angle given by $\Omega_{\rm A} = \frac{\pi}{4\rm{ln}2}\theta_{\rm maj}\theta_{\rm min}$ with $\theta_{\rm maj}$ and $\theta_{\rm min}$ being the major and minor axes of the synthesized beam; $\mu_{\rm H_2}$ is the mean molecular weight per hydrogen molecule ($\mu_{\rm H_2}\approx 2.76$ for the LMC, \citealt{remyruyer2014}); $m_{\rm H}$ is the mass of the hydrogen atom; $\kappa_{\nu, d}$ is the dust opacity per unit mass; and $B_{\nu}(T)$ is the Planck function.

We use a more convenient version of the above equation provided by \citet{kauffmann2008} to calculate the column densities for our sources,
\begin{eqnarray}
N({\rm H_2}) = 2.02\cdot10^{20} (e^{1.439(\lambda/{\rm mm})^{-1}(T/{\rm 10\,K})^{-1}}-1) \\ \nonumber
\cdot \left(\frac{\kappa_{{\rm \nu, d}}/R_{\rm gd}}{0.01\,{\rm cm^2\,g^{-1}}}\right)^{-1}\,\left(\frac{I_{\rm \nu}^{\rm beam}}{{\rm mJy\,beam^{-1}}}\right) \\ 
\cdot \left(\frac{\theta_{\rm HPBW}}{{\rm 10\,arcsec}}\right)^{-2}\,\left(\frac{\lambda}{{\rm mm}}\right)^3 {\rm cm^{-2}}, \nonumber 
\end{eqnarray}
where $\lambda$ is the wavelength of the continuum image; $I_{\rm \nu}^{\rm beam}$ is the continuum intensity; and $\theta_{\rm HPBW}= \sqrt{\theta_{\rm maj}\theta_{\rm min}}$.

Assuming that the dust and gas are well-coupled such that $T_{\rm d} \sim T_{\rm g} \sim T$, we adopt the CH$_3$OH rotational temperature as the dust temperature in Eq. 4. This assumption is valid in high-density regions ($n_{\rm H_2} \gtrsim 10^5$ cm$^{-3}$) similar to those we study here (e.g., \citealt{goldsmith1978, ceccarelli1996, kaufman1998}; see below). 

For sources A, C, and D we use the archival 870 $\mu$m continuum image to determine $I_{\rm \nu}^{\rm beam}$. At this wavelength, it is expected that the total continuum emission is from dust thermal emission, with little to no contribution from the free-free emission. For $I_{\rm \nu}^{\rm beam}$, we adopt the mean intensity of the area enclosed by 50\% of the peak 1.2 mm continuum emission contour (the same region used to extract the spectra). Given our extraction region is based on the 1.2 mm continuum image, we want to ensure that the 870 $\mu$m beam matches the 1.2 mm. To do this we regridded and smoothed the 870 $\mu$m image to the 1.2 mm one. 

The field of view of the 870 $\mu$m image is smaller than that of the 1.2 mm image and it therefore does not cover the entirety of source E. We instead use the 1.2 mm continuum emission to estimate $I_{\rm \nu}^{\rm beam}$ for this source. Though the 870 $\mu$m image is better suited to this analysis, as there is no expected contribution from the free-free emission, the 1.2 mm image is a suitable replacement for source E given there is no detected radio emission in that region (see Figure \ref{fig:ATCA}). 

The dust opacity per unit mass for 870 $\mu$m was taken from \cite{Ossenkopf1994} for the model with an MRN distribution \citep{mathis1977} with thin ice mantles at a gas density of $10^6$ cm$^{-3}$. We use a dust opacity value of 1.89 cm$^2$ g$^{-1}$ for the 870 $\mu$m calculations, which we get by interpolating values from \cite{Ossenkopf1994} Table 1. For 1.2 mm the dust opacity is 0.993 cm$^2$ g$^{-1}$. For the gas-to-dust mass ratio, we follow the same process as described in \cite{sewilo2022}. The Galactic gas-to-dust mass ratio is scaled by the broken-power-law relationship between gas-to-dust mass ratio and metallicity from \cite{remyruyer2014}. This gives a gas-to-dust mass ratio estimate of 316 for the LMC.

We can additionally calculate the source masses using \citep{kauffmann2008}:
\begin{eqnarray}
M = 0.12\, M_{\odot} (e^{1.439(\lambda/{\rm mm})^{-1}(T/{\rm 10\,K})^{-1}}-1) \\
\cdot \left(\frac{\kappa_{{\rm \nu, d}}/R_{\rm gd}}{0.01\,{\rm cm^2\,g^{-1}}}\right)^{-1}\,\left(\frac{F_{\rm \nu}}{{\rm Jy}}\right) \nonumber 
 \left(\frac{D}{{\rm 100\,pc}}\right)^{2} \,\left(\frac{\lambda}{{\rm mm}}\right)^3,  \nonumber
\end{eqnarray}
where $F_{\nu}$ is the flux density and D is the distance to the LMC. Flux densities ($F_{\nu}$) were taken to be the total intensity, using the same areas used for $I_{\nu}$, divided by the solid angle of the beam $\Omega = \pi \theta_{\rm HPBW}^2/(4\,\rm{ln}2)$ in squared pixels.

We estimate source sizes at 1.2 mm by calculating the effective radius, $R_{\rm eff} = \sqrt{A/\pi}$, where $A$ is the area within the 50\% of the 1.2 mm continuum peak contour. We then calculate the source size as $\rm{FWHM_{eff}}=2\cdot R_{\rm eff}$. By assuming that the sources can be modeled as Gaussian profiles, we can calculate the deconvolved source sizes using $\rm{FWHM_{eff,deconv}}=\sqrt{\rm{FWHM_{eff}}^2 - \theta_{\rm HPBW}^2}$. Intensities, source sizes, fluxes, masses, and column densities are all shown in Tables \ref{t:1.2mm} and \ref{t:datanh2}. 

The H$_2$ number density $n_{\rm H_2}$ can be estimated by $n_{\rm H_2} = N(\rm{H}_2)/\rm{FWHM_{eff,deconv}}$. Sources A, C, D, and E all have H$_2$ number densities of a several times $10^5$ cm$^{-3}$. The typical values for $n_{\rm H_2}$ in hot cores are on the order of $10^6$ cm$^{-3}$. 

\begin{deluxetable*}{cccccccccc}
\centering
\rotate
\tablecaption{Source Intensities and Sizes \label{t:1.2mm}}
\tablewidth{0pt}
\tablehead{
\colhead{Source} &
\colhead{RA (J2000)} &
\colhead{Dec (J2000)} &
\colhead{$I_{1.2\rm mm, peak}$\tablenotemark{\footnotesize a}} &
\colhead{$I_{1.2\rm mm, mean}$\tablenotemark{\footnotesize b}} &
\colhead{$I_{870\rm \mu m, peak}$\tablenotemark{\footnotesize a}} &
\colhead{$I_{870\rm \mu m, mean}$\tablenotemark{\footnotesize b}} &
\colhead{$Area$\tablenotemark{\footnotesize b}} &
\colhead{FWHM$_{\rm eff}$} &
\colhead{FWHM$_{\rm eff, deconv}$} \\
\colhead{N\,160A--mm} &
\colhead{($^{\rm h}$~$^{\rm m}$~$^{\rm s}$)} &
\colhead{($^{\rm \circ}$ $'$ $''$)} &
\multicolumn{4}{c}{(mJy beam$^{-1}$)} &
\colhead{(arcsec$^{2}$)} &
\colhead{($''$/pc)} &
\colhead{($''$/pc)}}
\startdata
A & 05:39:38.94 & $-$69:39:10.83 & 15.307 & 10.978 & 26.806 & 20.766 & 0.305 & 0.62/0.15 & 0.39/0.09 \\
B & 05:39:38.76 & $-$69:39:04.20 & 2.410 & 1.709 & 4.065 & 3.136 & 0.474 & 0.78/0.19 & 0.61/0.14 \\
C & 05:39:37.70 & $-$69:39:06.59 & 1.969 & 1.350 & 5.819 & 4.102 & 0.584 & 0.86/0.21 & 0.71/0.17 \\
D & 05:39:39.27 & $-$69:39:05.40 & 1.737 & 1.212 & 5.615 & 3.986 & 0.838 & 1.03/0.25 & 0.91/0.22 \\ 
E & 05:39:36.33 & $-$69:39:03.93 & 0.561 & 0.402 & \nodata & \nodata & 0.821 & 1.02/0.24 & 0.90/0.21 \\ 
F & 05:39:37.00 & $-$69:38:54.73 & 0.829 & 0.616 & \nodata & \nodata & 0.194 & 0.50/0.12 & 0.11/0.03 
\enddata
\tablenotetext{a}{$I_{\rm peak}$ is the observed continuum intensity peak for 1.2 mm and 870 $\mu$m.}
\tablenotetext{b}{$I_{\rm mean}$ is the continuum intensity averaged over the area ($Area$) within the contour corresponding to the 50\% of the 1.2 mm continuum peak. This is the same area used to extract spectra for the analysis (see Section~\ref{ss:Spectral Analysis}). The beam area for the observations is 0.23 arcsec$^{2}$.}
\end{deluxetable*}

\begin{deluxetable}{ccccc}
\centering
\tablecaption{Source Parameters \label{t:datanh2}}
\tablewidth{0pt}
\tablehead{
\colhead{Source} &
\colhead{$F_{\rm 50}$\tablenotemark{\footnotesize a}} &
\colhead{$M_{\rm 50, gas}$\tablenotemark{\footnotesize a}} &
\colhead{$N({\rm H_2})$\tablenotemark{\footnotesize b}} \\
\colhead{N\,160A--mm} &
\colhead{(mJy)} &
\colhead{($M_{\odot}$)} &
\colhead{(10$^{23}$ cm$^{-2}$)} 
}
\startdata
A & 23.7 & $72 \pm 7.8$ & $1.80 \pm 0.19$ \\
B & 5.6 & \nodata & \nodata \\
C & 9.0 & $231 \pm 33$ & $3.01 \pm 0.43$ \\
D & 12.5 & $386 \pm 54$ & $3.52\pm 0.49$ \\ 
E & 1.2 & $554 \pm 105$ & $5.15 \pm 0.97$ \\ 
F & 0.45 & \nodata & \nodata 
\enddata
\tablenotetext{a}{$F_{\rm 50}$ and $M_{\rm 50, gas}$ are flux densities and masses, respectively, calculated for the area above the 50\% of the 1.2 mm peak intensity. For sources A--D this is from archival 870 $\mu$m data and for E-F this is from the 1.2 mm data.}
\tablenotetext{b}{$N({\rm H_2})$ calculated assuming $T = T_{\rm rot} ({\rm CH_{3}OH})$ (see Section~\ref{xclass} for details).}
\end{deluxetable}

The small size (0.09 pc) and high ${\rm H_2}$ number density ($6.3\times 10^5$ cm$^{-3}$), combined with the high rotational temperature ($>$100 K for CH$_3$OH and SO$_2$; Section~\ref{xclass}) and the detection of COMs (CH$_3$OH and CH$_3$CN; Section~\ref{s:Line identification}) support the classification of N\,160A--mm A as a bona fide hot core. This marks the eighth detection of a hot core in the LMC.

\section{Discussion} \label{s:Discussion}

\subsection{ALMA Results for Individual Sources in the Context of Previous Observations}

\begin{figure*}
    \centering
    \includegraphics[width=1\textwidth]{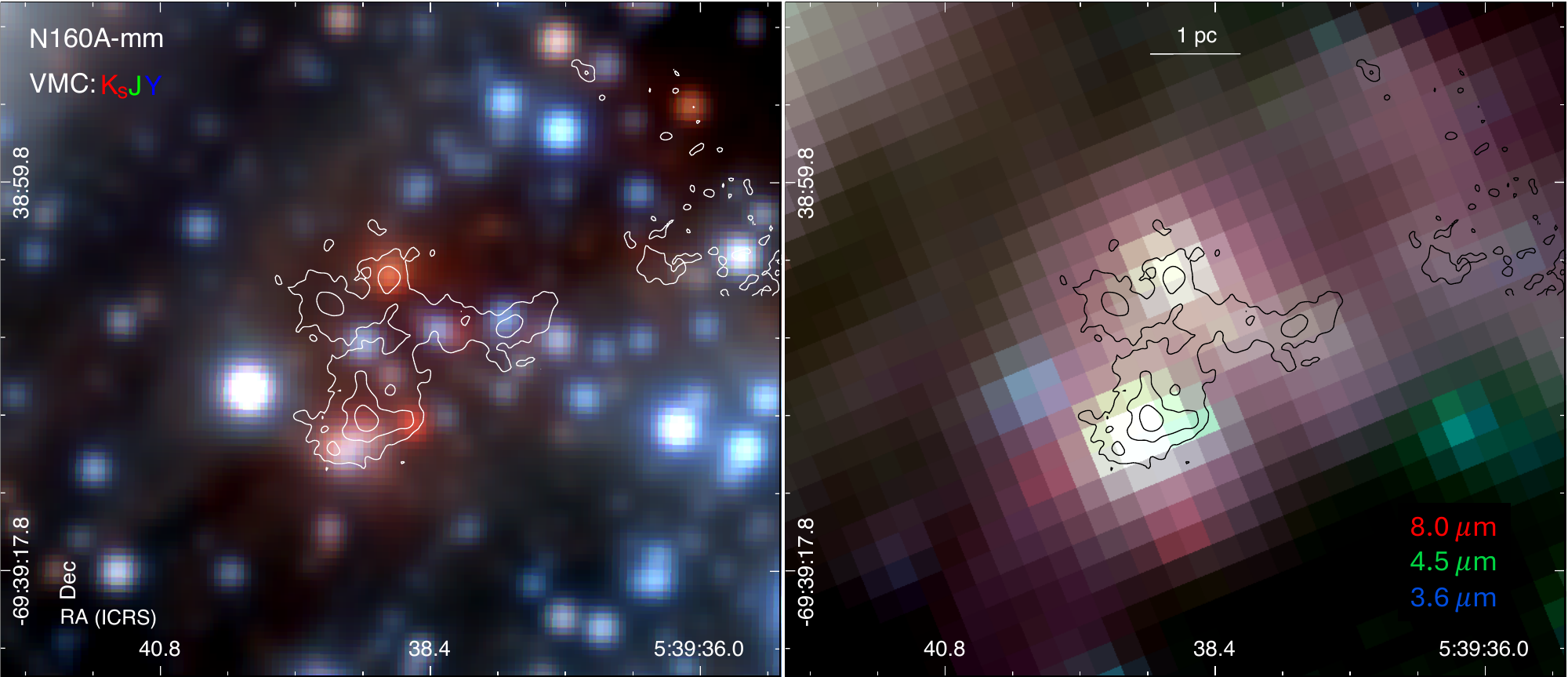}
    \caption{Three-color images of N\,160A--mm combining the VMC $K_s$ (red), $J$ (green), and $Y$ (blue) images in the left panel, and the SAGE {\it Spitzer}/IRAC 8.0 $\mu$m (red), 4.5 $\mu$m (green), and 3.6 $\mu$m (blue) images in the right panel. The field of view in both images is the same. The 1.2 mm continuum contours with contour levels of 1\%, 5\%, and 20\% of the continuum peak are overlaid for reference.}
    \label{fig:IRimages}
\end{figure*}

As N\,160 is one of the most prominent and well-studied star-forming regions in the LMC, a wealth of ancillary data exist for our ALMA field N\,160A--mm (see Section~\ref{s:intro} and Figures~\ref{fig:Ha_8.6_CO}, \ref{fig:N160RGB}, \ref{fig:IRimages}). In Sections \ref{sss: hot core} and \ref{sss: cold cores}, we summarize our findings on sources N\,160A--mm A and B--F, respectively, and discuss these results in the context of previous multi-wavelength observations.

\subsubsection{Hot Core: N\,160A--mm\,A} \label{sss: hot core}

Newly detected hot core N\,160A--mm\,A is the brightest and most complex (in terms of the morphology and chemical makeup) of the continuum sources in the ALMA field N\,160A--mm. The most important detections for the purpose of identifying hot cores are COMs. While methanol has been found toward all the sources in N\,160A--mm, CH$_3$CN has only been detected toward source A. Among the detected species listed in Table~\ref{t:detections}, the tentative detection of deuterated water (HDO) is noteworthy since to date, extragalactic HDO has only been detected toward two hot cores in the star-forming region N\,105 in the LMC (N\,105--2\,A and B; \citealt{sewilo2022hdo}). 

The detection of the hydrogen recombination line H36$\beta$ toward source A indicates that the massive protostar has already started ionizing its surrounding, forming an ultracompact (UC) H\,{\sc ii} region. Although not identified as a detection by our criteria, the H41$\gamma$ line can be seen in both the spectrum and the integrated intensity map (see Figure \ref{fig:h41gamma}). This would mark the second detection of extragalactic higher order H-recombination line at mm wavelengths ($\Delta n >2 $, where $n$ is the principal quantum number), following the detections toward source N\,105--1\,A by \citet{sewilo2022hdo}. The H40$\alpha$ line at 99.02295 GHz has been detected toward N\,160A--mm\,A in the archival Band 3 observations. The H40$\alpha$, H36$\beta$, and H41$\gamma$ recombination line emission coincides with the 1.2 mm continuum peak (see Figures~\ref{fig:Mom0Amaps} and \ref{fig:h41gamma}). 

The UC H\,{\sc ii} region associated with N\,160A--mm\,A, was identified by \cite{indebetouw2004} based on the ATCA 4.8 GHz and 8.6 GHz observations (B0540-6940(4); see Figures~\ref{fig:Ha_8.6_CO} and \ref{fig:ATCA}). \cite{indebetouw2004} provide a spectral type of O8.5\,V for B0540-6940(4) which corresponds to the luminosity of $\sim$7$\times$10$^4$ $L_{\odot}$ and the spectroscopic mass of $\sim$20 $M_{\odot}$ (\citealt{martins2005}).

N\,160A--mm\,A / UC H\,{\sc ii} region B0540-6940(4) also coincides with the positions of the H$_2$O, CH$_3$OH, and OH masers (see Figure~\ref{continuum}); specifically, H$_2$O masers at 22 GHz \citep{whiteoak1983, whiteoak1986,lazendic2002, ellingsen2010}, CH$_3$OH masers at 6.7 GHz \citep{green2008,ellingsen2010}, and OH masers at 1.665 GHz \citep{caswell1981, gardner1985, brooks1997} and 6.035 GHz \citep{caswell1995, green2008}. 

\begin{figure*}
    \centering
    \includegraphics[width=0.75\textwidth]{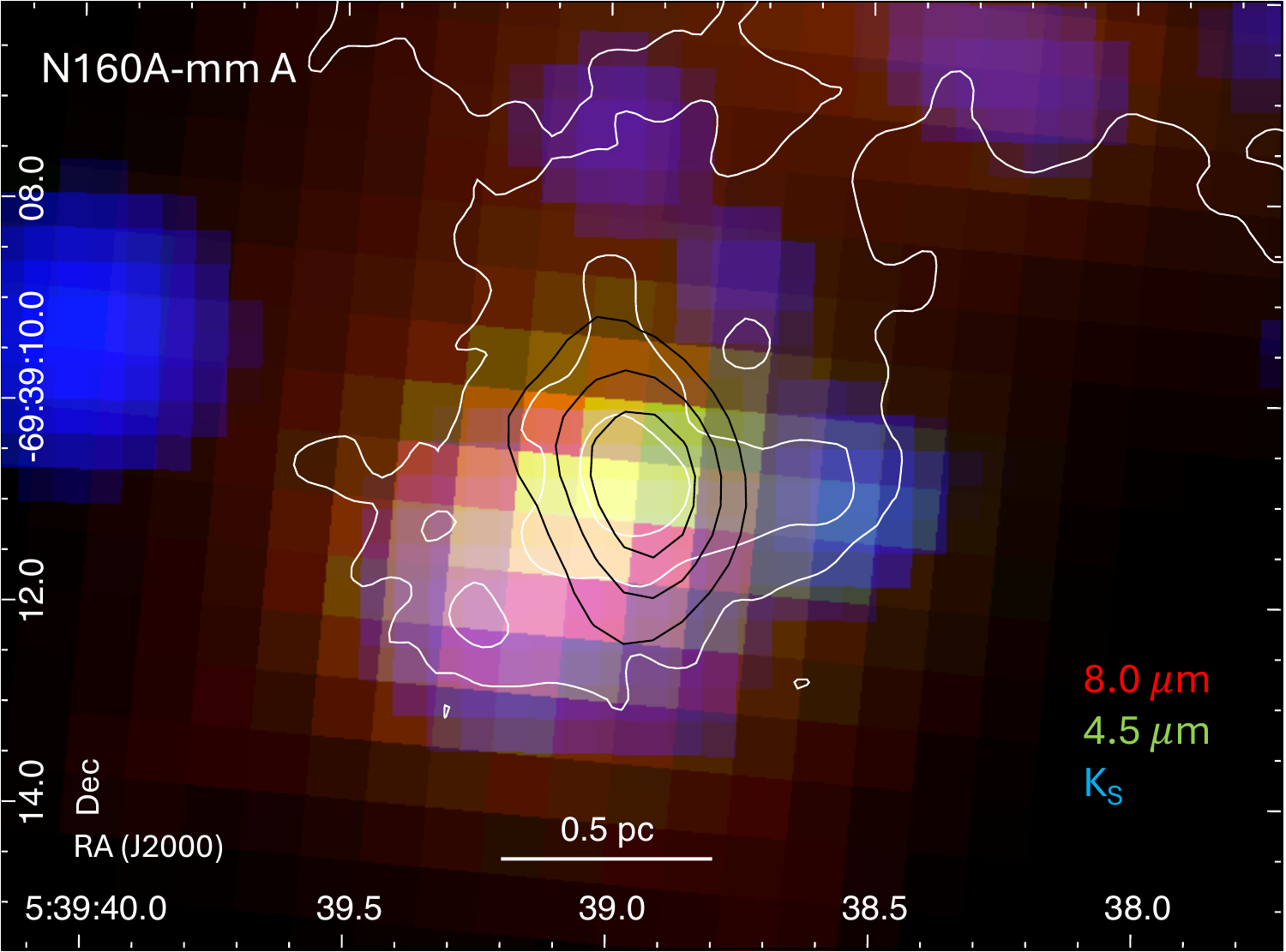}
    \caption{Three-color image of N\,160A--mm\,A combining the SAGE \textit{Spitzer}/IRAC 8.0 $\mu$m (red), 4.5 $\mu$m (green), and the VMC $K_s$ band (blue) images. The white contours are the 1.2 mm continuum emission with contour levels of 1\%, 5\%, and 20\% of the continuum peak of 15.3 mJy beam$^{-1}$. The black contours are the H40$\alpha$ emission with contour levels of 25\%, 50\%, and 75\% of the H40$\alpha$ peak emission of 441.6 mJy beam$^{-1}$ km s$^{-1}$.}
    \label{fig:IRAC-VMC}
\end{figure*}

There is no clear detection of a near-IR source at the position of source A in the $YJK_s$ images from the Visible and Infrared Survey Telescope for Astronomy (VISTA) survey of the Magellanic Clouds system (VMC; \citealt{cioni2011}), the highest resolution and sensitivity near-IR survey of the LMC/SMC (see Figure~\ref{fig:IRimages}; a typical seeing of $\sim$0$\rlap.{''}$8). The Spitzer/IRAC images (0$\rlap.{''}$6-pixel tiles) indicate that the mid-IR source coinciding with N\,160A--mm\,A is unresolved from the nearby source, a spectroscopically confirmed YSO located to the southeast of the 1.2 mm continuum peak of source A (053939.02-693911.4, \citealt{gruendl2009,seale2009,jones2017}; see Figure~\ref{fig:IRAC-VMC}).

The Spitzer YSO 053939.02-693911.4 is classified as a `Group PE' YSO by \citet{seale2009}, indicating the presence of the strong PAH emission features and fine-structure lines in its Spitzer Infrared Spectrograph (IRS) spectrum (5--37 $\mu$m), as can be seen in Figure \ref{fig:spectra}. Based on the analysis of the same spectrum, \citet{jones2017} classified the source (SSID 4661) as `H\,{\sc ii}/YSO3.' Both classifications indicate a relatively evolved YSO with an emerging ultracompact H\,{\sc ii} region. However, a careful examination of the Spitzer/IRS spectrum shown in Figure~\ref{fig:spectra} reveals the presence of a faint CO$_2$ ice feature at 15.2 $\mu$m (at a 2$\sigma$ level), typical for embedded YSOs. It should be noted that the Spitzer/IRS slit width ranges from 3$\rlap.{''}$6 to 11$\rlap.{''}$1 (from $\sim$0.9 to $\sim$2.7 pc) between the shortest to longest wavelengths, and encompasses the entirety of the region surrounding N\,160A--mm\,A (up to 1\% of the maximum 1.2 mm continuum emission peak). Consequently, the Spitzer/IRS spectrum likely combines spectral features originating in different sources: the embedded YSO, UC H\,{\sc ii} region, and their wider environment. 

The three-color VMC image in Figure~\ref{fig:IRimages} reveals the near-IR emission farther toward southeast from the Spitzer YSO (1$\rlap.{''}$7/0.4 pc from the 1.2 mm continuum peak of N\,160A--mm\,A) with the morphology indicating the presence of at least two unresolved sources. We use the archival Very Large Telescope (VLT) Multi Object Spectrograph (VLT/KMOS) \textit{K}-band data to inspect this region in the near-IR at higher resolution. The KMOS observations with the $2\rlap.{''}8\times2\rlap.{''}8$ (0.68 pc $\times$ 0.68 pc) field of view and the pixel scale of $0\rlap.{''}2$ are part of the larger YSO survey (PI J. L. Ward; see \citealt{sewilo2019,sewilo2022} for more details). For our qualitative analysis, we utilize the $K$-band continuum, Br$\gamma$ (2.166 $\mu$m), and H$_2$ (2.1218 $\mu$m) images. The three-color mosaic combining KMOS images is compared to the VMC $K_{s}$ image in Figure \ref{fig:KMOS}. The KMOS observations reveal at least three sources in the small field of view, two of which are still not fully resolved.

The Br$\gamma$ emission is commonly associated with accretion; the Br$\gamma$ luminosity has been used to determine the accretion luminosity toward the Galactic (e.g., \citealt{calvet2004}) and Magellanic YSOs (\citealt{ward2016,ward2017}). In the KMOS field in N\,160A--mm, the brightest Br$\gamma$ emission is compact and coincides with two of the $K$-band continuum sources, indicating that it traces accretion, and thus the detected sources are YSOs. A fainter Br$\gamma$ emission peak is located toward northwest and is likely associated with the Spitzer YSO.

The H$_2$ emission is present over the entire KMOS field with the brightest emission forming an incomplete ring around the region with the bright Br$\gamma$ emission. The H$_2$ emission peaks coincide with regions at and around the 1.2 mm continuum peak of N\,160A--mm\,A and toward southeast from the $K$-band continuum sources. The H$_2$ emission can be produced by photodissociation or shocks; without a detailed analysis of the H$_2$ lines (out of scope of this paper), it is not possible to determine a relative contribution of each mechanism. 

The morphology of the 1.2 mm continuum and dense gas (e.g., CS) emission around N\,160A--mm\,A is suggestive of the presence of filaments. Recently, \citet{Tokuda2023} identified filaments around 30 YSOs in the LMC based on the 0.87 mm continuum emission ($\sim$0.1 pc resolution) using the FilFinder (\citealt{koch2015} software. Following the method described in \citet{Tokuda2023}, we identified filaments toward N\,160A--mm\,A; the spines of several intersecting filaments (a `hub-filament system') are overplotted on the 0.87 mm continuum image in Figure \ref{fig:Filaments}. The inspection of the integrated intensity maps in Figures~\ref{fig:Mom0COMs} and \ref{fig:Mom0Maps} reveals that the filamentary structure identified with FilFinder is traced by the CS, CH$_3$OH, H$^{13}$CO$^{+}$, and SO extended emission over the full length of the filaments. The position of the emission peaks along the filaments differ for different species; however, they coincide with regions where filaments intersect (see Figure \ref{fig:Filaments}).

Our results combined with the existing multi-wavelength data indicate that hot core N\,160A--mm\,A is likely the most massive member of the protocluster forming at the intersection of the filaments. 

 \subsubsection{N\,160A--mm B, C, D, E, and F} \label{sss: cold cores}

{\it N\,160A--mm\,B}: Source B is the second brightest continuum source in the field, though its spectrum lacks any molecular emission lines aside from CS and tentatively H$^{13}$CO$^{+}$. The CS and H$^{13}$CO$^{+}$ integrated intensity maps show that these species do not have emission peaks coinciding with the 1.2 mm continuum peak and therefore the detected lines likely originate in the background. Due to the lack of line detections, no spectral line analysis was performed for source B. 

N\,160A--mm\,B is one of the two sources in our ALMA field associated with a Spitzer YSO (053938.73-693904.3, \citealt{gruendl2009,seale2009}; SSID 4660 in \citealt{jones2017}). Based on the Spitzer IRS spectroscopic observations, this YSO was classified as Group PE YSO by \cite{seale2009} and H\,{\sc ii}/YSO--3 by \cite{jones2017}; both classifications indicate it is likely an evolved YSO. These results are consistent with the presence of an ultracompact (UC) H\,{\sc ii} region at the position of the YSO / the 1.2 mm continuum emission peak (B0540-6940(5) in \citealt{indebetouw2004}; see Figures~\ref{fig:Ha_8.6_CO} and \ref{fig:ATCA}). However, as for source A, our inspection of the Spitzer IRS spectrum revealed a faint CO$_2$ ice feature suggesting that the source could be younger (see Figure \ref{fig:spectra} and Section \ref{sss: hot core}). Alternately, a younger and a more evolved YSO are present but remained unresolved with a relatively low angular resolution provided by Spitzer ($\sim$2$''$ or 0.5 pc for IRAC bands). Finally, the peak of source B's 1.2 mm continuum emission coincides with the positions of the near-IR (VMC K$_s$ and J bands) and mid-IR (Spitzer 3.6--8.0 $\mu$m; see Figure \ref{fig:IRimages}) sources. 

{\it N\,160A--mm\,C}: We detected CH$_3$OH (multiple transitions -- 10 detections, and 4 tentative detections), HC$^{15}$N, H$^{13}$CN, H$^{13}$CO$^{+}$, SO, H$_2$CS, CS, C$^{33}$S, and SO$_2$ (3 transitions) toward source C. The XCLASS analysis revealed that the source is cold ($\sim$30 K; see Section~\ref{xclass}). The Spitzer 3-color image shown in Figure~\ref{fig:IRimages} indicates a possible detection of a faint mid-IR source at the position of source C; however, the source is not included in any of the Spitzer catalogs (SAGE, \citealt{meixner2006,sage}). No VMC near-IR source coincides with the 1.2 mm continuum peak of source C, and no radio emission has been detected by \citet{indebetouw2004}. The detection of SiO indicates the presence of shocks (possibly a protostellar outflow). More observations are needed to reveal the nature N\,160A--mm\,C. 

{\it N\,160A--mm\,D}: In terms of detected species, N\,160A--mm\,D is similar to source C. Out of the species detected toward source C, only SiO is not detected toward source D. We also classify N\,160A--mm\,D as a cold core based on the XCLASS analysis ($\sim$25 K). No near- or mid-IR sources (thus no known YSO), radio or maser emission coincide with source D. Based on the 0.87 mm continuum emission, \citet{Tokuda2023} identified filaments in the vicinity of their target YSO Lh08 (our N\,160A--mm\,B). The longest filament nearest to Lh08 that was analyzed in detail is associated with source D; it extends from northeast to southwest where it connects to a shorter filament, forming a hub-filament system. 

{\it N\,160A--mm\,E}: Source E is the faintest of the 1.2 mm continuum sources in N\,160A--mm; the continuum signal-to-noise ratio is only above 5$\sigma$, i.e., below the 10$\sigma$ cutoff we adopted for source identification (see Section~\ref{ss:Continuum identification}). However, N\,160A--mm\,E was selected for the spectral analysis because the 1.2 mm continuum peak coincides with the emission peaks of several molecular species (including CH$_3$OH). Even though N\,160A--mm\,E lies close to the edge of the ALMA field resulting in more noise in the spectra, we detected 7 transitions of CH$_3$OH, as well as H$^{13}$CO$^{+}$, SO, H$_2$CS, CS, and C$^{33}$S. The results of the XCLASS analysis reveal that source E is a cold core ($\sim$9 K). As for source C, the Spitzer images indicate the presence of a faint mid-IR source at the position of source E, but it is not included in any of the existing catalogs (see Figure~\ref{fig:IRimages}). No near-IR or radio emission is detected toward N\,160A--mm\,E. 

{\it N\,160A--mm\,F}: Similarly to source E, source F is a $<$10$\sigma$ detection in the continuum, but it is associated with the molecular emission peaks. It is located at the edge of the ALMA field and thus it is heavily impacted by the noise. Only two molecular species have been detected toward source F: CS and tentatively SO. As for source B, it was not possible to characterize source F through the spectral analysis. 

Of these five sources, CH$_3$OH is detected toward three of them (sources C, D, and E) but the CH$_3$OH temperature is cold, all below 30 K. This is seen in other searches for hot cores in the LMC; \cite{sewilo2022} detected CH$_3$OH toward all twelve cores, but seven had temperatures $\lesssim30$ K and four showed a hot $(\sim100$ K) core surrounded by cold ($\lesssim15$ K) extended emission. \cite{Golshan2024} detected CH$_3$OH toward 47 cores but 23/47 had temperatures $<30$ K, 10/47 had temperatures between 30 K and 70 K, and the remaining four were hot with surrounding extended emission. 

CH$_3$OH is formed during the cold prestellar phase through CO hydrogenation on dust grains \citep{herbst2009} but the temperatures in sources C, D, and E are too low to release the CH$_3$OH through sublimation. These instances of cold CH$_3$OH are also seen in the Galaxy toward dark clouds and cold cores (e.g. \citealt{vastel2014}). Other suggested release mechanisms include photodesorption, explosion of UV-irradiated ices, grain heating from cosmic-ray impacts, cosmic-ray sputtering, reactive desorption, and grain-grain collisions (see our previous paper \citealt{sewilo2022} for further details). Very recent quantum-mechanical calculations of methanol molecules binding to a water ice surface have found that a small fraction of surface sites ($\approx 2 \%$) have much lower binding energies for physisorption (1240 K) than the majority ($\approx$ 4200 K), suggesting that, if sufficiently abundant in ices, thermal desorption could be a source of cold methanol even at dust temperatures of around 10 K \citep{Bariosco2025}. 

In addition to the compact cold CH$_3$OH detected toward these three sources, extended CH$_3$OH emission surrounds these peaks as well as the hot CH$_3$OH core of source A. This extended emission is most noticeable in the lowest upper state energies detected for CH$_3$OH indicating it is cold. This is consistent with what we see toward other LMC cores \citep{shimonishi2020, sewilo2022} and the SMC hot cores \citep{Shimonishi2023}.

\subsection{Molecular Abundances in N\,160A--mm} \label{ss:N160Abundances}
\begin{figure*}
    \centering
    \includegraphics[width=\textwidth]{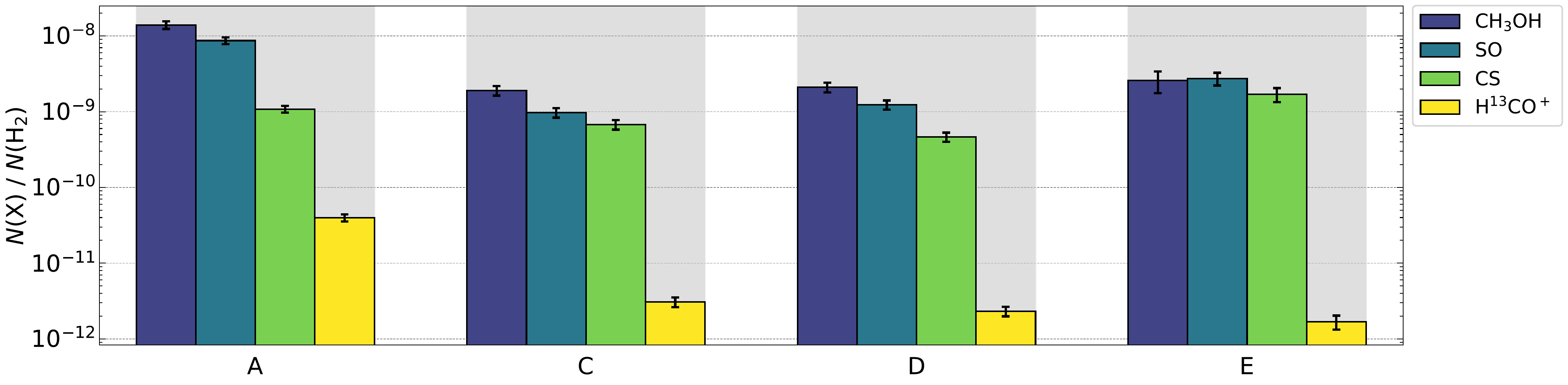}
    \caption{Comparison of molecular abundances of species detected toward all sources analyzed with XCLASS: CH$_3$OH, SO, CS, and H$^{13}$CO$^+$. The CH$_3$OH abundance for source A was measured assuming LTE; non-LTE conditions were assumed for sources C, D, and E.}
    \label{fig:N160MolsAll}
\end{figure*}

\begin{figure*}
    \centering
    \includegraphics[width=\textwidth]{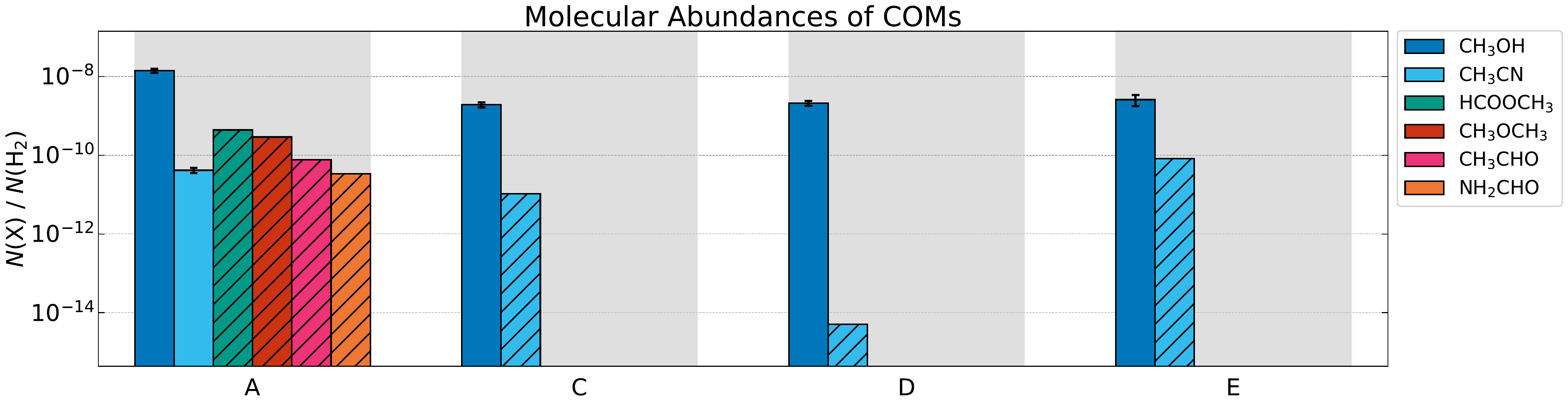} \\
    \includegraphics[width=\textwidth]{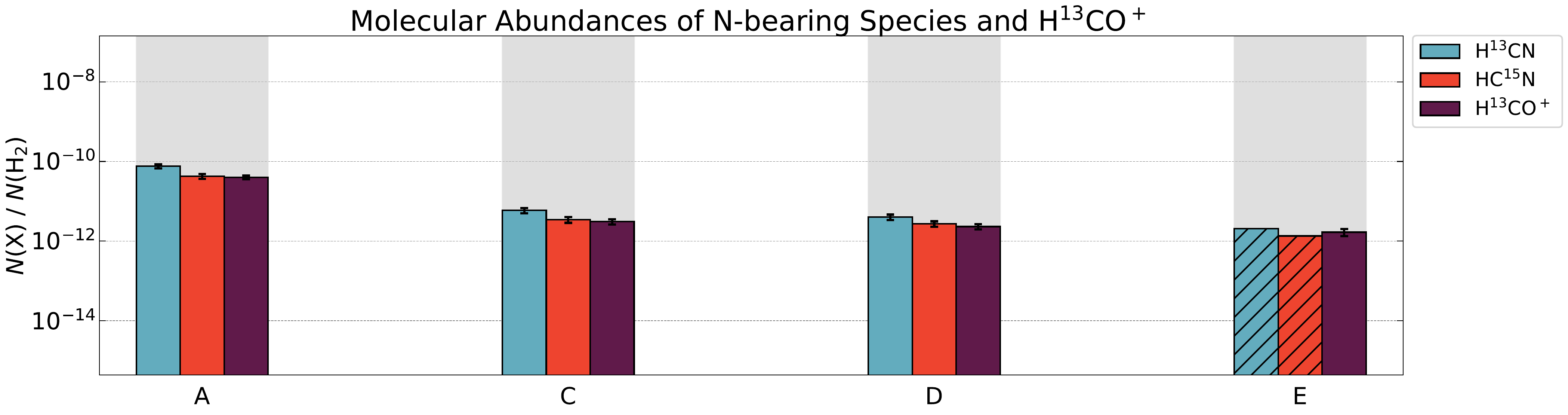} \\
    \includegraphics[width=\textwidth]{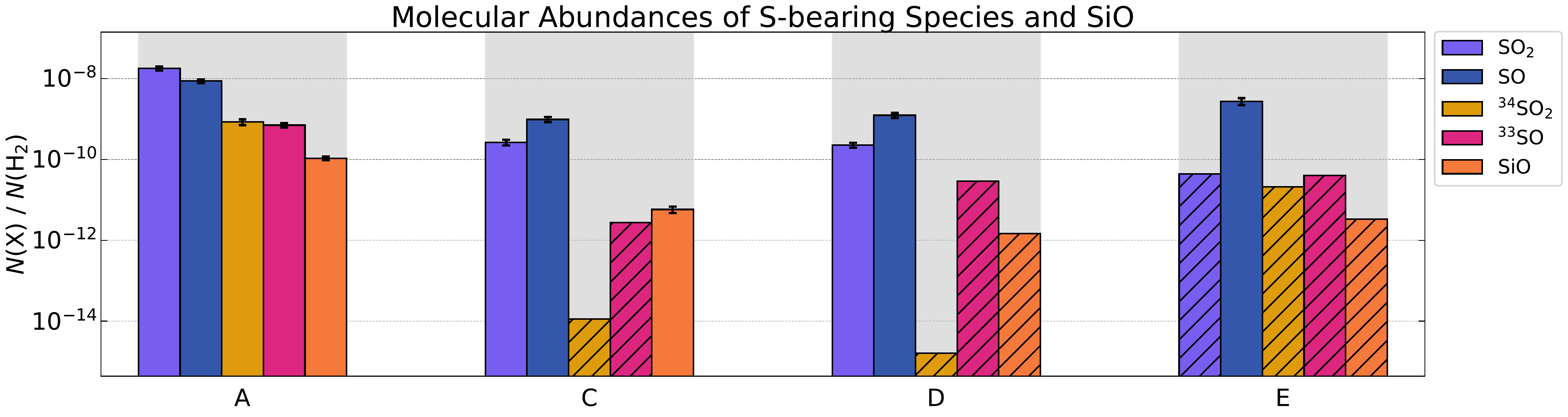} \\
    \includegraphics[width=\textwidth]{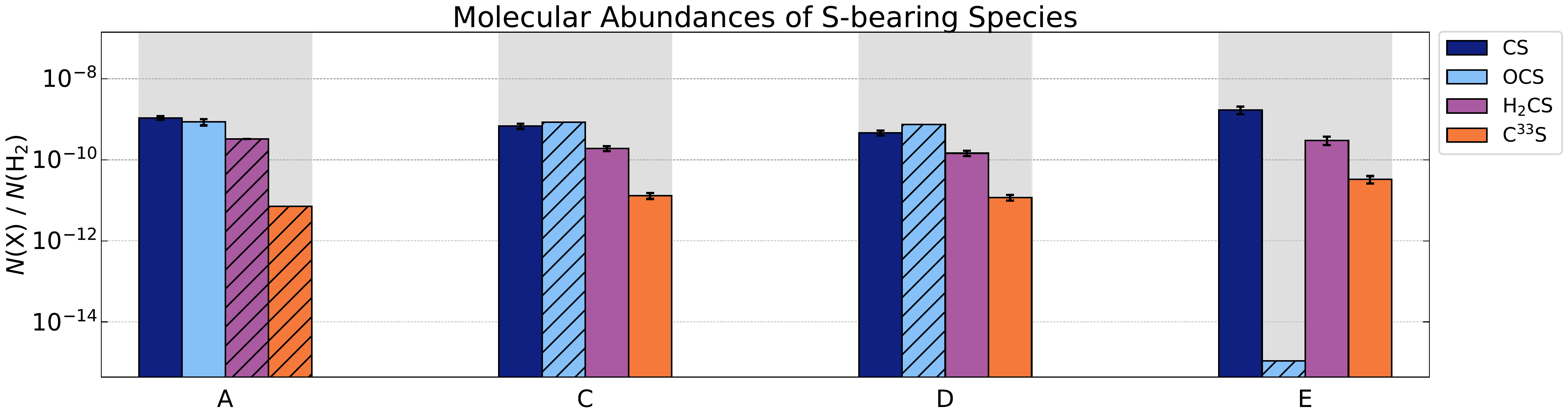}
    \caption{Comparison of molecular abundances between N\,160A--mm A, C, D, and E. Bars with hatch marks indicate upper limits. The SO$_2$ abundance for source A is from the hot component. The hot core N\,160A--mm\,A features the highest abundances. }
    \label{fig:N160Abundances}
\end{figure*}

In Figures \ref{fig:N160MolsAll} and \ref{fig:N160Abundances} we compare the molecular abundances with respect to H$_2$ toward the 1.2 mm sources that were analyzed with XCLASS. In Figure \ref{fig:N160MolsAll}, we compare the abundances of molecules that were detected toward all analyzed sources: CH$_3$OH, SO, CS, and H$^{13}$CO$^+$. Source A has the highest abundances of all species except CS. Sources C, D, and E have comparable abundances for a given species. Although sources B and F were not analyzed with XCLASS, CS was detected toward both sources, while H$^{13}$CO$^+$ and SO were tentatively detected toward source B and F, respectively. These molecules being the most frequently detected is consistent with previous hot core searches in the LMC \citep{sewilo2022, Golshan2024}.

Figure \ref{fig:N160Abundances} shows the abundances toward molecules detected toward at least one source in N\,160A--mm. In addition, for source A (the hot core), we include upper limits for COMs observed toward other LMC hot cores (HCOOCH$_3$, CH$_3$OCH$_3$, CH$_3$CHO, NH$_2$CHO; no COMs other than methanol have been detected in the SMC). Newly identified hot core N\,160A--mm\,A has the highest abundance of CH$_3$OH and all the other detected species except CS; it is the only source with the detection of CH$_3$CN. The CS emission line is present in the spectrum of source A; however, no CS emission peak coincides with the 1.2 mm continuum peak indicating that it may originate in the background. H$_2$CS and C$^{33}$S are not detected toward source A; the upper limits are comparable to H$_2$CS and C$^{33}$S abundances observed toward sources C--E. It is evident from Figure \ref{fig:N160Abundances} that the abundances of N-bearing species are lower than those of S-bearing species. This result is consistent with other observations of the LMC hot cores.

\subsection{Hot Core Abundances: N\,160A--mm\,A vs. other Magellanic and Galactic Hot Cores}
\begin{figure*}
    \centering
    \includegraphics[width=\textwidth]{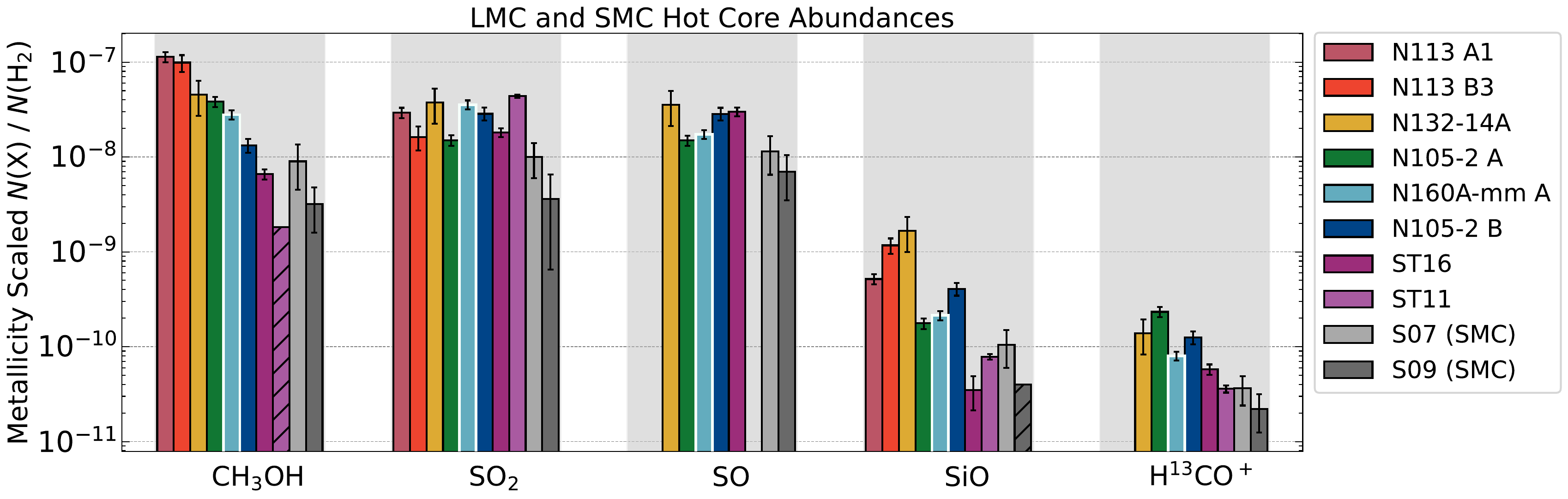} \\
    \includegraphics[width=\textwidth]{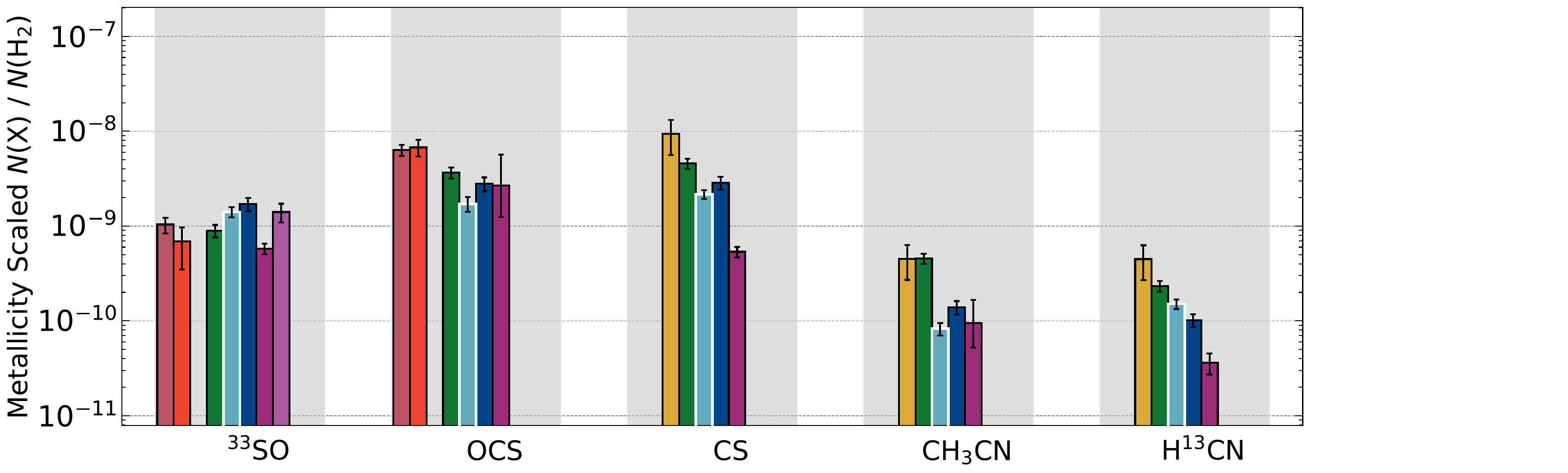}
    \caption{Comparison of metallicity-scaled molecular abundances in known hot cores in the LMC (ST11 \citep{shimonishi2016b}, N\,113 A1 and B3 \citep{sewilo2018}, ST16 \citep{shimonishi2020}, N\,105--2\,A and 2\,B \citep{sewilo2022}, N\,132-14A \citep{Golshan2024}) and SMC (S07 and S09 \citep{Shimonishi2023}). To account for their differing metallicities, we multiply the LMC abundances by a factor of 2 and the SMC abundances by a factor of 5. Upper limits are indicated with diagonal hatch marks. The newly-identified LMC hot core, N\,160A--mm\,A (light blue with white border) has molecular abundances within the range covered by previously known extragalactic hot cores.}
    \label{fig:HotCoresLMCSMCOverview}
\end{figure*}

In Figure \ref{fig:HotCoresLMCSMCOverview}, we compare the molecular fractional abundances with respect to H$_2$ measured toward the hot core N\,160A--mm\,A, $N({\rm X})/N({\rm H_2})$, to the abundances of seven previously detected hot cores in the LMC (ST11, \citealt{shimonishi2016b}; N\,113\,A1 and B3, \citealt{sewilo2018}; ST16, \citealt{shimonishi2020}; N\,105-2\,A and B, \citealt{sewilo2022}; N\,132-14A, \citealt{Golshan2024}) and the two hot cores detected in the SMC (S07 and S09, \citealt{Shimonishi2023}). We compare the LMC and SMC hot core abundances scaled to the same (Galactic) metallicity; they are multiplied by a factor of $1/Z_{\rm LMC} = 2$ and $1/Z_{\rm SMC} = 5$ for the LMC and SMC, respectively \citep{rolleston2002}. This simple scaling accounts for lower atomic abundances of elements such as C, N, and O in the LMC and SMC compared to what is observed in the Solar neighborhood. Taking into account the difference in abundances of specific metals and how that would translate to the production of molecules would require detailed physicochemical modeling that is not necessary for our analysis.

The molecular abundances for N\,160A--mm\,A are listed in Table \ref{t:XCLASS Params}. For hot core ST16 \citep{shimonishi2020}, we use molecular abundances recalculated in our previous study \citep{sewilo2022} for $T($CH$_3$CN$)$ and the same gas-to-dust ratio used in this paper. For consistency, we recalculate the abundances of hot core ST11 \citep{shimonishi2016b} using the same gas-to-dust ratio; since CH$_3$OH or CH$_3$CN were not detected, we use the same (dust) temperature as the authors. To calculate column densities for hot core N\,132-14A, \cite{Golshan2024} used the same gas-to-dust ratio. For this hot core, we recalculated $N({\rm H_2})$ using the same method as in this paper given the Band 7 data were available for 14A in the same program.

All molecular abundances measured toward N\,160A--mm\,A (including CH$_3$OH) fit within the abundance ranges reported toward previously identified hot cores. The CH$_3$OH abundance is higher than that of N\,105-2\,B, ST16, and ST11 though still lower than both N\,113 hot cores, N\,132-14A, and N\,105-2\,A. For the LMC hot cores alone and across the entire sample of the Magellanic hot cores, the CH$_3$OH abundances show significantly larger variations than the abundances of S-bearing species such as SO$_2$ and SO. \citet{Shimonishi2023} suggested that different formation processes of these molecules could explain this result. CH$_3$OH forms on dust grains by CO hydrogenation during the cold prestellar phase \citep{herbst2009}; it is then released into the gas through sublimation in the later hot core phase. The formation of CH$_3$OH would therefore be heavily influenced by the physical environment of this earlier phase. In particular, higher dust temperatures could make the production of CH$_3$OH less efficient during this stage \citep{shimonishi2016a,shimonishi2020}. The SO$_2$ and SO abundances do not depend as strongly on the interstellar conditions during this earlier cold phase, as these molecules form in the gas-phase reactions from S-bearing species that were sublimated during the hot core phase. 

Another potential explanation that has been previously discussed for why we observe large CH$_3$OH abundance variations in the LMC hot cores is metallicity inhomogeneities in the disk of the LMC \citep{sewilo2022, Golshan2024}. These inhomogeneities could be the result of a tidal interaction between the LMC and SMC 0.2 Gyr ago as suggested by both hydrodynamical simulations \citep{fujimoto1990, bekki2007a, yozin2014} and observations \citep{luks1992}. During this encounter, the H\,{\sc i} gas was stripped from both galaxies with some falling back into the LMC. Collisions of this infalling gas with the LMC disk is thought to have triggered star formation in 30 Dor and the H\,{\sc i} Ridge \citep{fukui2017}, as well as the region toward N\,44 \citep{tsuge2019}. The infalling gas would be the LMC's stripped H\,{\sc i} gas mixed with the metal-poor gas of the SMC. This suggests that regions that are impacted by these collisions would have their gas replenished with lower metallicity gas with higher dust temperatures, leading to a decrease in the production of COMs. One could expect hot cores in these regions to resemble the hot cores found in the SMC. In general, the SMC hot cores have lower abundances than hot cores in the LMC, but for certain molecules, particularly CH$_3$OH, SiO, and H$^{13}$CO$^+$, the SMC hot cores do have similar abundances to the LMC hot core ST11 which is nearby one of these sites impacted by the LMC and SMC encounter. 

Prior to the detection of N\,160A--mm\,A, all hot cores with COM detections had been found in the bar (see Figure \ref{fig:bar}) of the LMC which was not strongly influenced by the LMC-SMC encounter. Hot core N\,160A--mm\,A is found in the Molecular Ridge, which, along with the CO Arc, makes up the H\,{\sc i} Ridge where infalling gas from the LMC-SMC interaction triggered star formation. This could explain why the COM abundances toward N\,160A--mm\,A are not as high as the majority of the other hot cores. The N\,160 complex is located only $\sim100$ pc from the N\,159 region, where the formation of filaments and subsequent high-mass star formation triggered by H\,{\sc i} gas flows has already been suggested (\citealt{fukui2019,tokuda2019,tokuda2022}). The idea that similar gas flows may have influenced the N\,160 region could lend further support to the above interpretation. 

In addition to metallicity variations and differences in the dust temperature in the earlier cold stage (the result of variations in the degree of shielding and local radiation field strength),  other possible explanations for an observed chemical diversity in the LMC hot cores include differences in the evolutionary stage and the presence and strength of shocks.

\begin{figure*}
    \centering
    \includegraphics[width=\textwidth]{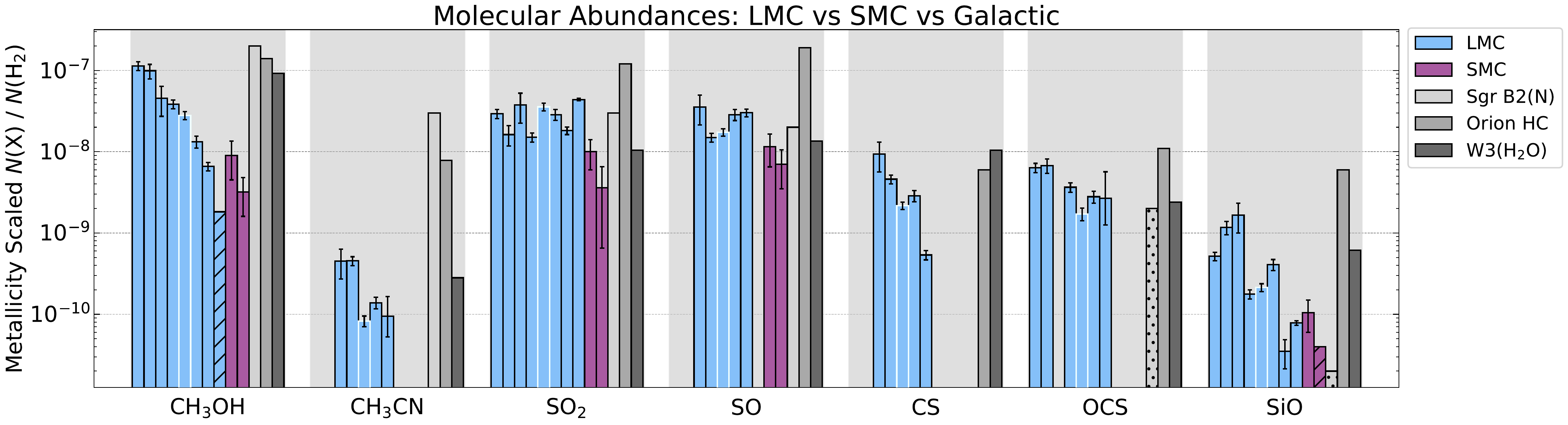}
    \caption{Comparison of metallicity scaled molecular abundances of LMC and SMC hot cores with a sample of Galactic hot cores (Orion \citep{sutton1995} and \citep{blake1987}, W3(H$_2$O) \citep{helmich1997}, Sgr B2(N) \citep{nummelin2000}). Upper limits are shown with diagonal hatch marks and lower limits are shown with dotted bars. The bars corresponding to N\,160A--mm\,A are outlined in white.}
    \label{fig:HotCoresGalactic}
\end{figure*}

The metallicity-scaled abundances with respect to H$_2$ measured toward the LMC and SMC hot cores are compared to those reported for a sample of Galactic hot cores in Figure \ref{fig:HotCoresGalactic}. The Galactic hot core abundances were obtained in single-dish observations of the Orion Hot Core, W3(H$_2$O), and SgrB2(N). These observations probe similar physical size scales as our 12-m Array observations of N\,160A--mm\,A. The Orion data comes from \cite{sutton1995} except for CH$_3$CN abundance which are taken from \cite{blake1987}. W3(H$_2$O) and SgrB2(N) abundances come from \cite{helmich1997} and \cite{nummelin2000}, respectively.

There is a wide range of abundances in both the extragalactic and Galactic hot cores. In some instances there is over an order of magnitude difference between the highest and lowest abundance in our sample, in Figure \ref{fig:HotCoresGalactic} see CH$_3$CN or SiO for Galactic cores and CH$_3$OH or SiO for extragalactic cores. The largest discrepancies between the extragalactic and Galactic hot cores appears in the COM abundances, with the Galactic cores having higher abundances than the LMC and SMC cores' metallicity scaled abundances. Though it should be noted that the LMC hot cores N\,113\,A1 and B3 have comparable CH$_3$OH abundances to the Galactic hot cores when scaled for metallicity.

\section{Summary and Conclusions} \label{s:Conclusions}
We utilized the data from the ALMA observations of the field in the star-forming region N\,160 in the LMC referred to as N\,160A--mm, to search for hot cores.  N\,160A--mm is one of the seven fields in the LMC included in our ALMA Band 6 program aimed at increasing the sample of known LMC hot cores (see \citealt{sewilo2022}); the observations cover four $\sim$2 GHz spectral windows between $\sim$241 GHz and $\sim$261 GHz.  Based on the ALMA and ancillary multi-wavelength data, we performed a detailed analysis of the physical and chemical properties of the sources detected in N\,160A--mm.  We used the XCLASS software to determine the rotational temperatures and total column densities for molecular species with the detection of multiple transitions, and column densities only for the remaining species. The main results are as follows:

\begin{enumerate}
     \item We identified the brightest out of six 1.2 mm continuum sources in the ALMA field N\,160A--mm as {\it a bona fide hot core}: N\,160A--mm\,A.  COMs CH$_3$OH and CH$_3$CN were detected toward N\,160A--mm\,A along with other species that are commonly found in Galactic hot cores (SO$_2$, SO, H$^{13}$CN, SiO, HC$^{15}$N, $^{33}$SO, $^{34}$SO$_2$, OCS, CS, and H$^{13}$CO$^+$).  Both the CH$_3$OH and SO$_2$ rotational temperatures are $>$100 K. The small size (0.09 pc) and high H$_2$ number density ($6.3\times10^5$ cm$^{-3}$) further support the hot core classification of N\,160A--mm\,A. This marks the eighth detection of a hot core in the LMC and the tenth extragalactic hot core.

    \item Our XCLASS analysis indicates that three out of the remaining five 1.2 mm continuum sources in N\,160A--mm are cold cores:  N\,160A--mm\,C, D, and E. Sources B and F were not analyzed with XCLASS due to an insufficient number of line detections. 
    
      \item The hot core N\,160A--mm\,A has the largest number of molecular line detections.  Methanol, SO, CS, and H$^{13}$CO$^+$  are the only species detected toward all spectrally analyzed sources (A, C, D, E). The extended emission from all these species is detected throughout the ALMA field with the CS emission being most ubiquitous.  The origin of the cold CH$_3$OH emission is uncertain. These results are consistent with those we obtained for the three ALMA fields in the star-forming region N\,105  (see \citealt{sewilo2022}).  
    
    \item The molecular abundances with respect to H$_2$ of the hot core N\,160A--mm A are higher than those measured toward sources C, D, and E except for CS; the CS emission peak is offset from the source A's 1.2 mm continuum peak toward northeast.

    \item We compared the molecular abundances with respect to H$_2$  measured toward the newly identified hot core  N\,160A--mm\,A to those reported for other LMC (7) and the SMC (2) hot cores, and a sample of Galactic hot cores.  The CH$_3$OH abundance for N\,160A--mm\,A falls roughly in the middle of the range of other known LMC and SMC hot cores with COMs and thus it fits the  previously observed trends, i.e., the large CH$_3$OH abundance variations over the 0.2--1 $Z_{\odot }$ metallicity range, with significantly smaller variations observed for S-bearing species such as SO and SO$_2$.  

    \item The CH$_3$OH abundance toward N\,160A--mm\,A falls roughly in the middle of the previously known LMC hot core abundances. The lower value of the CH$_3$OH abundance (compared to most of the known LMC hot cores with measured CH$_3$OH abundances) could be explained by N\,160A--mm\,A being located in the Molecular Ridge which is thought to have been impacted by a tidal interaction between the LMC and SMC depositing the SMC's lower metallicity gas into the LMC. In a lower-metallicity environment the dust temperature is higher, resulting is the less efficient CH$_3$OH formation. Other potential factors that may impact COM abundances include evolutionary age, the presence and strength of shocks, or local UV radiation field strength.

    \item The intensity weighted velocity maps of the ALMA field N\,160A--mm reveal complex kinematics, in particular toward the hot core, source A.  The complex velocity structure toward N\,160A--mm\,A  indicates the presence of the bulk motions and/or multiple velocity components and is reflected in the observed asymmetric and multi-peaked spectral lines.  Additionally,  the spectral lines detected toward source A are broader than those observed toward other sources in the field.

    \item Based on the detailed analysis of the multi-wavelength data (from the near-IR to radio regime) we conclude that the newly detected hot core N\,160A--mm\,A is the most massive member of the protocluster forming at the intersection of the filaments. 

\end{enumerate}

Efforts to increase the number of known extragalactic hot cores continue in the LMC and SMC. It is important to build a more representative sample of the Magellanic hot cores to better understand the impact of the subsolar metallicity and higher UV radiation field on the production of COMs.

\begin{acknowledgments}
We thank the anonymous reviewer who provided insightful and helpful comments which allowed us to improve our manuscript. We thank Alberto Bolatto and Elizabeth Tarantino for making the ALMA ACA Band 6 mosaic covering the star-forming region N\,160 available to us. This material is based upon work supported by the National Science Foundation Graduate Research Fellowship Program under Grant No. DGE 2236417. Any opinions, findings, and conclusions or recommendations expressed in this material are those of the author(s) and do not necessarily reflect the views of the National Science Foundation. The material is based upon work supported by NASA under award number 80GSFC24M0006 (M.S.). RI was partially supported for this work by NSF AAG award 2009624. The National Radio Astronomy Observatory is a facility of the National Science Foundation operated under cooperative agreement by Associated Universities, Inc. This paper makes use of the following ALMA data: ADS/JAO.ALMA\#2019.1.01720.S, ADS/JAO.ALMA\#2019.1.01770.S, and \\ ADS/JAO.ALMA\#2017.1.00093.S. ALMA is a partnership of ESO (representing its member states), NSF (USA) and NINS (Japan), together with NRC (Canada), NSC and ASIAA (Taiwan), and KASI (Republic of Korea), in cooperation with the Republic of Chile. The Joint ALMA Observatory is operated by ESO, AUI/NRAO and NAOJ. This research made use of APLpy, an open-source plotting package for Python \citep{robitaille2012}.
\end{acknowledgments}

\clearpage

\appendix
\counterwithin{figure}{section}
\section{The XCLASS Fitting Results} \label{a:xclass}

Table \ref{t:XCLASS Params} lists the XCLASS best fit parameters for all molecular species (X) detected toward sources A, C, D, and E in the ALMA field N\,160A--mm: the rotational temperature ($T_{\rm rot}$), total column density ($N$), velocity ($v_{\rm LSR}$), and line width ($\Delta v_{\rm FWHM}$). Also included is the estimated abundance with respect to H$_2$ for all species, $N$(X)/N(H$_2$). The detailed discussion on the XCLASS fitting and the H$_2$ column density determination can be found in Sections~\ref{xclass} and \ref{ss:h2columndensity}, respectively.

\clearpage
\startlongtable
\begin{deluxetable*}{cccccccc}
\centering
\tablecaption{Results of XCLASS Modeling and Fractional Abundances with Respect to H$_2$\label{t:XCLASS Params}}
\tablewidth{0pt}
\tablehead{
\colhead{Source} &
\colhead{Species, X} &
\colhead{$T_{\rm rot}$(X)\tablenotemark{\footnotesize a}} &
\colhead{$N$(X)}&
\colhead{$v_{\rm LSR}$ \tablenotemark{\footnotesize a}} & 
\colhead{$\Delta v_{\rm FWHM}$ \tablenotemark{\footnotesize a}} & 
\colhead{$N$(X)/N(H$_2$)} \\
\colhead{} &
\colhead{} & 
\colhead{(K)} &
\colhead{(cm$^{-2}$)} &
\colhead{(km s$^{-1}$)} &
\colhead{(km s$^{-1}$)} &
\colhead{}
}
\startdata
 A & CH$_3$OH & 188.36 $\pm5.48$ & (2.51 $\pm0.12$) $\times$ $10^{15}$ & 233.62 $\pm0.06$& 5.43 $\pm0.15$& (1.40 $\pm 0.16$) $\times$ $10^{-8}$ \\
  & CH$_3$CN & 83.64 $\pm19.15$ & (7.44 $\pm0.80$) $\times$ $10^{12}$ & 233.22 $\pm0.24$& 4.20 $\pm0.53$& (4.13 $\pm 0.62$) $\times$ $10^{-11}$ \\
  & SO$_2$ & 178.16 $\pm6.71$ & (3.20 $\pm0.11$) $\times$ $10^{15}$ & 235.57 $\pm0.08$& 5.88 $\pm0.17$& (1.78 $\pm 0.20$) $\times$ $10^{-8}$ \\
  & SO$_2$ (cold) & 35.17 $\pm2.06$ & (3.06 $\pm0.27$) $\times$ $10^{14}$ & 236.48 $\pm0.05$& 3.25 $\pm0.12$& (1.70 $\pm 0.23$) $\times$ $10^{-9}$ \\
  & $^{34}$SO$_2$ & 120.27 $\pm19.12$ & (1.51 $\pm0.19$) $\times$ $10^{14}$ & 236.16 $\pm0.10$& 3.16 $\pm0.26$& (8.39 $\pm 1.36$) $\times$ $10^{-10}$ \\
  & SO & 178 & (1.564 $\pm0.007$) $\times$ $10^{15}$ & 236.82 $\pm0.01$& 4.54 $\pm0.02$& (8.68 $\pm 0.91$) $\times$ $10^{-9}$ \\
  & $^{33}$SO & 178 & (1.27 $\pm0.08$) $\times$ $10^{14}$ & 236.08 $\pm0.21$& 5.92 $\pm0.66$& (7.03 $\pm 0.86$) $\times$ $10^{-10}$ \\
  & CS & 188 & (1.95 $\pm0.02$) $\times$ $10^{14}$ & 236.72 $\pm0.02$& 5.74 $\pm0.06$& (1.08 $\pm 0.11$) $\times$ $10^{-9}$ \\
  & C$^{33}$S & 188 & $<1.27 \times10^{12}$ & 233.62 & 5.43 & $<7.04 \times10^{-12}$ \\
  & OCS & 188 & (1.54 $\pm0.22$) $\times$ $10^{14}$ & 234.92 $\pm0.43$& 5.62 $\pm1.03$& (8.57 $\pm 1.53$) $\times$ $10^{-10}$ \\
  & H$_2$CS & 188 & $<5.90 \times 10^{13}$ & 233.62 & 5.43 & $<3.28 \times10^{-10}$ \\
  & H$^{13}$CN & 188 & (1.36 $\pm0.07$) $\times$ $10^{13}$ & 235.87 $\pm0.12$& 5.19 $\pm0.30$& (7.56 $\pm 0.88$) $\times$ $10^{-11}$ \\
  & HC$^{15}$N & 188 & (7.62 $\pm0.65$) $\times$ $10^{12}$ & 234.87 $\pm0.31$& 7.20 $\pm0.77$& (4.23 $\pm 0.57$) $\times$ $10^{-11}$ \\
  & H$^{13}$CO$^+$ & 188 & (7.19 $\pm0.17$) $\times$ $10^{12}$ & 237.15 $\pm0.10$ & 3.90 $\pm0.10$& (3.99 $\pm 0.43$) $\times$ $10^{-11}$ \\
  & HDO & 188 & (1.65 $\pm0.36$) $\times$ $10^{14}$ & 235.88 $\pm0.26$ & 1.92 $\pm0.59$& (9.14 $\pm 2.19$) $\times$ $10^{-10}$ \\
  & SiO & 188 & (1.92 $\pm0.07$) $\times$ $10^{13}$ & 237.00 $\pm0.10$ & 5.62 $\pm0.24$& (1.07 $\pm 0.12$) $\times$ $10^{-10}$ \\
  & HCOOCH$_3$ & 188 & $< 7.89\times10^{13}$ & 233.62 & 5.43 & $< 4.38\times10^{-10}$ \\
  & CH$_3$OCH$_3$ & 188 & $< 5.22\times10^{13}$ & 233.62 & 5.43 & $< 2.90\times10^{-10}$ \\
  & NH$_2$CHO & 188 & $< 6.12\times10^{12}$ & 233.62 & 5.43 & $< 3.40\times10^{-11}$ \\
  & C$_2$H$_5$CN & 188 & $< 5\times10^{8}$ & 233.62 & 5.43 & $< 2.78\times10^{-15}$ \\
  & CH$_3$CHO & 188 & $< 1.38\times10^{13}$ & 233.62 & 5.43 & $< 7.68\times10^{-11}$ \\
  & C$_2$H$_5$OH & 188 & $< 5\times10^{8}$ & 233.62 & 5.43 & $< 2.78\times10^{-15}$ \\
  \hline
 C & CH$_3$OH\tablenotemark{\footnotesize b} & 28.76 $\pm2.22$ & (5.75 $\pm0.16$) $\times$ $10^{14}$ & 235.86 $\pm0.02$& 3.34 $\pm0.04$& (1.91 $\pm 0.28$) $\times$ $10^{-9}$ \\ 
   & CH$_3$CN & 29 & $ < 3.15\times10^{12}$ & 235.86 & 3.34 & $<1.04\times10^{-11}$ \\
   & SO$_2$ & 31.98 $\pm2.43$ & (7.89 $\pm0.58$) $\times$ $10^{13}$ & 235.38 $\pm0.24$& 6.40 $\pm0.58$& (2.62 $\pm 0.42$) $\times$ $10^{-10}$ \\
   & $^{34}$SO$_2$ & 32 & $< 3.40\times10^{9}$ & 235.38& 6.40 & $<1.13\times10^{-14}$ \\
   & SO & 32 & (2.93 $\pm0.03$) $\times$ $10^{14}$ & 235.82 $\pm0.02$& 3.57 $\pm0.04$& (9.74 $\pm 1.39$) $\times$ $10^{-10}$ \\
   & $^{33}$SO & 32 & $< 8.27\times10^{11}$ & 235.38& 6.40 & $<2.75\times10^{-12}$ \\
   & CS & 29 & (2.037 $\pm$0.006) $\times$ $10^{14}$ & 235.700 $\pm0.004$& 3.22 $\pm0.01$& (6.76 $\pm 0.96$) $\times$ $10^{-10}$ \\
   & C$^{33}$S & 29 & (3.91 $\pm0.35$) $\times$ $10^{12}$ & 235.62 $\pm0.18$& 3.17 $\pm0.12$& (1.30 $\pm0.22 $) $\times$ $10^{-11}$ \\
   & OCS & 29 & $ < 2.53\times10^{14}$ & 235.86 & 3.34 & $< 8.40\times10^{-10}$ \\
   & H$_2$CS & 29 & (5.73 $\pm0.19$) $\times$ $10^{13}$ & 235.60 $\pm0.06$& 2.67 $\pm0.07$& (1.90 $\pm 0.28$) $\times$ $10^{-10}$ \\
   & H$^{13}$CN & 29 & (1.78 $\pm0.10$) $\times$ $10^{12}$ & 235.70 $\pm0.11$& 2.05 $\pm0.11$& (5.90 $\pm 0.91$) $\times$ $10^{-12}$ \\
   & HC$^{15}$N & 29 & (1.04 $\pm0.09$) $\times$ $10^{12}$ & 235.44 $\pm0.23$& 2.63 $\pm0.13$& (3.44 $\pm 0.57$) $\times$ $10^{-12}$ \\
   & H$^{13}$CO$^+$ & 29 & (9.27 $\pm0.31$) $\times$ $10^{11}$ & 235.47 $\pm0.05$& 2.81 $\pm0.10$& (3.08 $\pm 0.45$) $\times$ $10^{-12}$ \\
   & HDO & 29 & $ < 1.79\times10^{9}$ & 235.86 & 3.34 & $<5.94\times10^{-15}$ \\
   & SiO & 29 & (1.74 $\pm0.19$) $\times$ $10^{12}$ & 235.46 $\pm0.23$& 3.16 $\pm0.20$& (5.78 $\pm 1.04$) $\times$ $10^{-12}$ \\
 \hline
 D & CH$_3$OH \tablenotemark{\footnotesize b} & 25.08 $\pm1.79$ & (7.39 $\pm0.29$) $\times$ $10^{14}$ & 237.71 $\pm0.01$& 3.71 $\pm0.03$& (2.10 $\pm 0.30$) $\times$ $10^{-9}$ \\
  & CH$_3$CN & 25 & $<1.79\times10^{9}$ & 237.71 & 3.71 & $<5.10\times10^{-15}$ \\
  & SO$_2$ & 25 & (7.94 $\pm0.32$) $\times$ $10^{13}$ & 237.07 $\pm0.08$& 3.14 $\pm0.09$& (2.26 $\pm 0.33$) $\times$ $10^{-10}$ \\
  & $^{34}$SO$_2$ & 25 & $<5.62\times10^{8}$ & 237.71 & 3.71 & $<1.60\times10^{-15}$ \\
  & SO & 25 & (4.35 $\pm0.03$) $\times$ $10^{14}$ & 237.43 $\pm0.01$& 3.48 $\pm0.03$& (1.24 $\pm 0.17$) $\times$ $10^{-9}$ \\
  & $^{33}$SO & 25 & $<1.02\times10^{13}$ & 237.71 & 3.71 & $<2.90\times10^{-11}$ \\
  & CS & 25 & (1.632 $\pm0.005$) $\times$ $10^{14}$ & 237.785 $\pm0.006$& 3.74 $\pm0.01$& (4.64 $\pm 0.65$) $\times$ $10^{-10}$ \\
  & C$^{33}$S & 25 & (4.12 $\pm0.36$) $\times$ $10^{12}$ & 238.38 $\pm0.19$& 2.99 $\pm0.15$& (1.17 $\pm 0.19$) $\times$ $10^{-11}$ \\
  & OCS & 25 & $<2.61\times10^{14}$ & 237.71 & 3.71 & $<7.42\times10^{-10}$ \\
  & H$_2$CS & 25 & (5.09 $\pm0.20$) $\times$ $10^{13}$ & 238.39 $\pm0.07$& 2.89 $\pm0.10$& (1.45 $\pm 0.21$) $\times$ $10^{-10}$ \\
  & H$^{13}$CN & 25 & (1.41 $\pm0.09$) $\times$ $10^{12}$ & 238.52 $\pm0.18$& 3.55 $\pm0.15$& (4.02 $\pm 0.62$) $\times$ $10^{-12}$ \\
  & HC$^{15}$N & 25 & (9.51 $\pm0.70$) $\times$ $10^{11}$ & 238.13 $\pm0.15$& 5.13 $\pm0.14$& (2.71 $\pm 0.43$) $\times$ $10^{-12}$ \\
  & H$^{13}$CO$^+$ & 25 & (8.16 $\pm0.31$) $\times$ $10^{11}$ & 237.59 $\pm0.07$& 3.99 $\pm0.16$& (2.32 $\pm 0.34$) $\times$ $10^{-12}$ \\
  & HDO & 25 & $<2.27\times10^{12}$ & 237.71 & 3.71 & $<6.46\times10^{-12}$ \\
  & SiO & 25 & $<5.18\times10^{11}$ & 237.71 & 3.71 & $<1.47\times10^{-12}$ \\
  \hline
 E & CH$_3$OH \tablenotemark{\footnotesize b} & 9.14 $\pm0.81$ & (1.33 $\pm0.34$) $\times$ $10^{15}$ & 238.05 $\pm0.02$& 1.68 $\pm0.05$& (2.58 $\pm 0.82$) $\times$ $10^{-9}$ \\
  & CH$_3$CN & 10 & $<4.18\times10^{13}$ & 238.05 & 1.68 & $<8.11\times 10^{-11}$ \\
  & SO$_2$ & 10 & $<2.23\times10^{13}$ & 238.05 & 1.68 & $<4.34\times 10^{-11}$ \\
  & $^{34}$SO$_2$ & 10 & $<1.08\times10^{13}$ & 238.05 & 1.68 & $<2.10\times 10^{-11}$ \\
  & SO & 10 & (1.41 $\pm0.05$) $\times$ $10^{15}$ & 237.91 $\pm0.03$& 2.02 $\pm0.08$& (2.74 $\pm 0.52$) $\times$ $10^{-9}$ \\
  & $^{33}$SO & 10 & $<2.08\times10^{13}$ & 238.05 & 1.68 & $<4.03\times 10^{-11}$ \\
  & CS & 10 & (8.73 $\pm0.78$) $\times$ $10^{14}$ & 237.97 $\pm0.01$& 1.50 $\pm0.04$& (1.69 $\pm 0.35$) $\times$ $10^{-9}$ \\
  & C$^{33}$S & 10 & (1.68 $\pm0.15$) $\times$ $10^{13}$ & 238.17 $\pm0.11$& 1.62 $\pm0.39$& (3.27 $\pm 0.69$) $\times$ $10^{-11}$ \\
  & OCS & 10 & $<5.61\times10^{8}$ & 238.05 & 1.68 & $<1.09\times 10^{-15}$ \\
  & H$_2$CS & 10 & (1.55 $\pm0.21$) $\times$ $10^{14}$ & 238.18 $\pm0.12$& 1.50 $\pm0.95$& (3.01 $\pm 0.70$) $\times$ $10^{-10}$ \\
  & H$^{13}$CN & 10 & $<1.06\times10^{12}$ & 238.05 & 1.68 & $<2.06\times 10^{-12}$ \\
  & HC$^{15}$N & 10 & $<6.95\times10^{11}$ & 238.05 & 1.68 & $<1.35\times 10^{-12}$ \\
  & H$^{13}$CO$^+$ & 10 & (8.64 $\pm0.79$) $\times$ $10^{11}$ & 237.88 $\pm0.10$& 2.00 $\pm0.26$& (1.68 $\pm 0.35$) $\times$ $10^{-12}$ \\
  & HDO & 10 & $<8.48\times10^{15}$ & 238.05 & 1.68 & $<1.65\times 10^{-8}$ \\
  & SiO & 10 & $<1.72\times10^{12}$ & 238.05 & 1.68 & $<3.34\times 10^{-12}$ \\
\enddata
\tablenotetext{a}{Values without errors are fixed as described in Section \ref{xclass}}
\tablenotetext{b}{XCLASS analysis was performed in non-LTE with H$_2$ as the collision partner. Collision partner number densities are $5.8 \times 10^6$ cm$^{-3}$, $3.9 \times 10^6$ cm$^{-3}$, and $4.5 \times 10^8$ cm$^{-3}$ for sources C, D, and E, respectively.}
\end{deluxetable*}

\section{Observed Spectra} \label{a:spectra}

\begin{figure*}
    \centering
    \includegraphics[width=\textwidth]{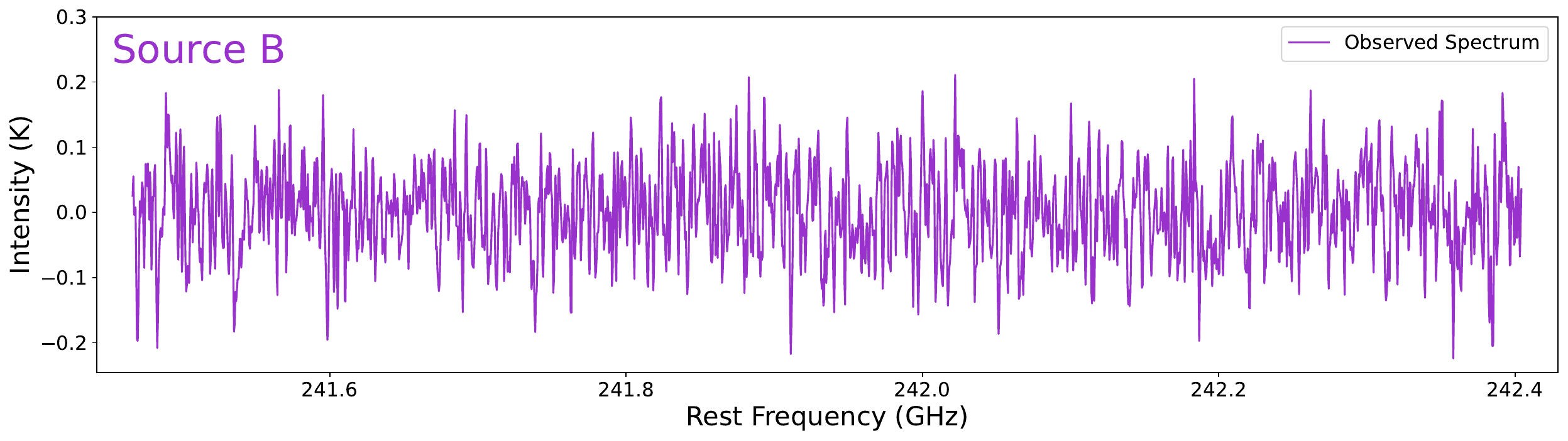} \\
    \includegraphics[width=\textwidth]{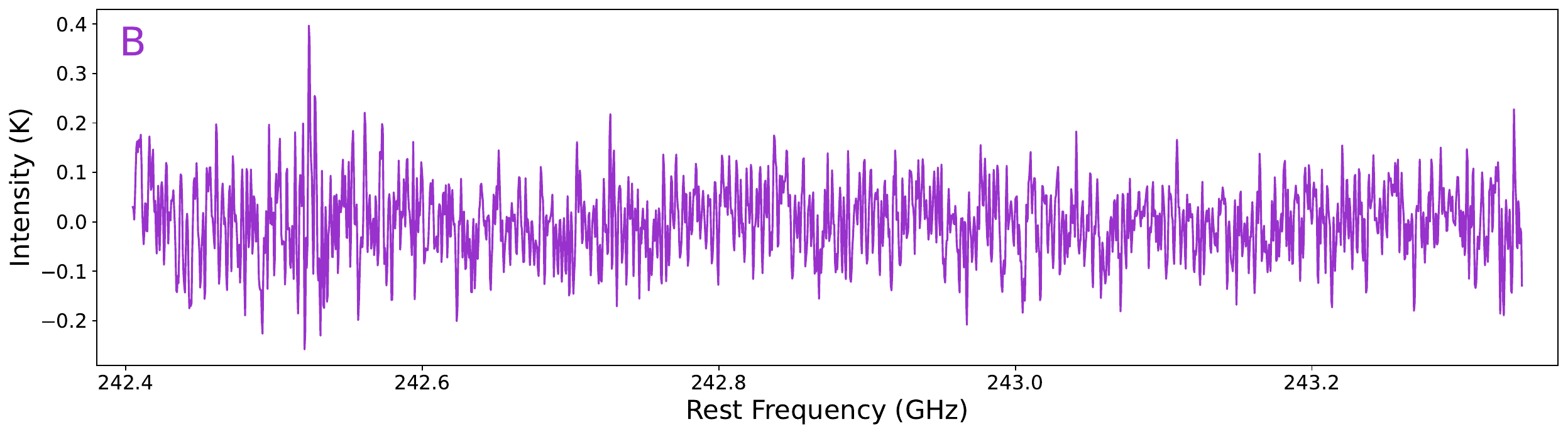} \\
    \includegraphics[width=\textwidth]{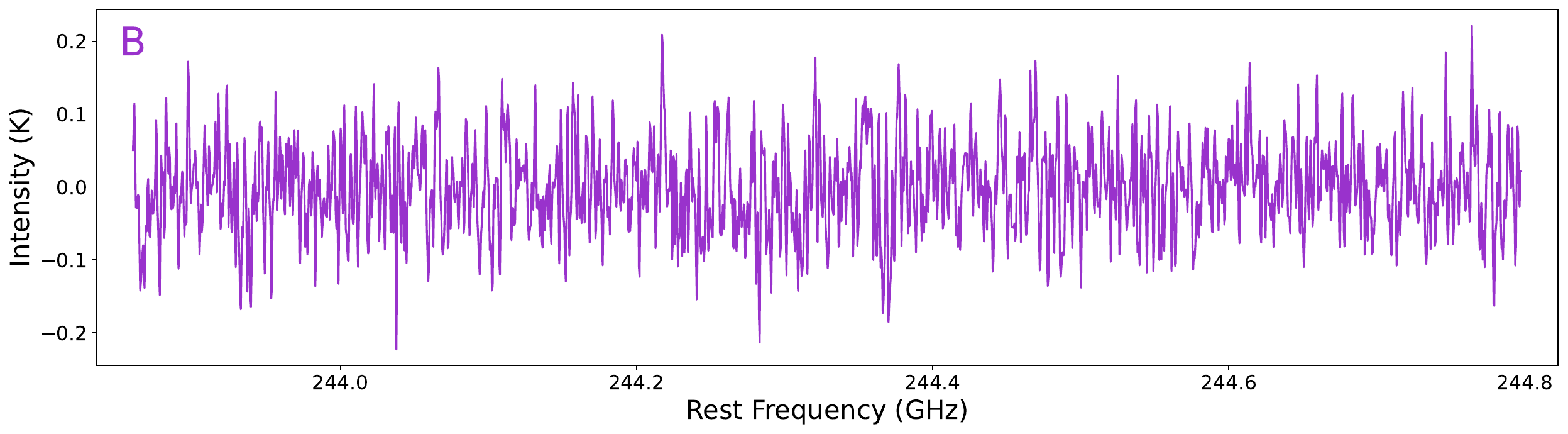} \\
    \includegraphics[width=\textwidth]{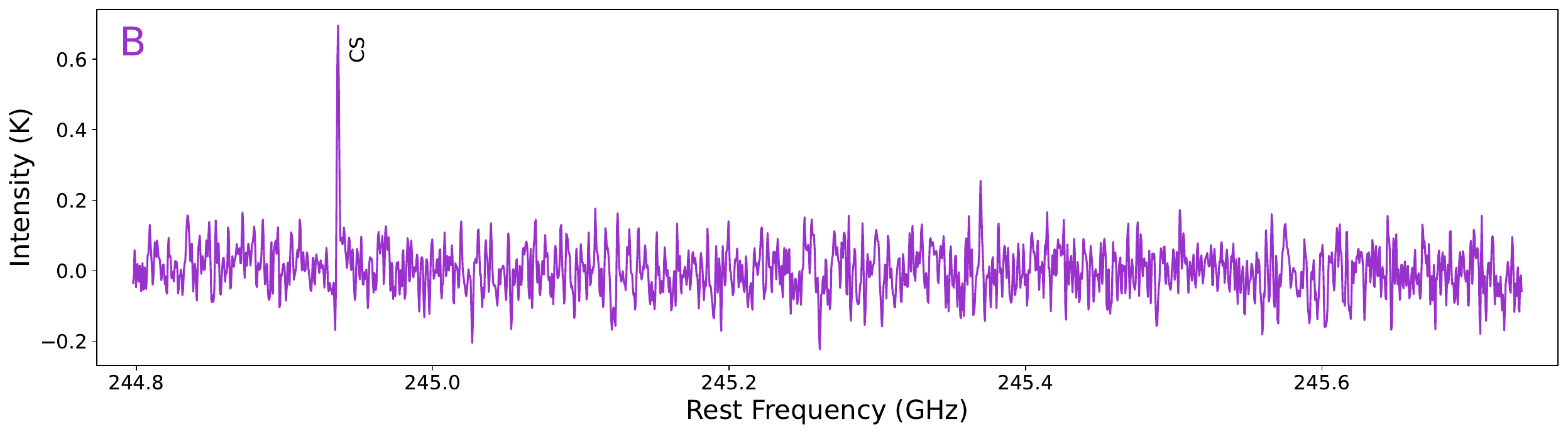} \\
    \caption{Spectra of N\,160A--mm B for spectral windows 242 GHz and 245 GHz. Detected lines above $5\sigma_{\rm s,i}$ are labeled with black solid lines, detected lines between $3\sigma_{\rm s,i}$ and $5\sigma_{\rm s,i}$ are labeled with black dashed lines, and tentative detections are labeled with gray dotted lines.}
    \label{fig:SpectraB1}
\end{figure*}

\begin{figure*}
    \centering
    \includegraphics[width=\textwidth]{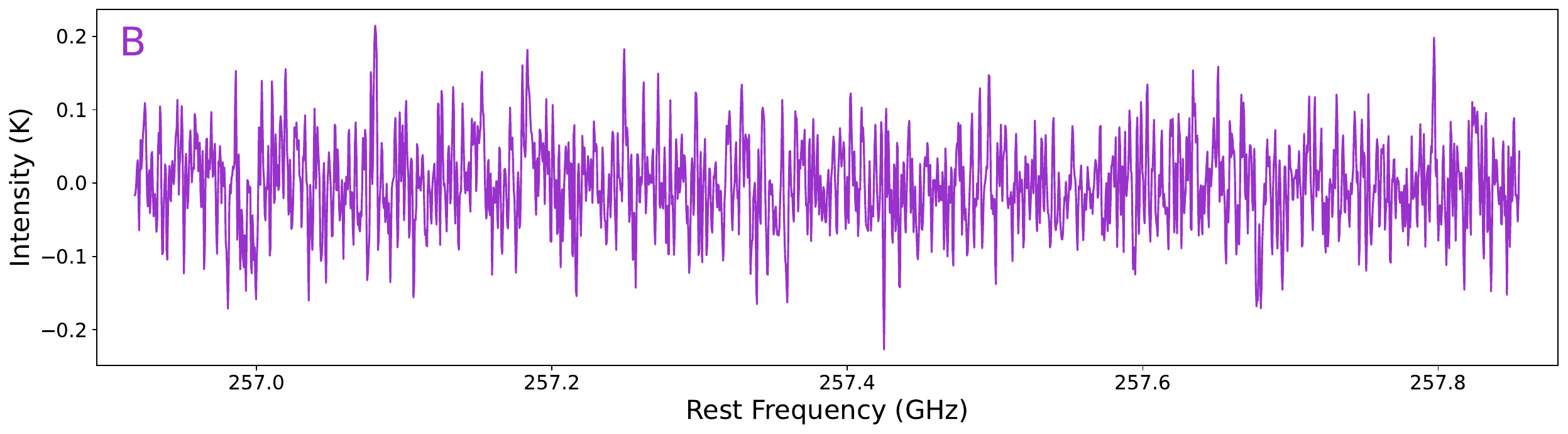} \\
    \includegraphics[width=\textwidth]{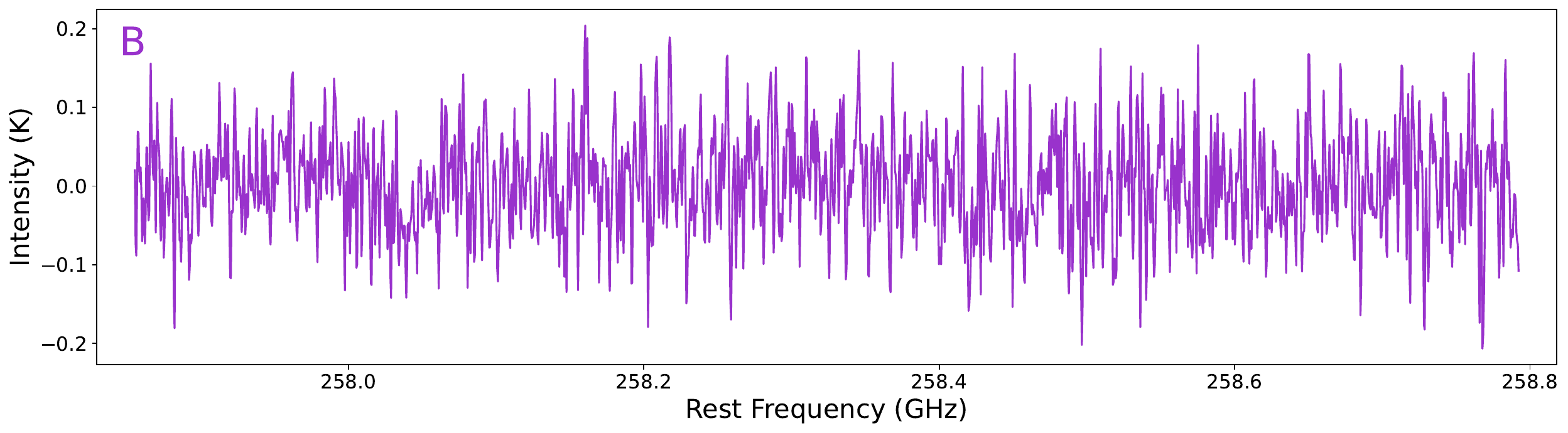} \\
    \includegraphics[width=\textwidth]{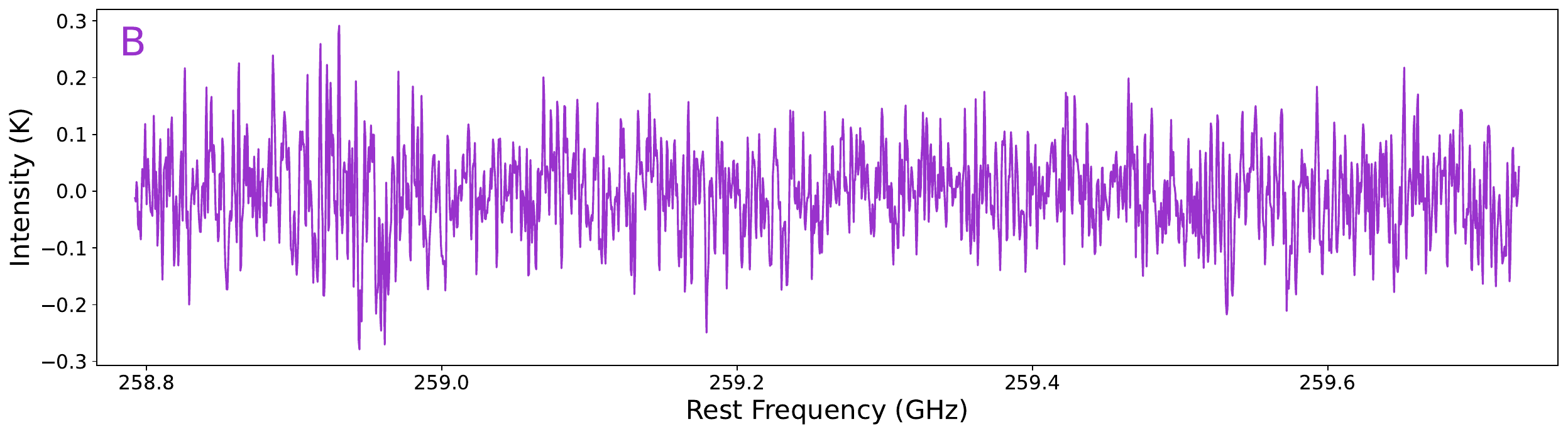} \\
    \includegraphics[width=\textwidth]{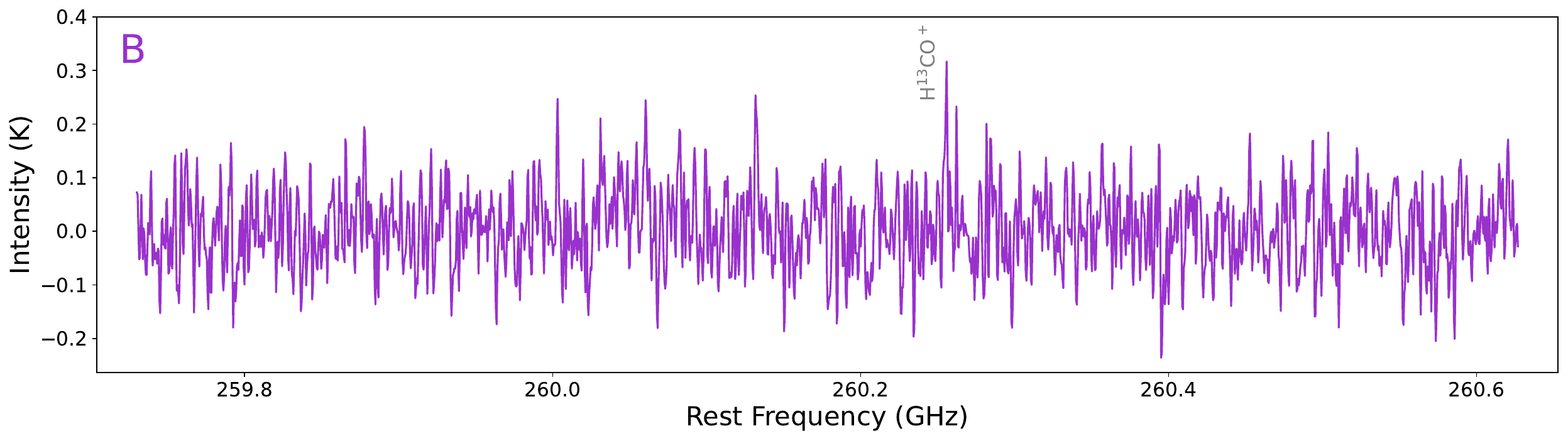} \\
    \caption{Spectra of N\,160A--mm B for spectral windows 258 GHz and 260 GHz.}
    \label{fig:SpectraB2}
\end{figure*}

\begin{figure*}
    \centering
    \includegraphics[width=\textwidth]{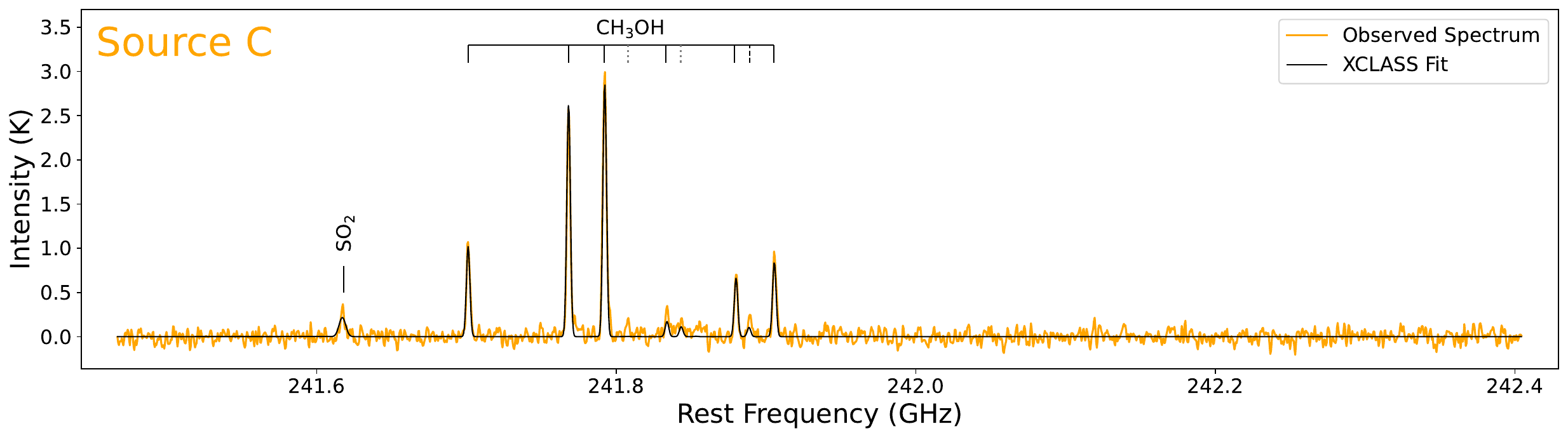} \\
    \includegraphics[width=\textwidth]{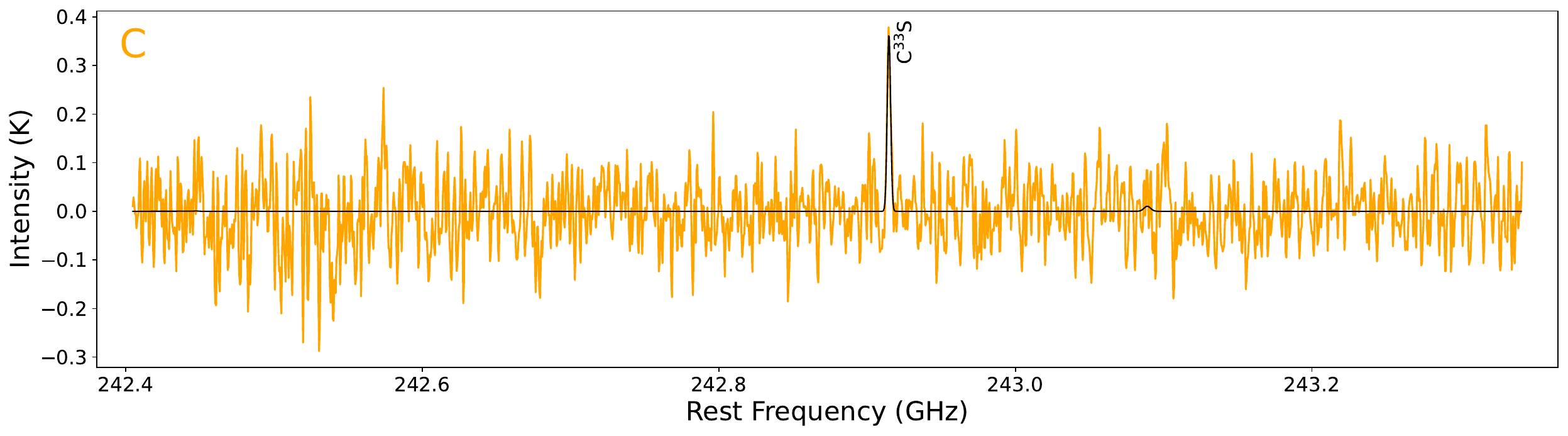} \\
    \includegraphics[width=\textwidth]{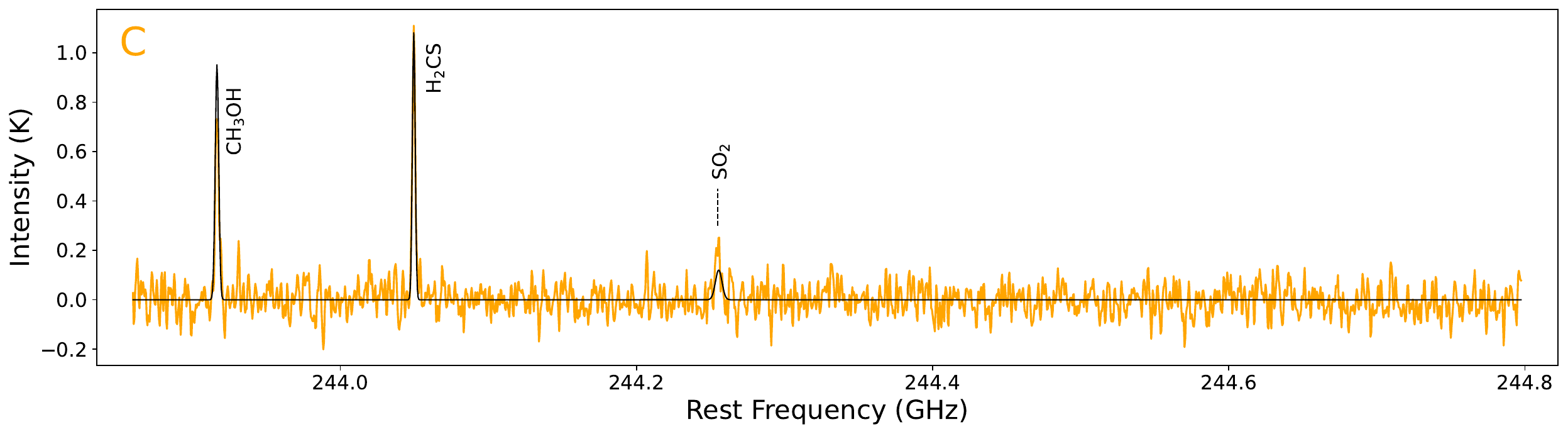} \\
    \includegraphics[width=\textwidth]{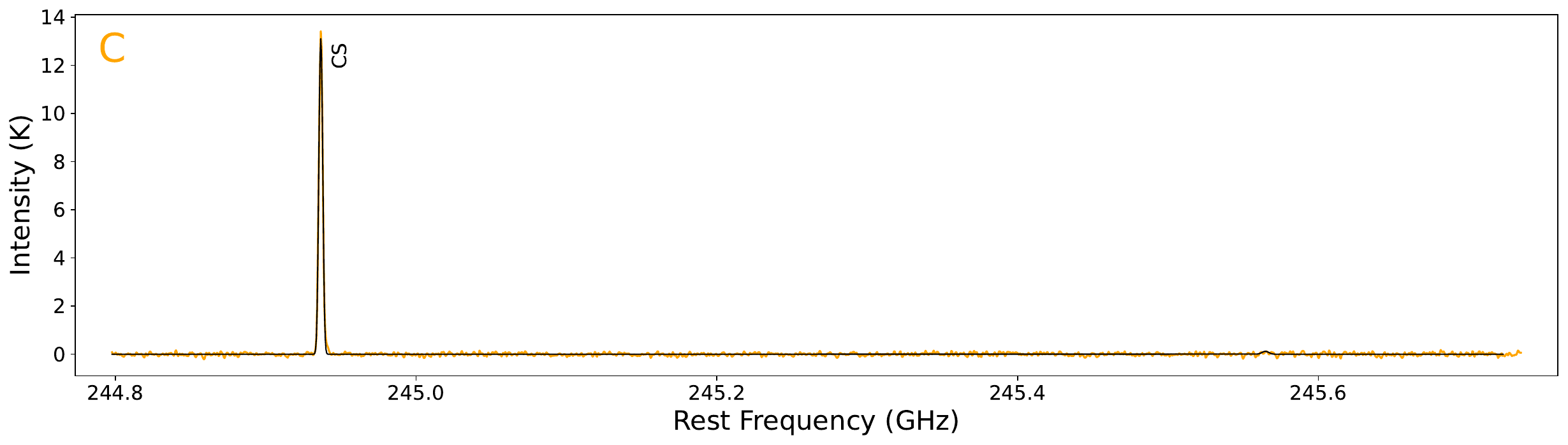} \\
    \caption{Spectra of N\,160A--mm C for spectral windows 242 GHz and 245 GHz. The yellow solid line shows the observed spectra and the black solid line shows the XCLASS best fit. Detected lines above $5\sigma_{\rm s,i}$ are labeled with black solid lines, detected lines between $3\sigma_{\rm s,i}$ and $5\sigma_{\rm s,i}$ are labeled with black dashed lines, and tentative detections are labeled with gray dotted lines.}
    \label{fig:SpectraC1}
\end{figure*}

\begin{figure*}
    \centering
    \includegraphics[width=\textwidth]{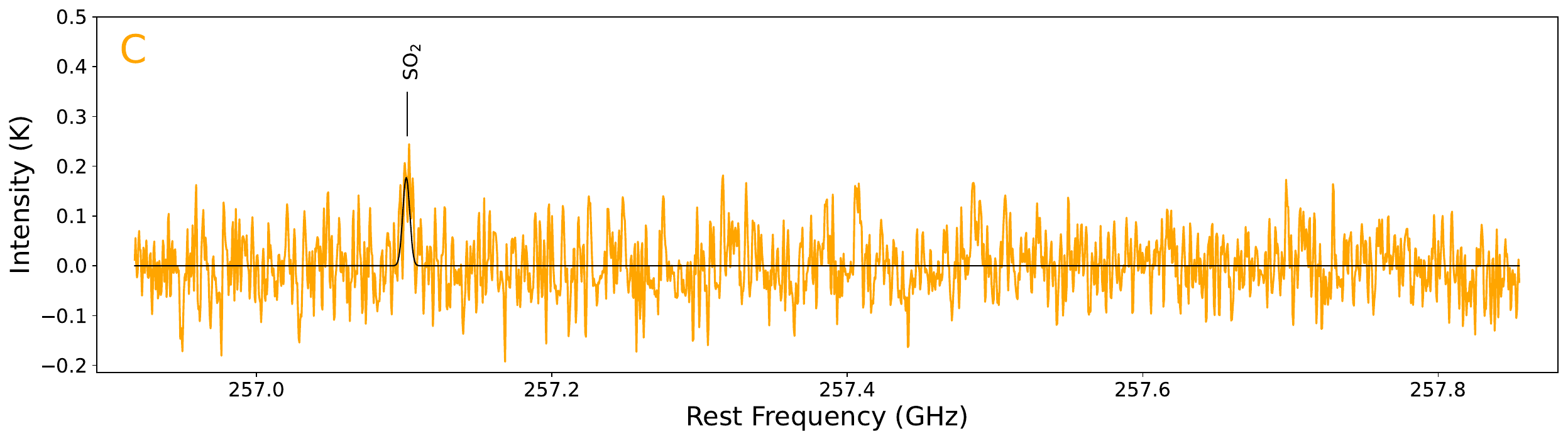} \\
    \includegraphics[width=\textwidth]{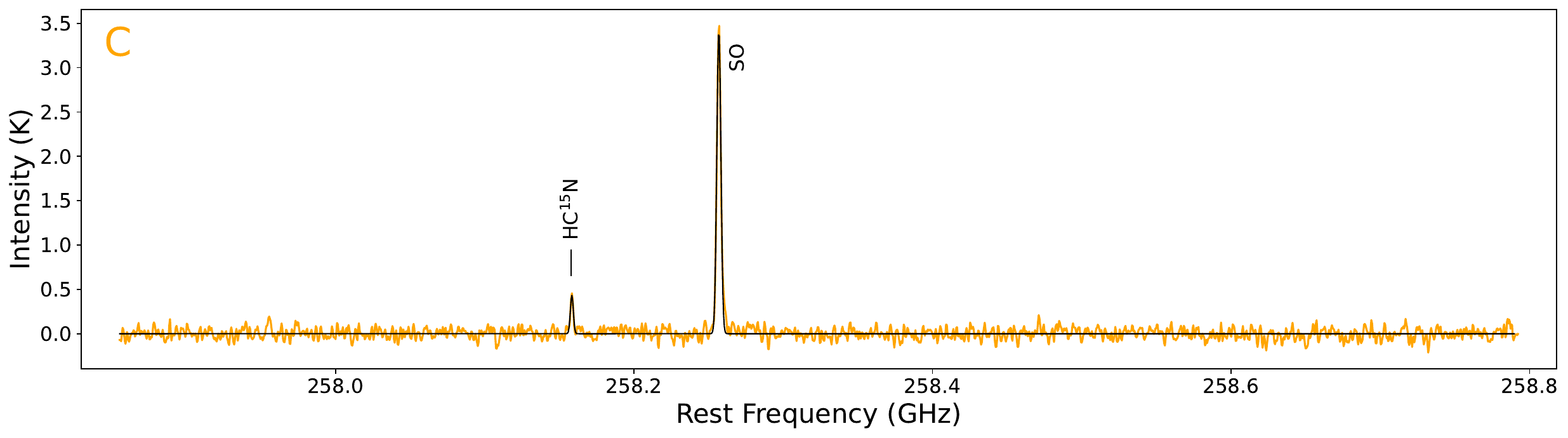} \\
    \includegraphics[width=\textwidth]{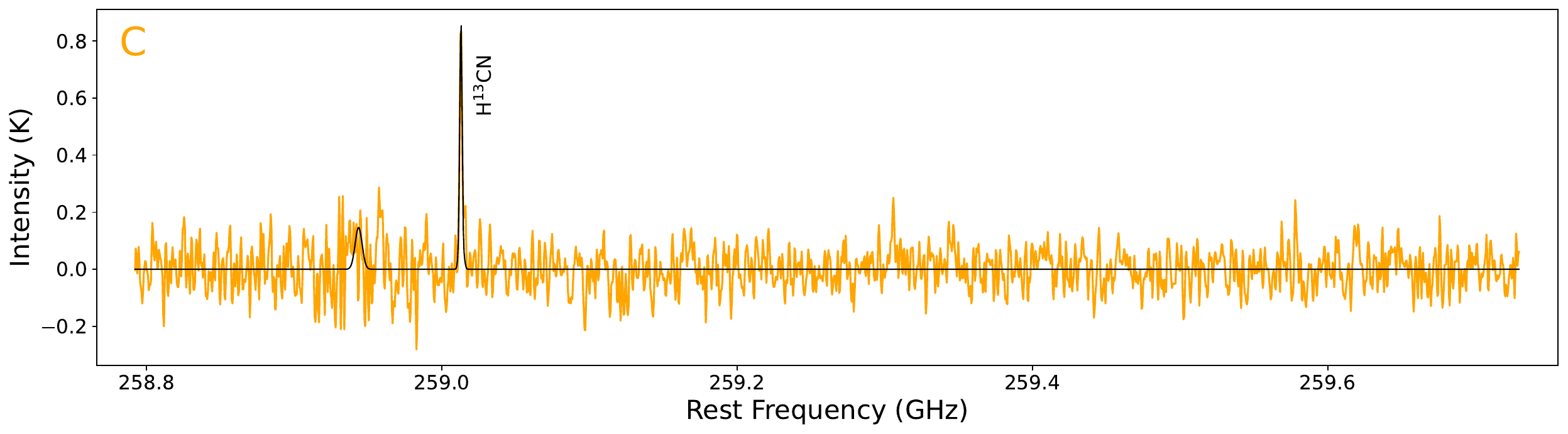} \\
    \includegraphics[width=\textwidth]{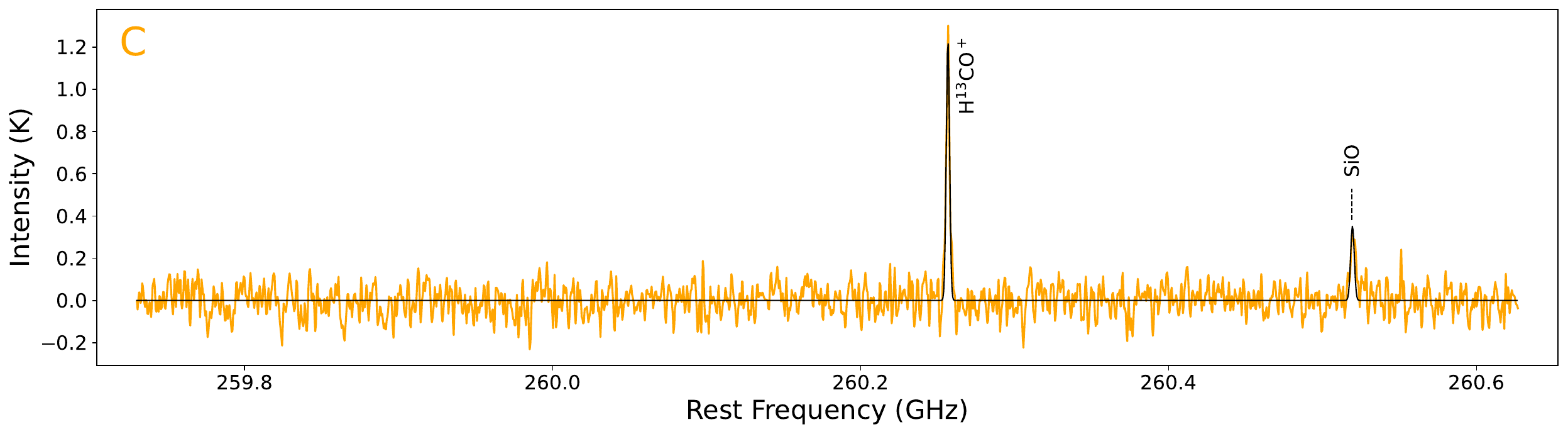} \\
    \caption{Spectra of N\,160A--mm C for spectral windows 258 GHz and 260 GHz.}
    \label{fig:SpectraC2}
\end{figure*}

\begin{figure*}
    \centering
    \includegraphics[width=\textwidth]{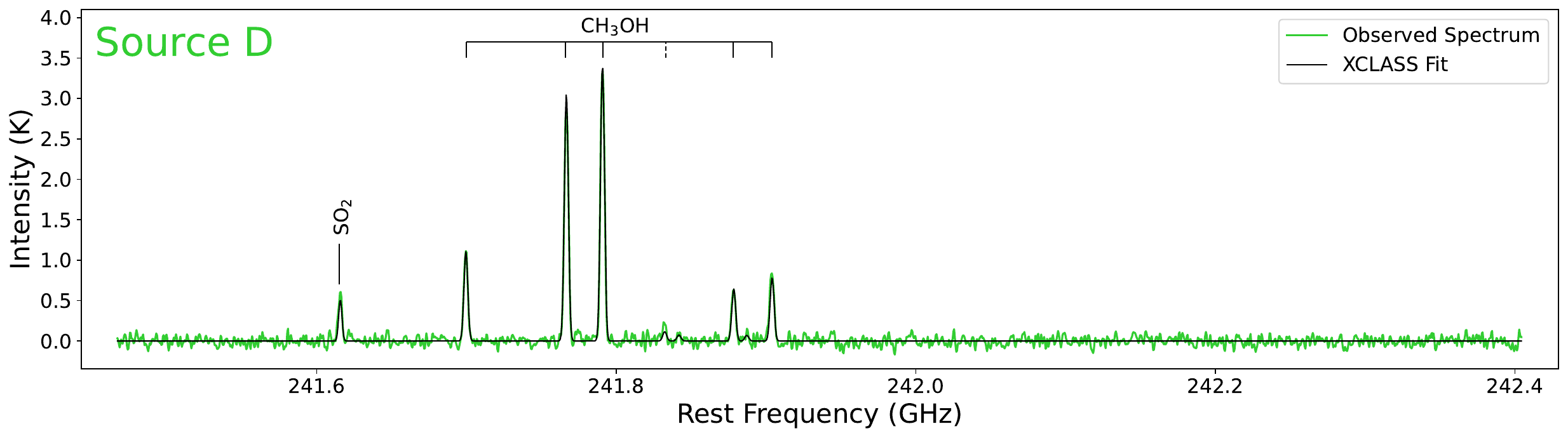} \\
    \includegraphics[width=\textwidth]{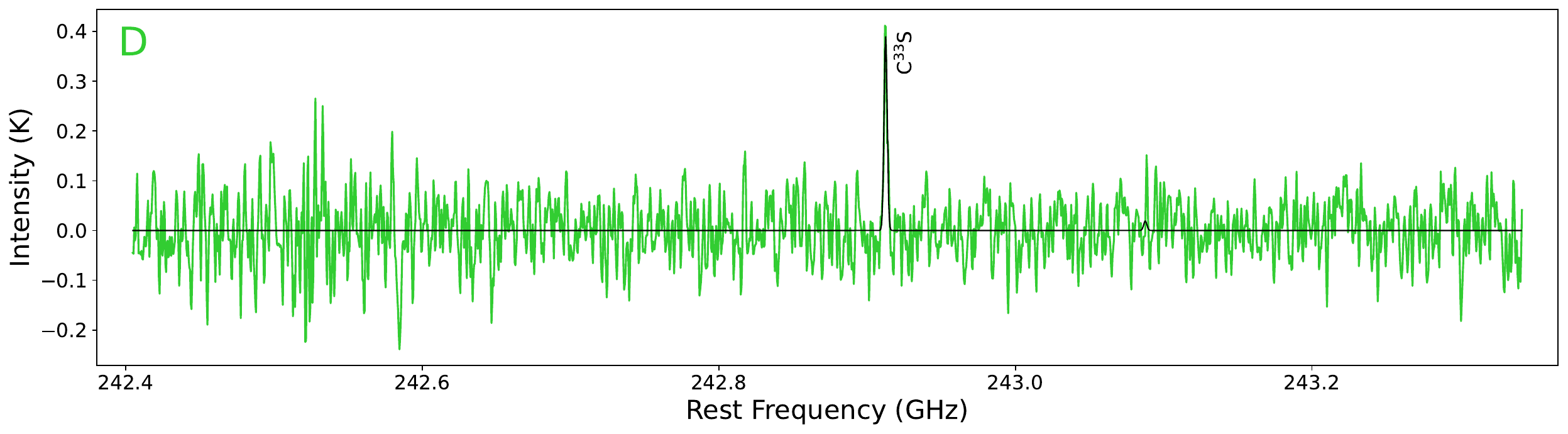} \\
    \includegraphics[width=\textwidth]{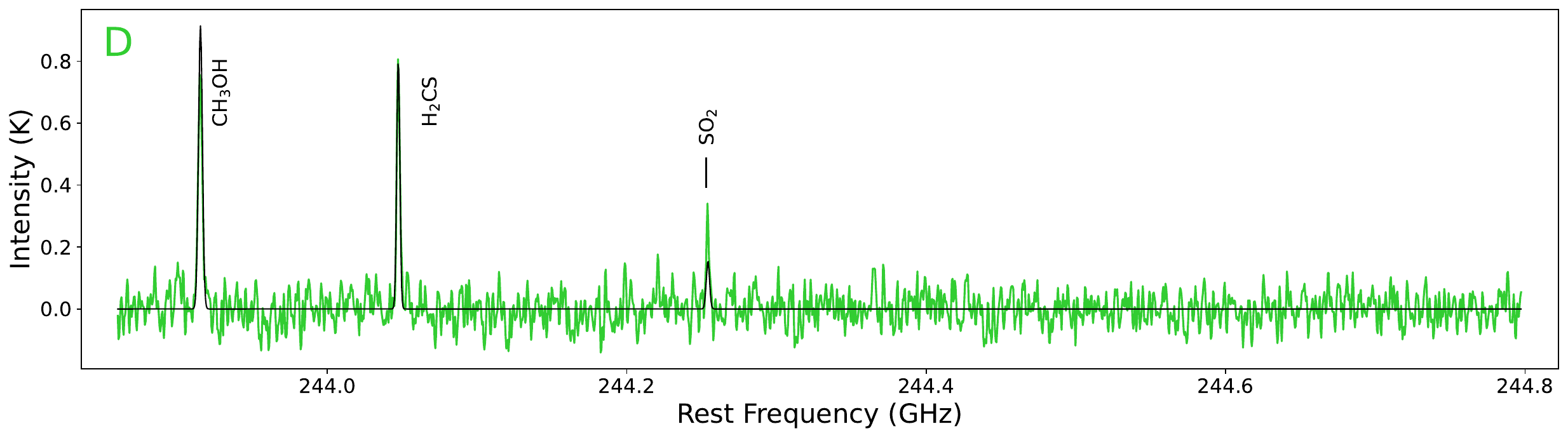} \\
    \includegraphics[width=\textwidth]{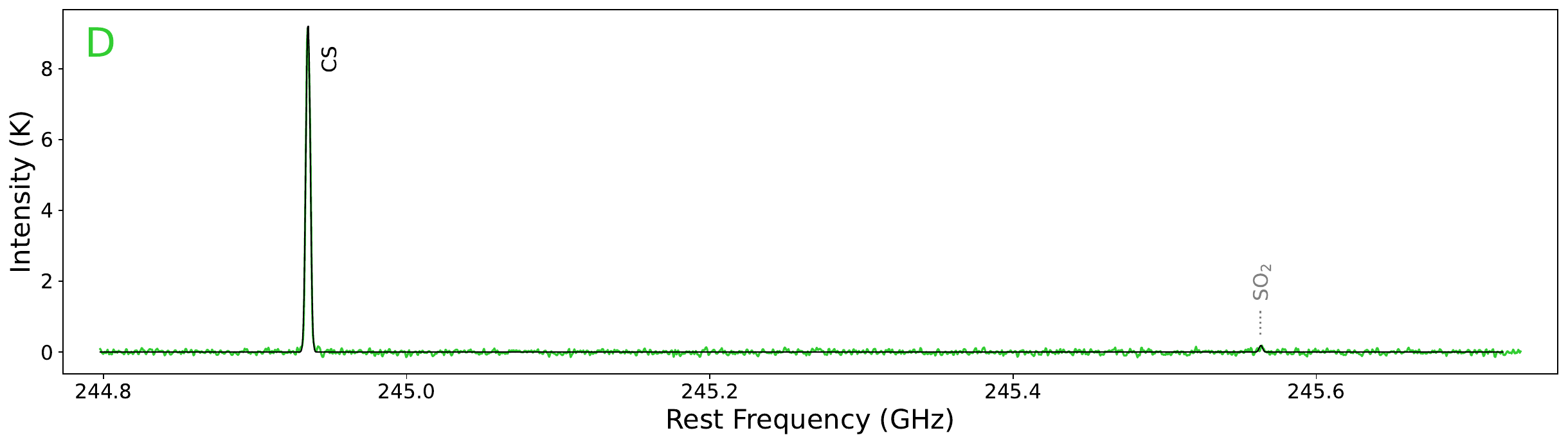} \\
    \caption{Same as Fig. \ref{fig:SpectraC1} but for N\,160A--mm\,D.}
    \label{fig:SpectraD1}
\end{figure*}

\begin{figure*}
    \centering
    \includegraphics[width=\textwidth]{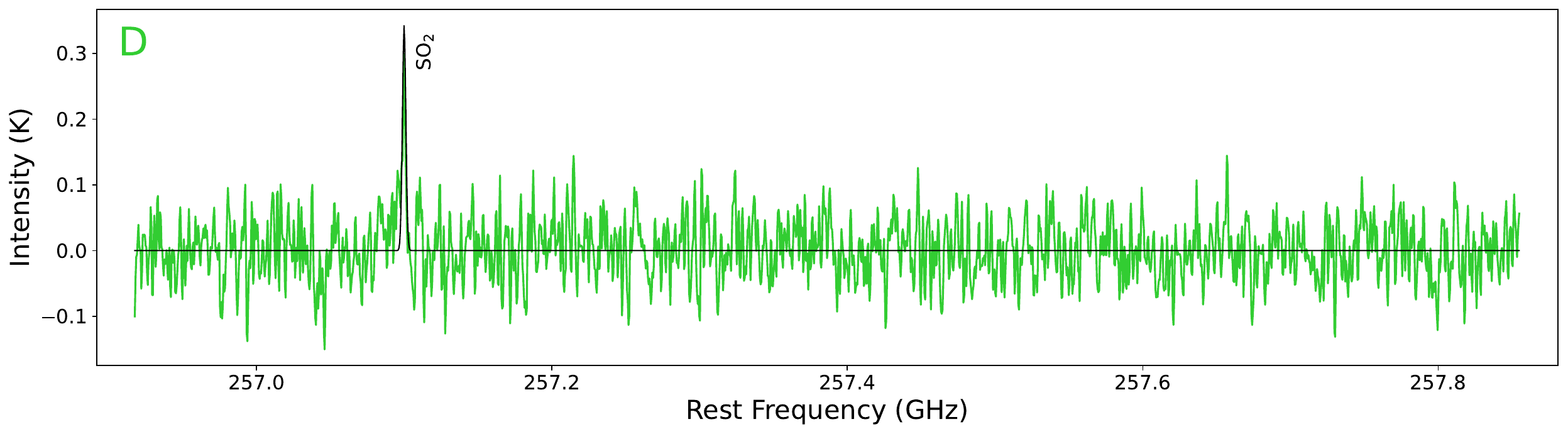} \\
    \includegraphics[width=\textwidth]{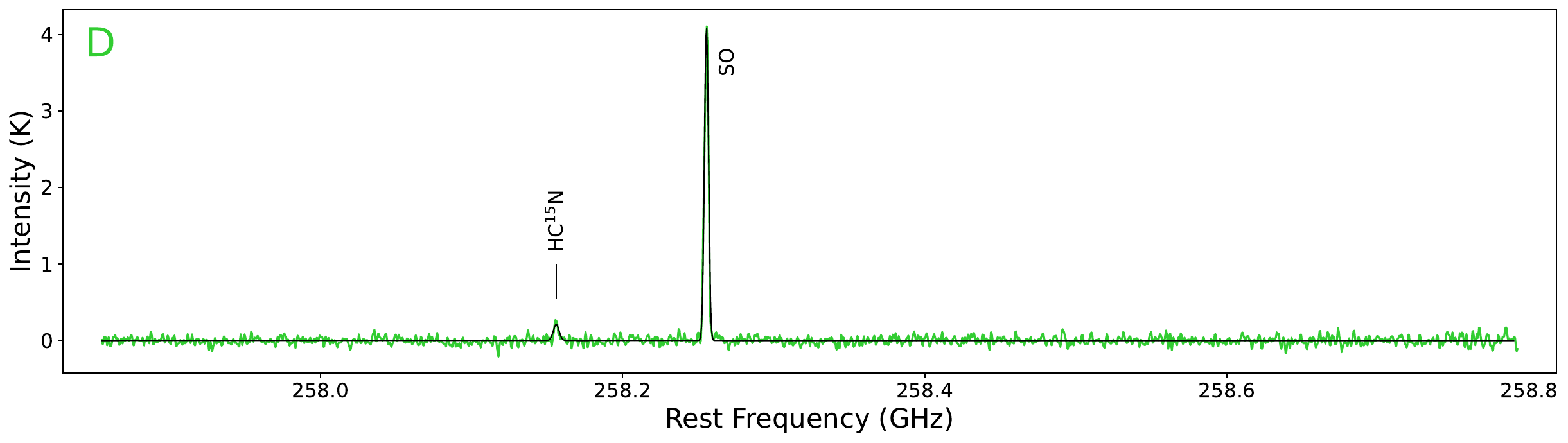} \\
    \includegraphics[width=\textwidth]{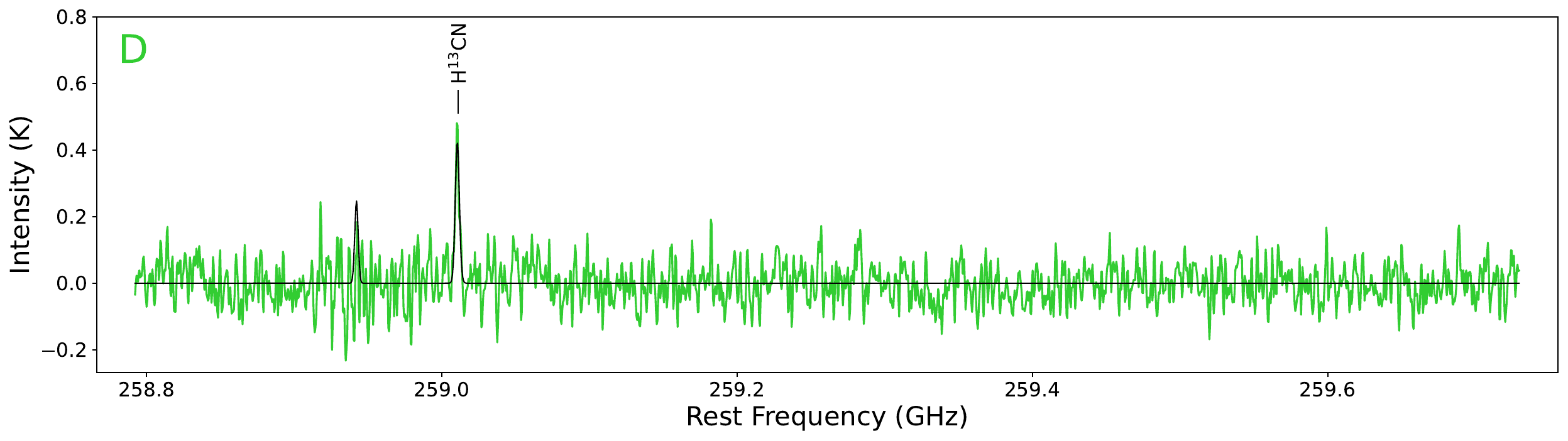} \\
    \includegraphics[width=\textwidth]{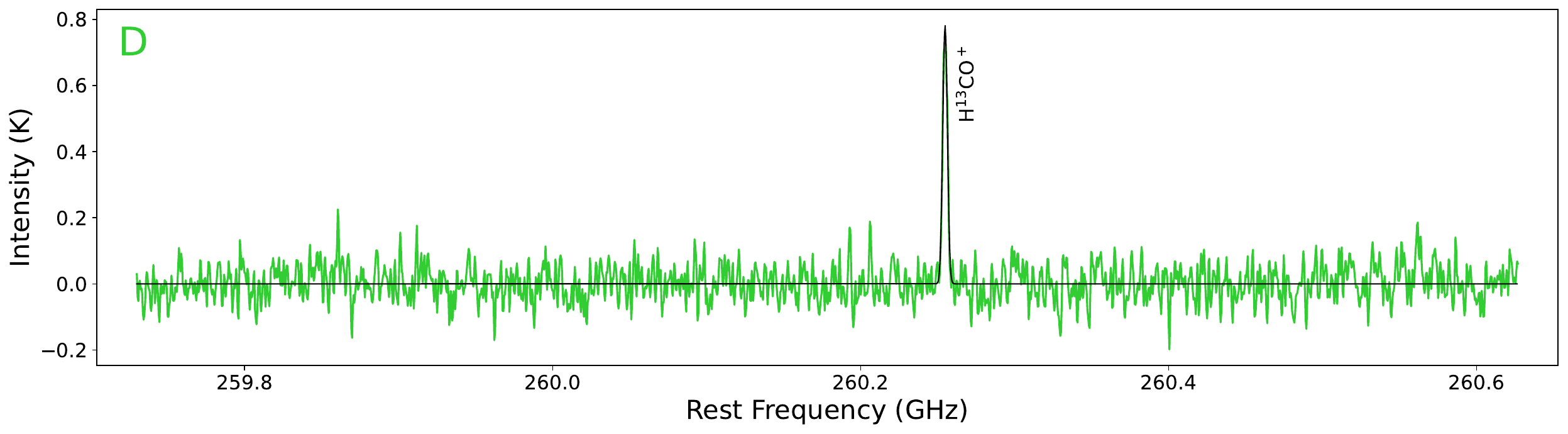} \\
    \caption{Same as Fig. \ref{fig:SpectraC2} but for N\,160A--mm\,D.}
    \label{fig:SpectraD2}
\end{figure*}

\begin{figure*}
    \centering
    \includegraphics[width=\textwidth]{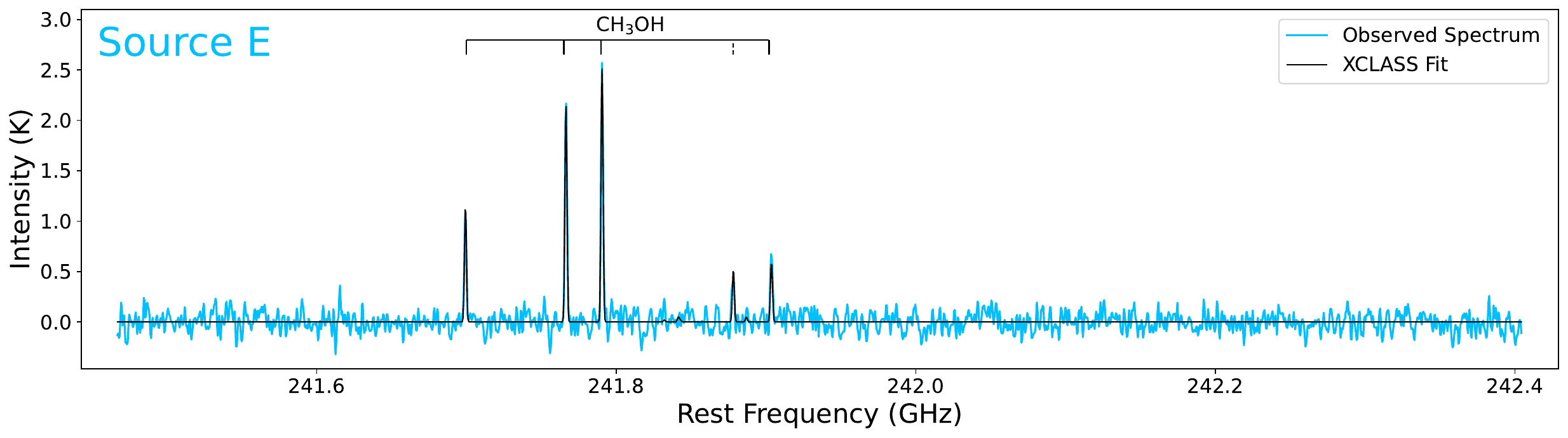} \\
    \includegraphics[width=\textwidth]{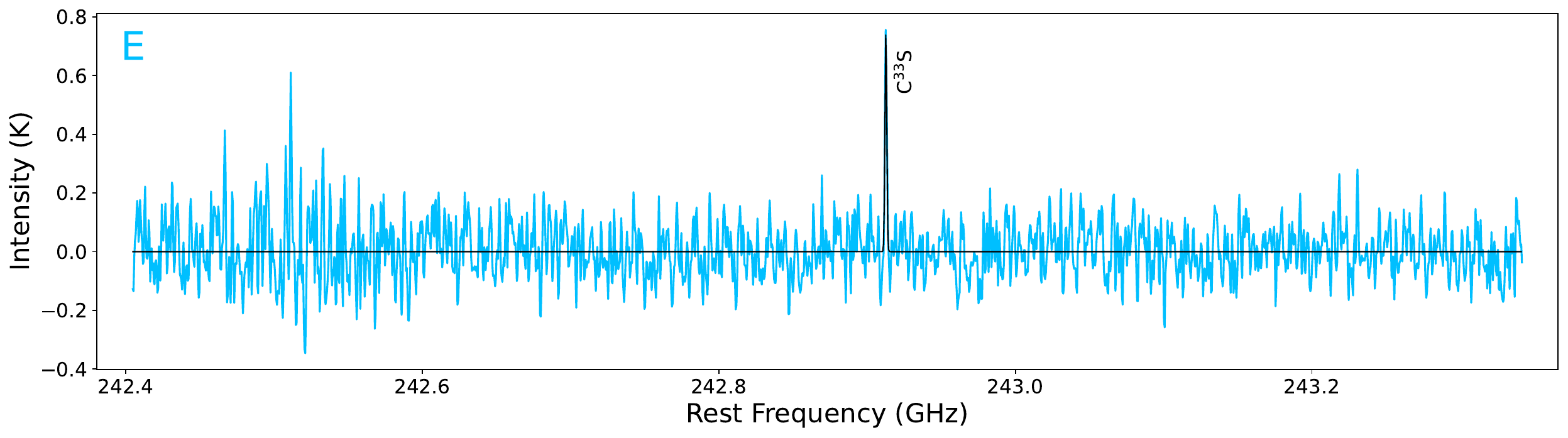} \\
    \includegraphics[width=\textwidth]{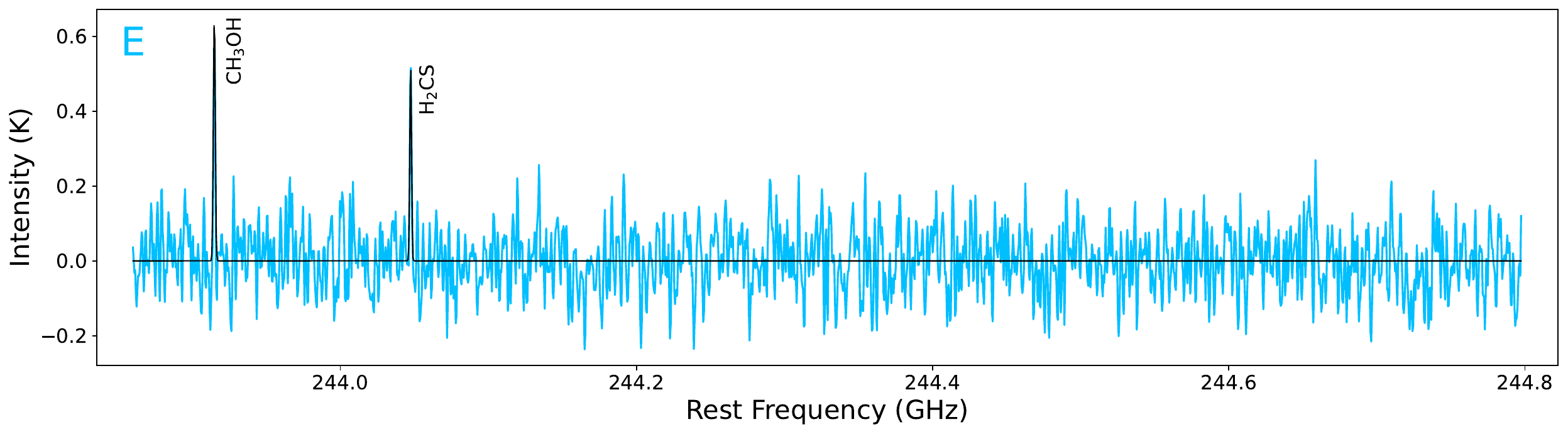} \\
    \includegraphics[width=\textwidth]{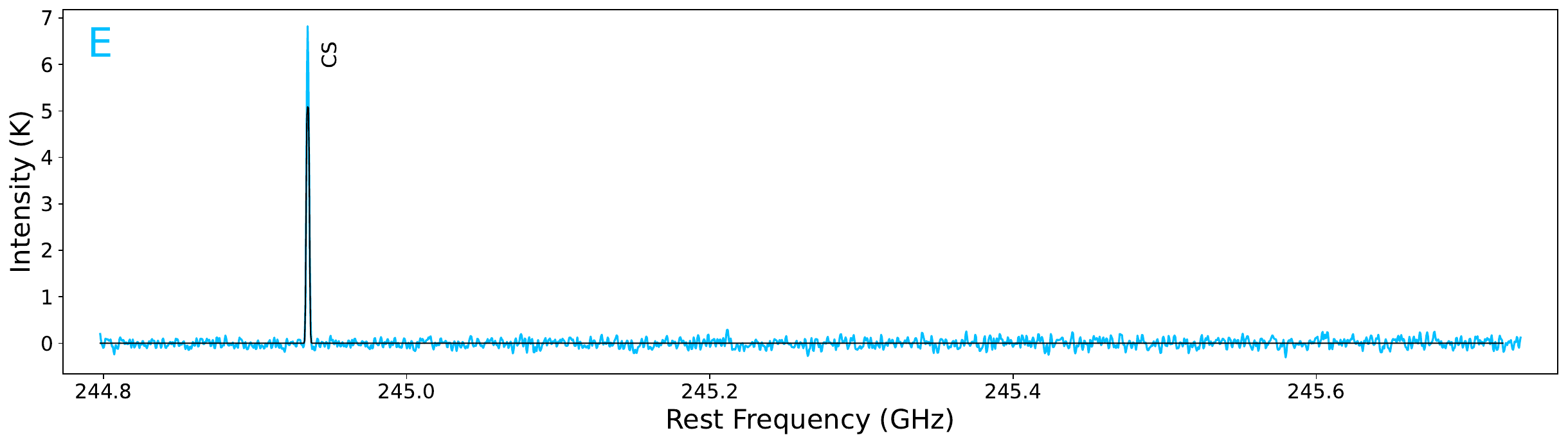} \\
    \caption{Same as Fig. \ref{fig:SpectraC1} but for N\,160A--mm\,E.}
    \label{fig:SpectraE1}
\end{figure*}

\begin{figure*}
    \centering
    \includegraphics[width=\textwidth]{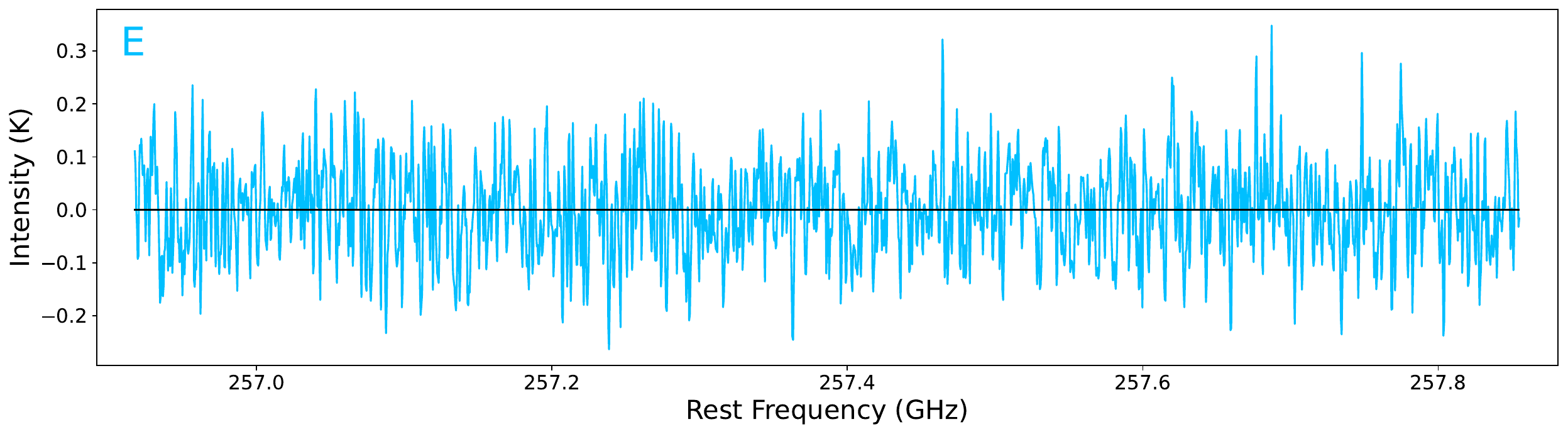} \\
    \includegraphics[width=\textwidth]{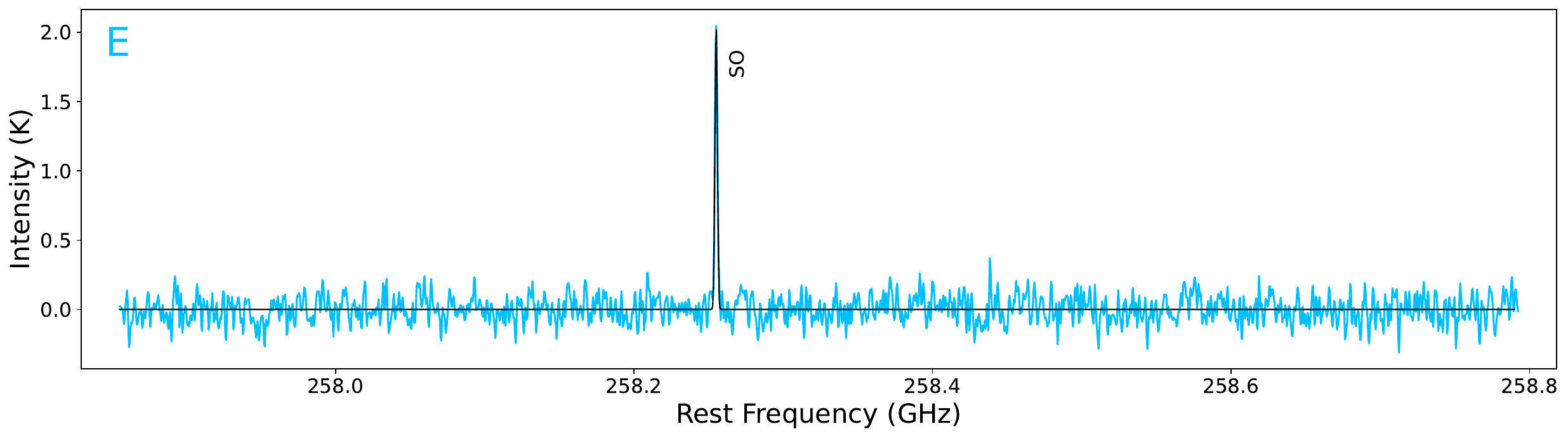} \\
    \includegraphics[width=\textwidth]{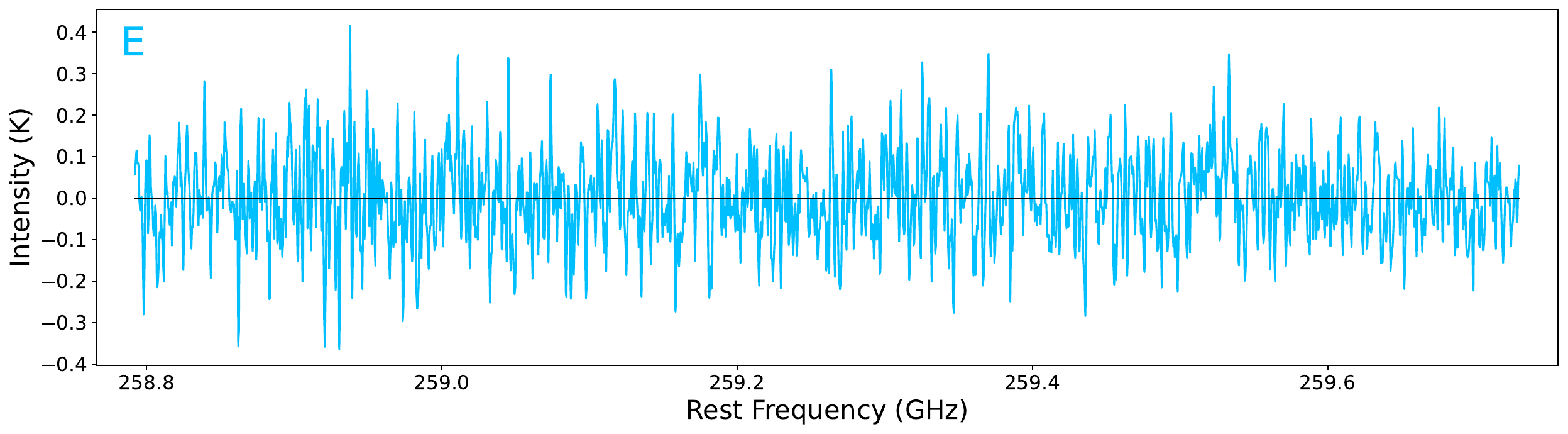} \\
    \includegraphics[width=\textwidth]{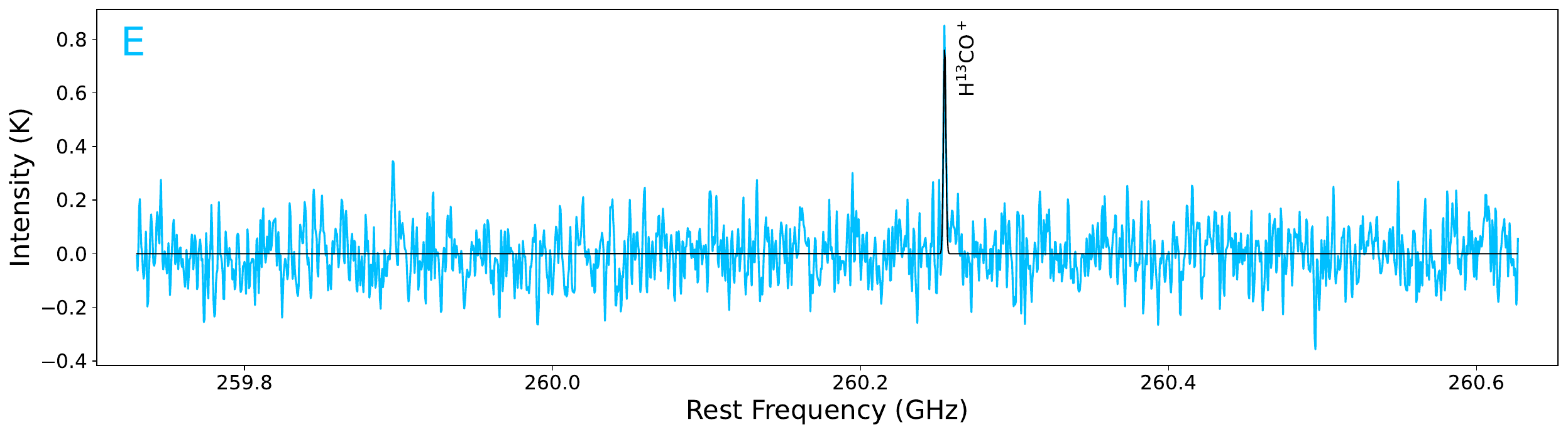} \\
    \caption{Same as Fig. \ref{fig:SpectraC2} but for N\,160A--mm\,E.}
    \label{fig:SpectraE2}
\end{figure*}

\begin{figure*}
    \centering
    \includegraphics[width=\textwidth]{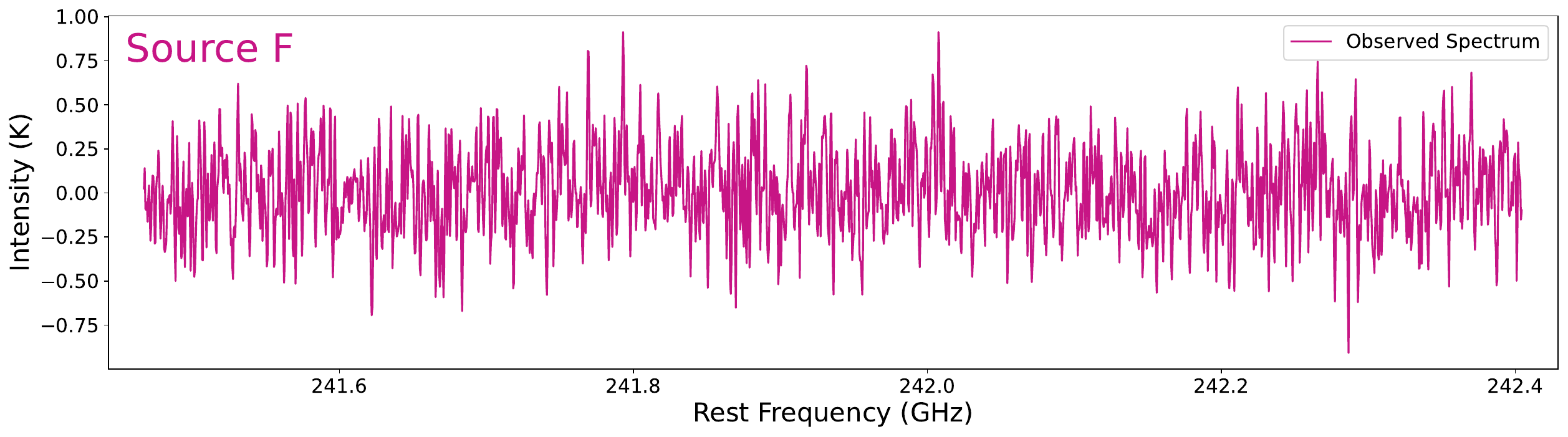} \\
    \includegraphics[width=\textwidth]{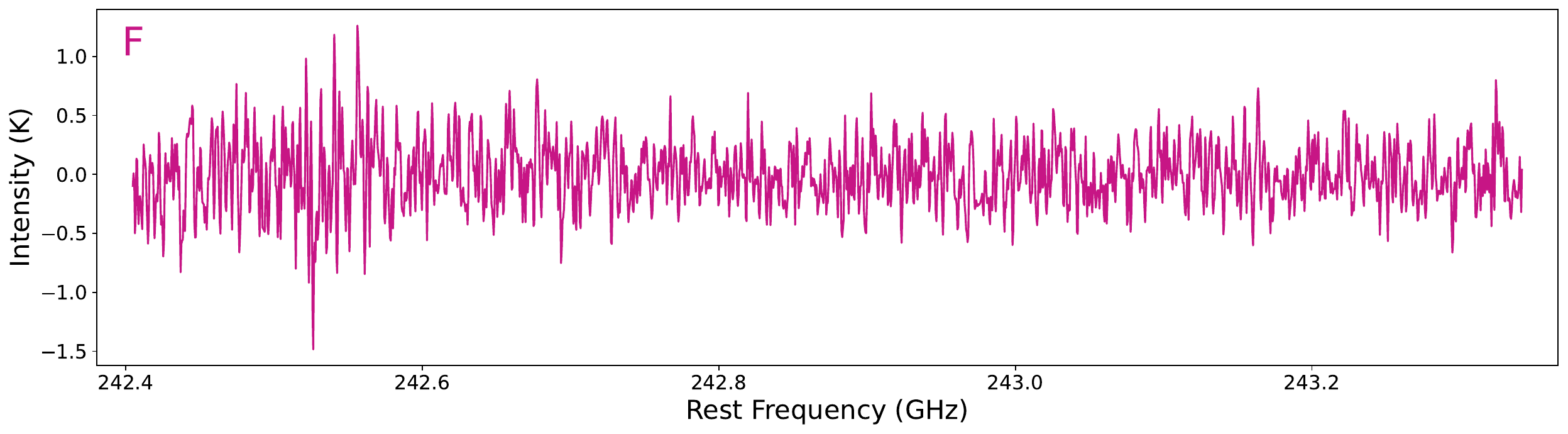} \\
    \includegraphics[width=\textwidth]{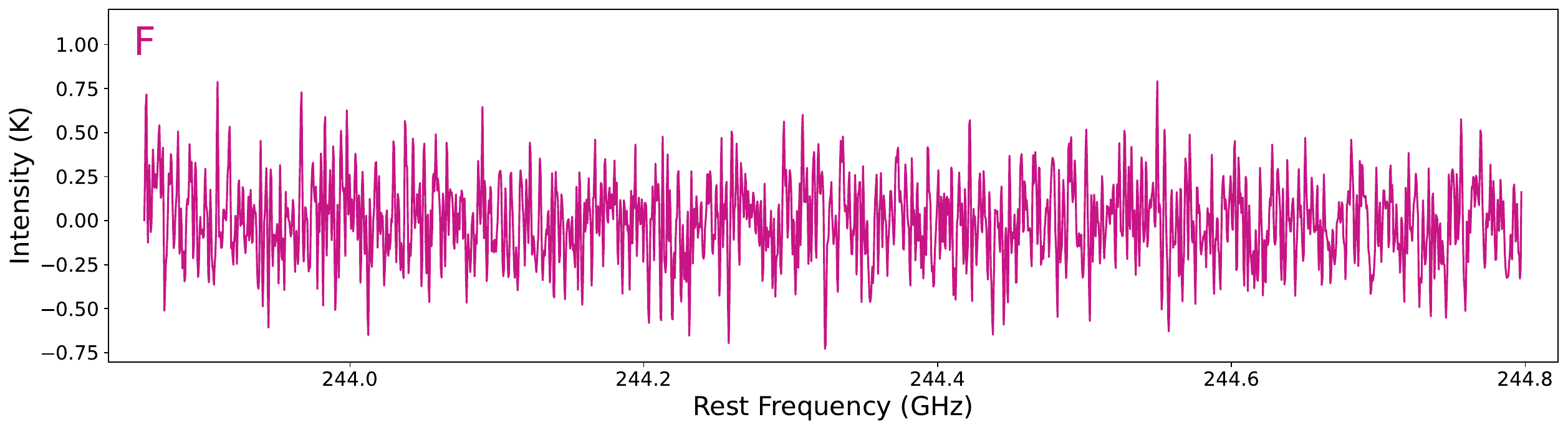} \\
    \includegraphics[width=\textwidth]{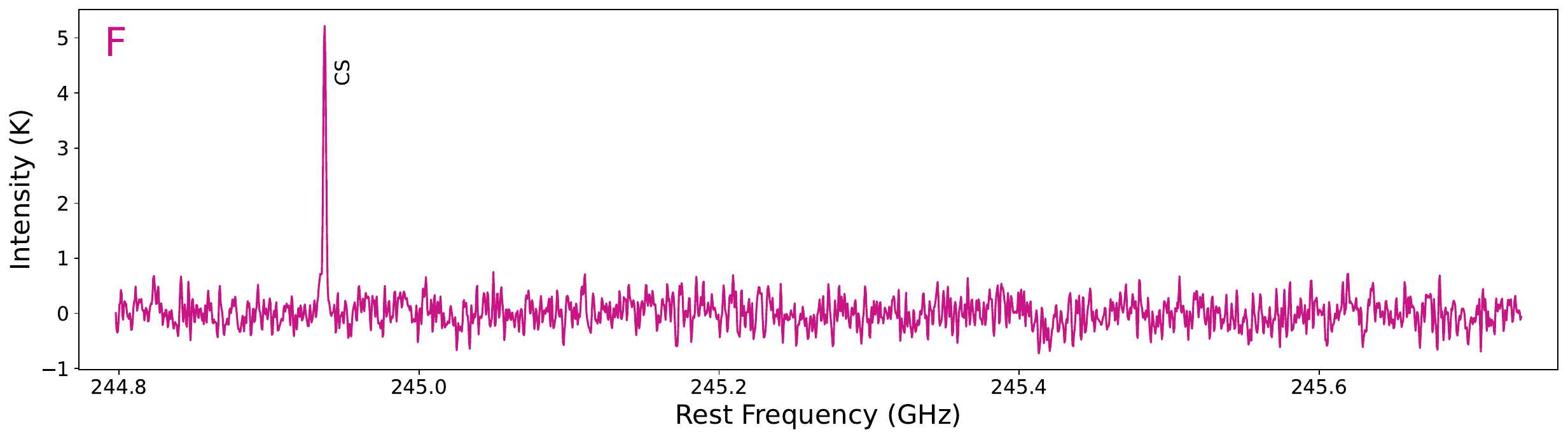} \\
    \caption{Same as Fig. \ref{fig:SpectraB1} but for N\,160A--mm\,F.}
    \label{fig:SpectraF1}
\end{figure*}

\begin{figure*}
    \centering
    \includegraphics[width=\textwidth]{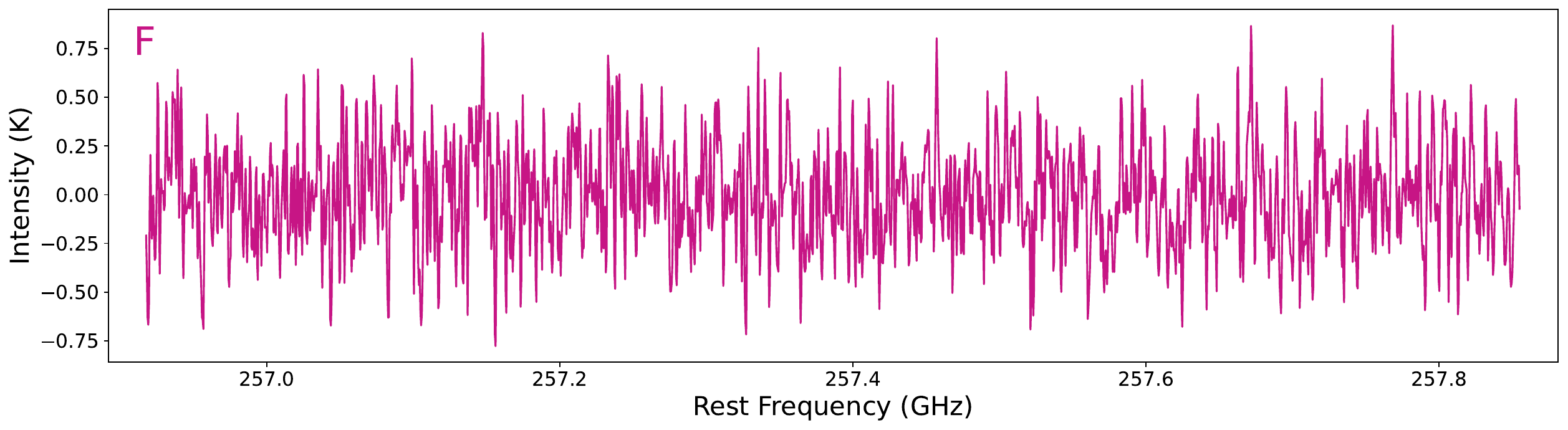} \\
    \includegraphics[width=\textwidth]{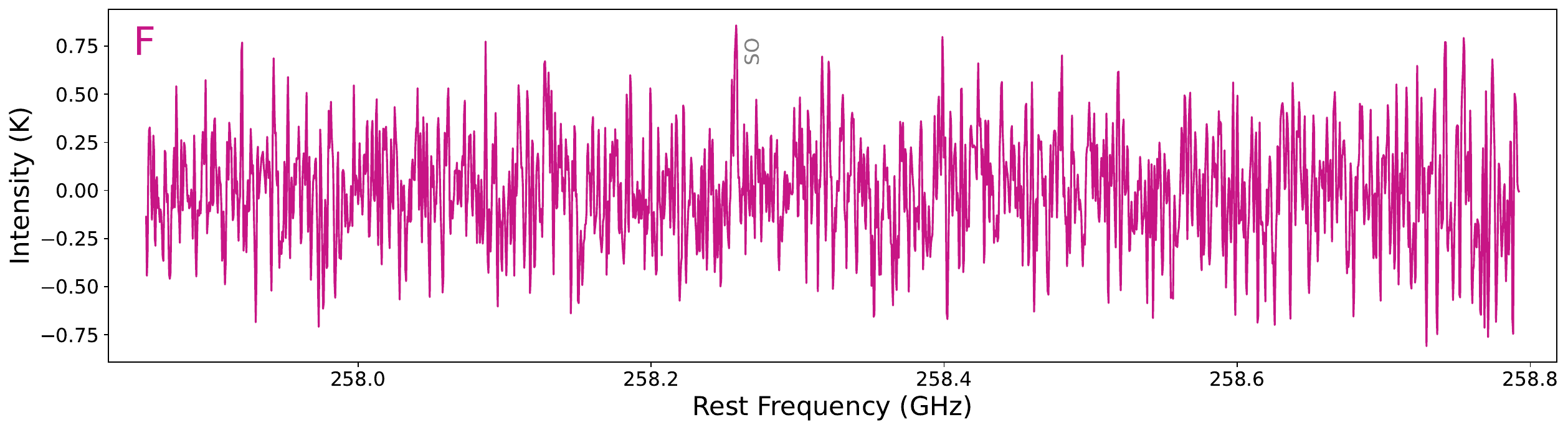} \\
    \includegraphics[width=\textwidth]{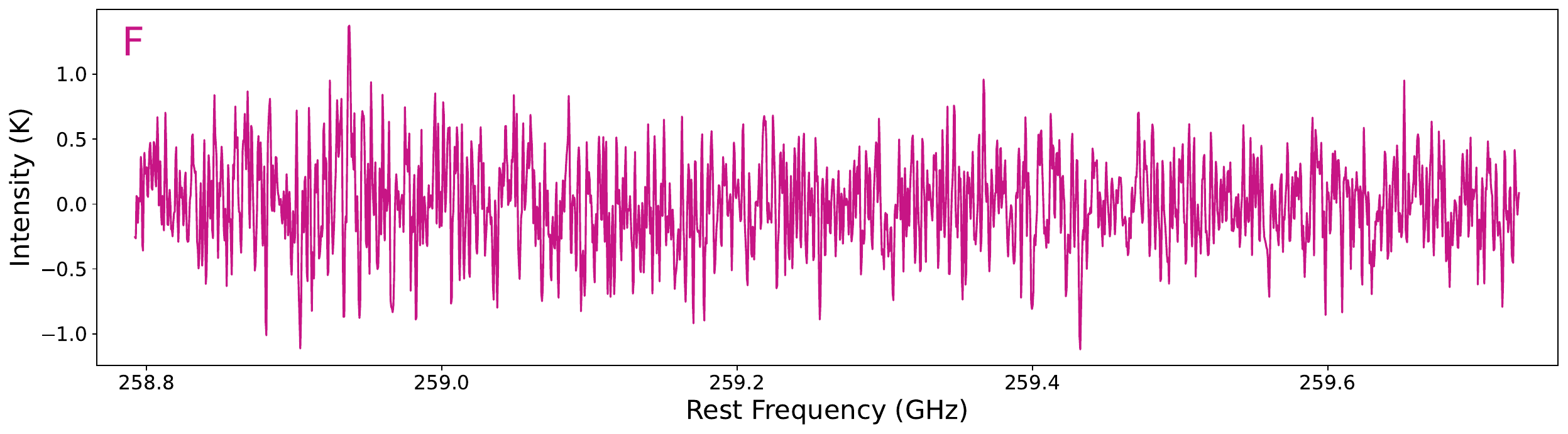} \\
    \includegraphics[width=\textwidth]{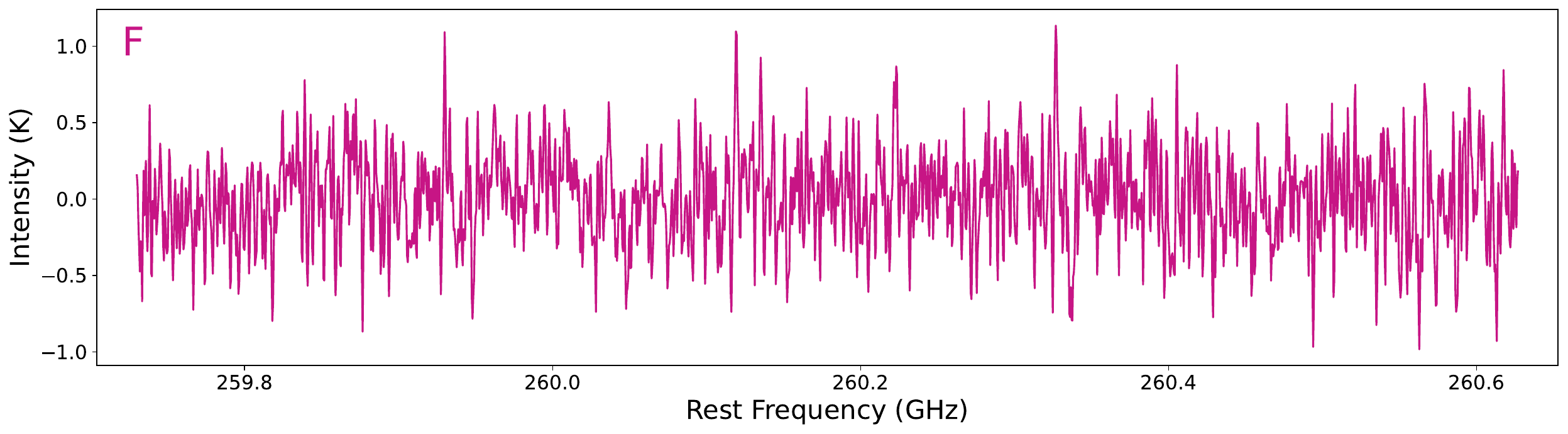} \\
    \caption{Same as Fig. \ref{fig:SpectraB2} but for N\,160A--mm\,F.}
    \label{fig:SpectraF2}
\end{figure*}

Figures~\ref{fig:SpectraB1}--\ref{fig:SpectraF2} show the ALMA Band 6 spectra for sources N\,160A--mm B--F for all four spectral windows. The spectra for the newly detected hot core, N\,160A--mm\,A, are presented in Figure~\ref{fig:SpectraA1}. The spectra were extracted as the mean intensity of the area enclosed by the 50\% contour of the peak 1.2 mm continuum emission of each source. For sources analyzed with XCLASS, the best-fit model spectra are overlaid on the observed spectra.

\section{Molecular Gas Kinematics} \label{a:kinematics}

\begin{figure*}
    \centering
    \includegraphics[width=0.48\textwidth]{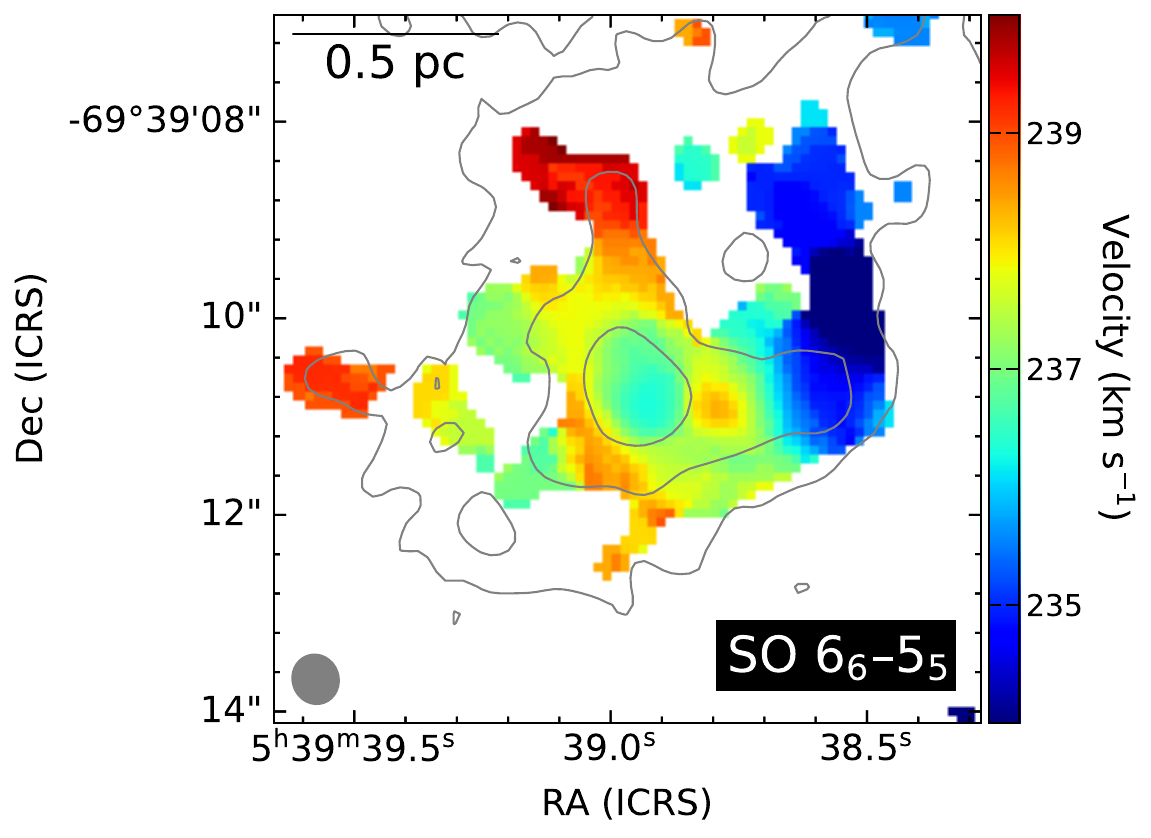} \includegraphics[width=0.48\textwidth]{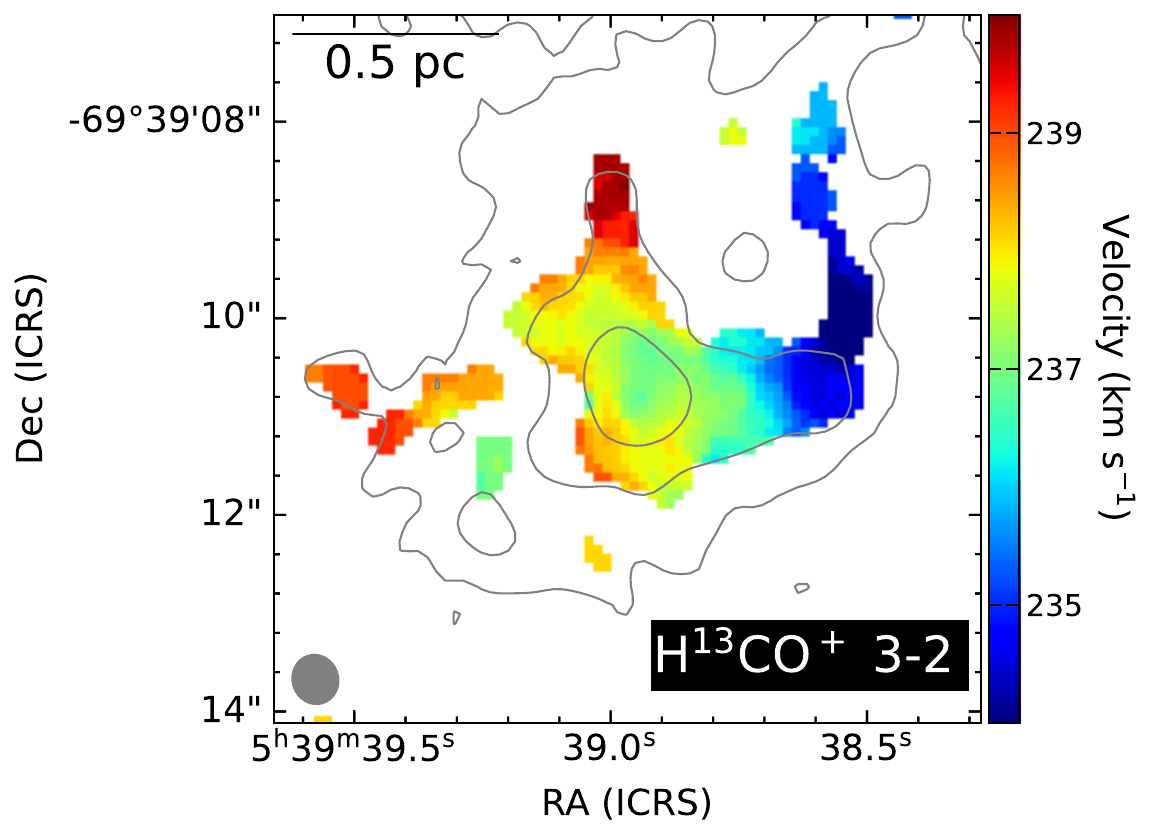}
    \caption{The SO and H$^{13}$CO$^+$ velocity maps of N\,160A--mm\,A. The 1.2 mm continuum contour levels are the same as in Figure \ref{fig:Mom0Amaps}.}
    \label{fig:vmaps}
\end{figure*}

In Figure \ref{fig:vmaps}, we provide the intensity weighted velocity (moment 1) maps of N\,160A--mm\,A for SO and H$^{13}$CO$^+$. These maps, as well as those presented in Figure~\ref{fig:mom1maps} (CS and CH$_3$OH), use a pixel mask where any pixel with intensity below $4\sigma$ in the corresponding integrated intensity map is excluded from the velocity map. A similar velocity structure is seen across all four velocity maps, with higher velocities to the north of the N\,160A--mm\,A's 1.2 mm continuum peak and lower velocities to the west.

\begin{figure*}
    \centering
    \includegraphics[width=1\textwidth]{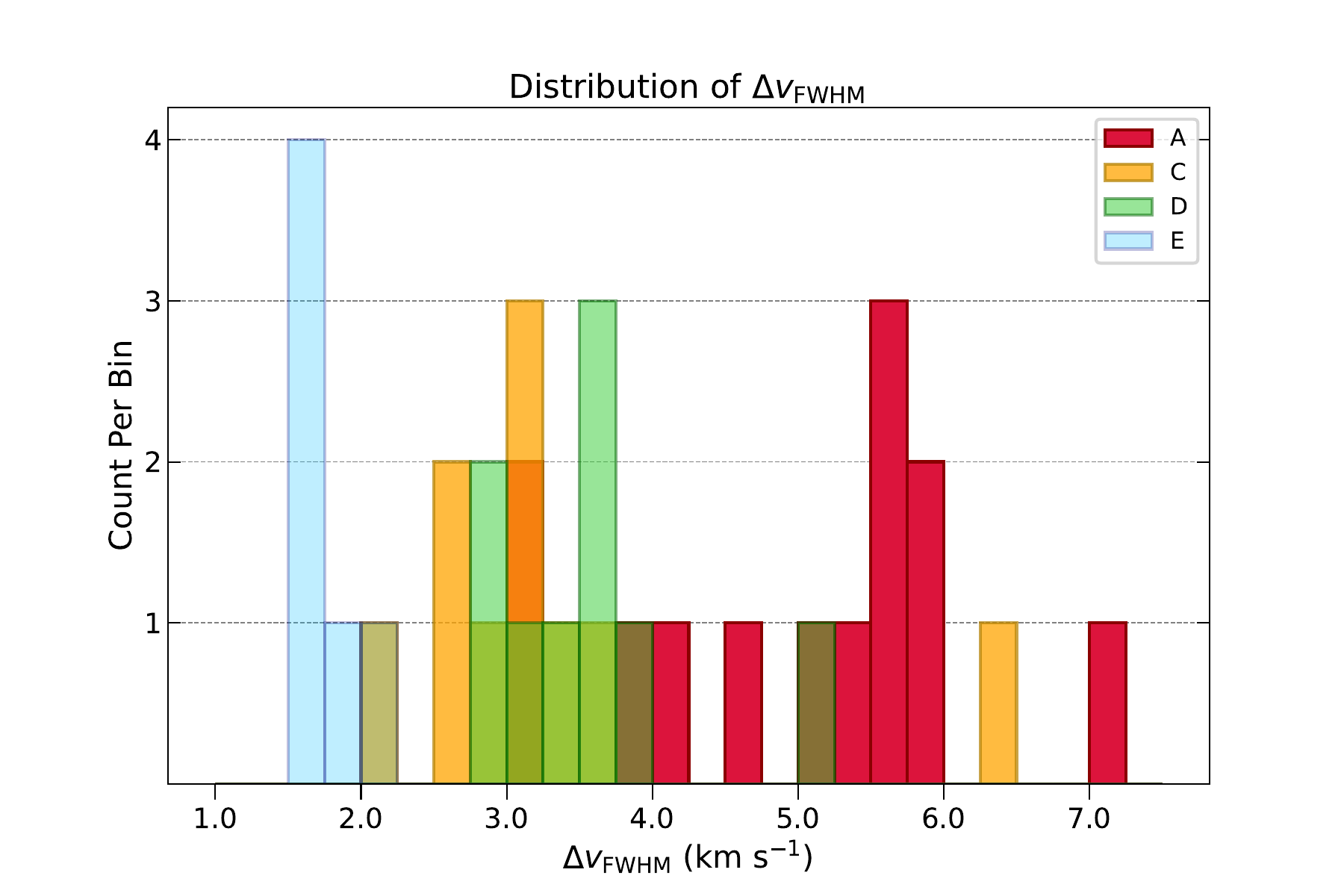}
    \caption{The distribution of line widths for all the spectrally analyzed sources in the ALMA field N\,160A--mm. The plot includes line widths for all the detected species (see Table~\ref{t:XCLASS Params}). The molecular lines detected toward source A are broader than those observed toward other sources, indicating a more complex kinematic structure in this source. The bin size is 0.25 km s$^{-1}$.}
    \label{fig:vfwhm}
\end{figure*}

Figure \ref{fig:vfwhm} shows the distribution of line widths for all sources in the ALMA field N\,160A--mm analyzed with XCLASS. For each source, the plot includes line widths for all the species listed in Table~\ref{t:XCLASS Params}. 

\section{The Multi-wavelength View of the ALMA field N\,160A--mm} \label{a:ancillarydata}

Here, we provide additional figures supporting the discussion in Sections~\ref{sss: hot core} and \ref{sss: cold cores}. 

\begin{figure*}
    \centering
    \includegraphics[width=0.49\textwidth]{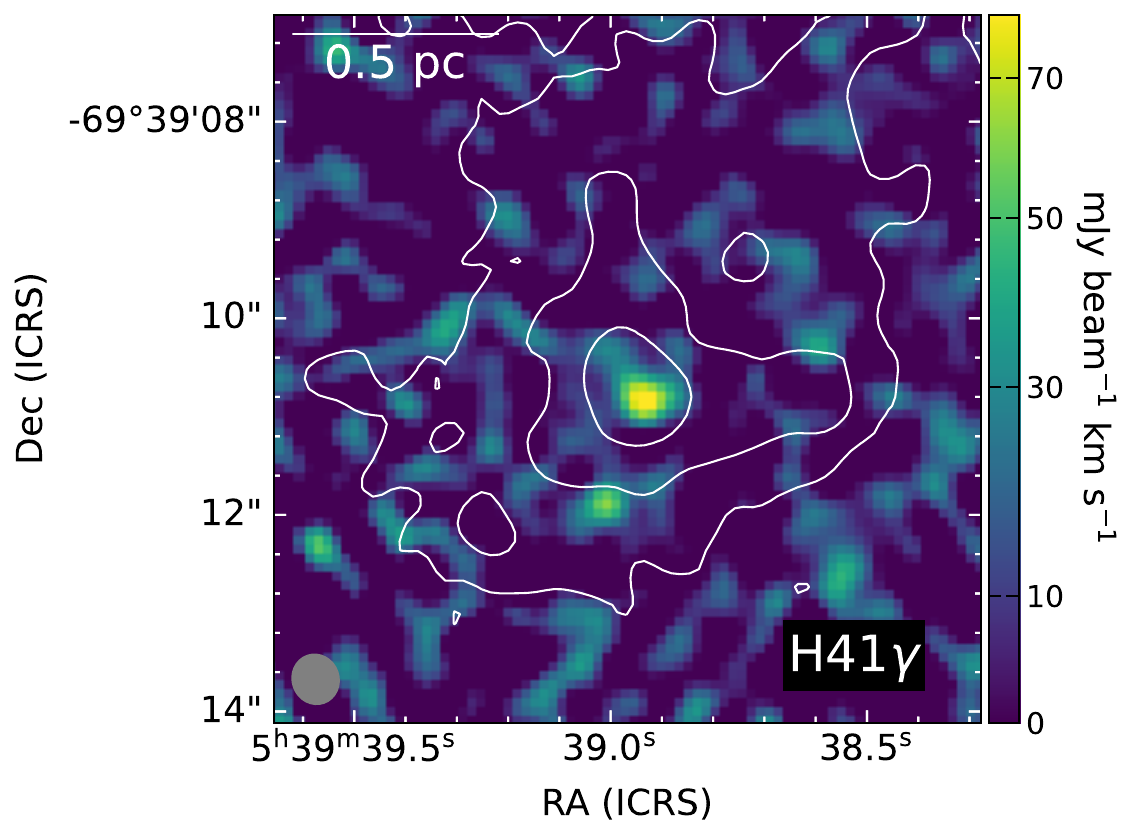}
    \includegraphics[width=0.49\textwidth]{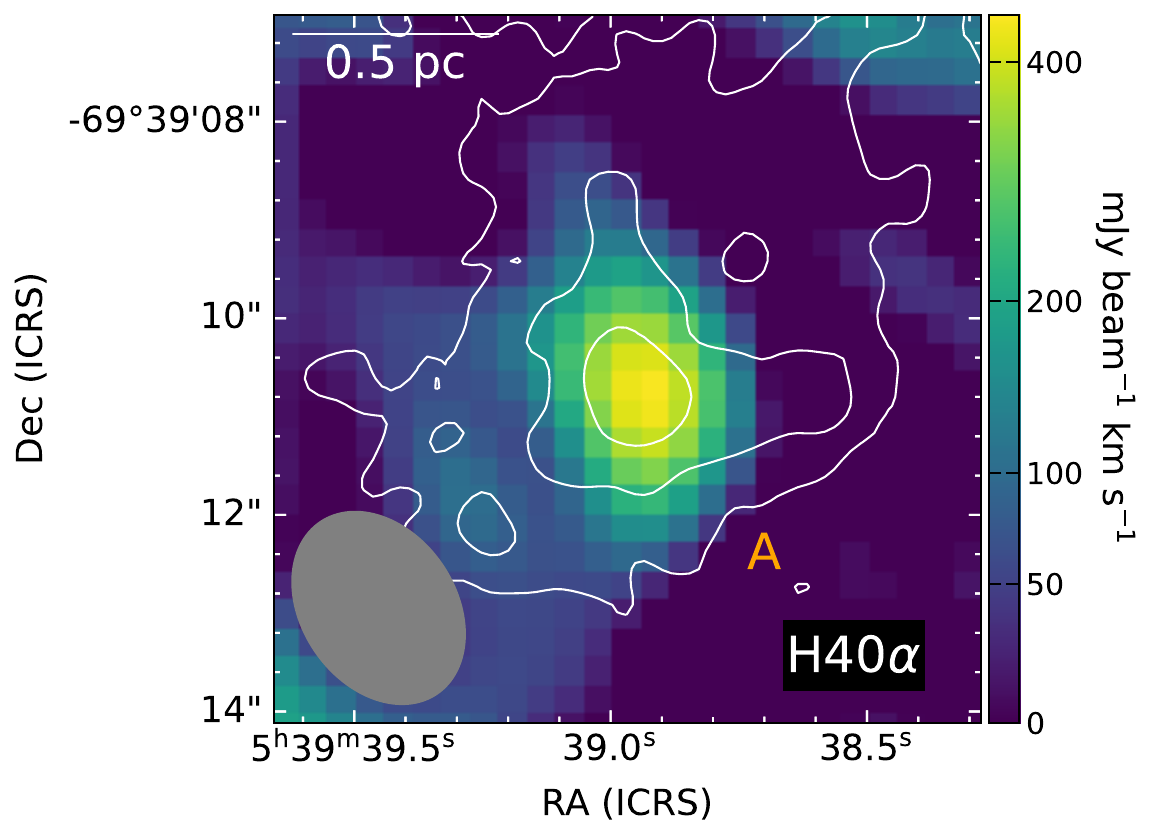}
    \caption{The H41$\gamma$ (the rest frequency of 257.63549 GHz) and H40$\alpha$ (99.02295 GHz) recombination line integrated intensity maps of N\,160A--mm\,A with the 1.2 mm continuum contours overlaid. The ALMA beam size is indicated in the lower left corner in each image: $0\rlap.{''}508\times0\rlap.{''}469$ for H41$\gamma$ and $2\rlap.{''}112\times1\rlap.{''}564$ for H40$\alpha$ observations. The H41$\gamma$ and H40$\alpha$ emission peaks coincide with the 1.2 mm continuum and molecular gas emission peaks (see Figure \ref{fig:Mom0Amaps}).}
    \label{fig:h41gamma}
\end{figure*}

In Figure \ref{fig:h41gamma}, we show the H41$\gamma$ and H40$\alpha$ recombination line integrated intensity maps of N\,160A--mm\,A based on our ALMA Band 6 and archival Band 3 observations, respectively. The H41$\gamma$ and H40$\alpha$ emission peaks coincide with the 1.2 mm continuum peak of source A, the newly detected hot core.

\begin{figure*}
    \centering
    \includegraphics[width=0.49\textwidth]{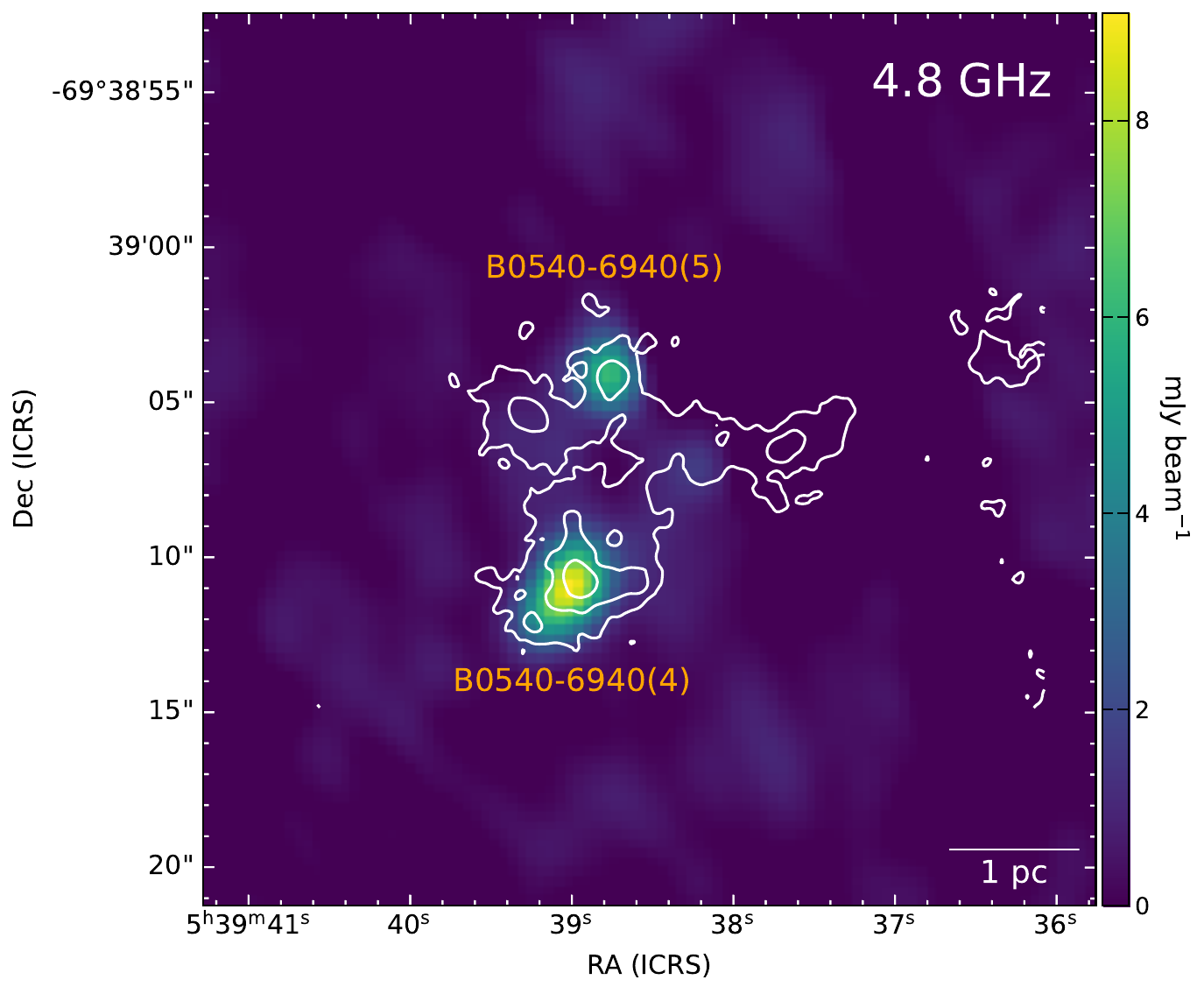} \includegraphics[width=0.49\textwidth]{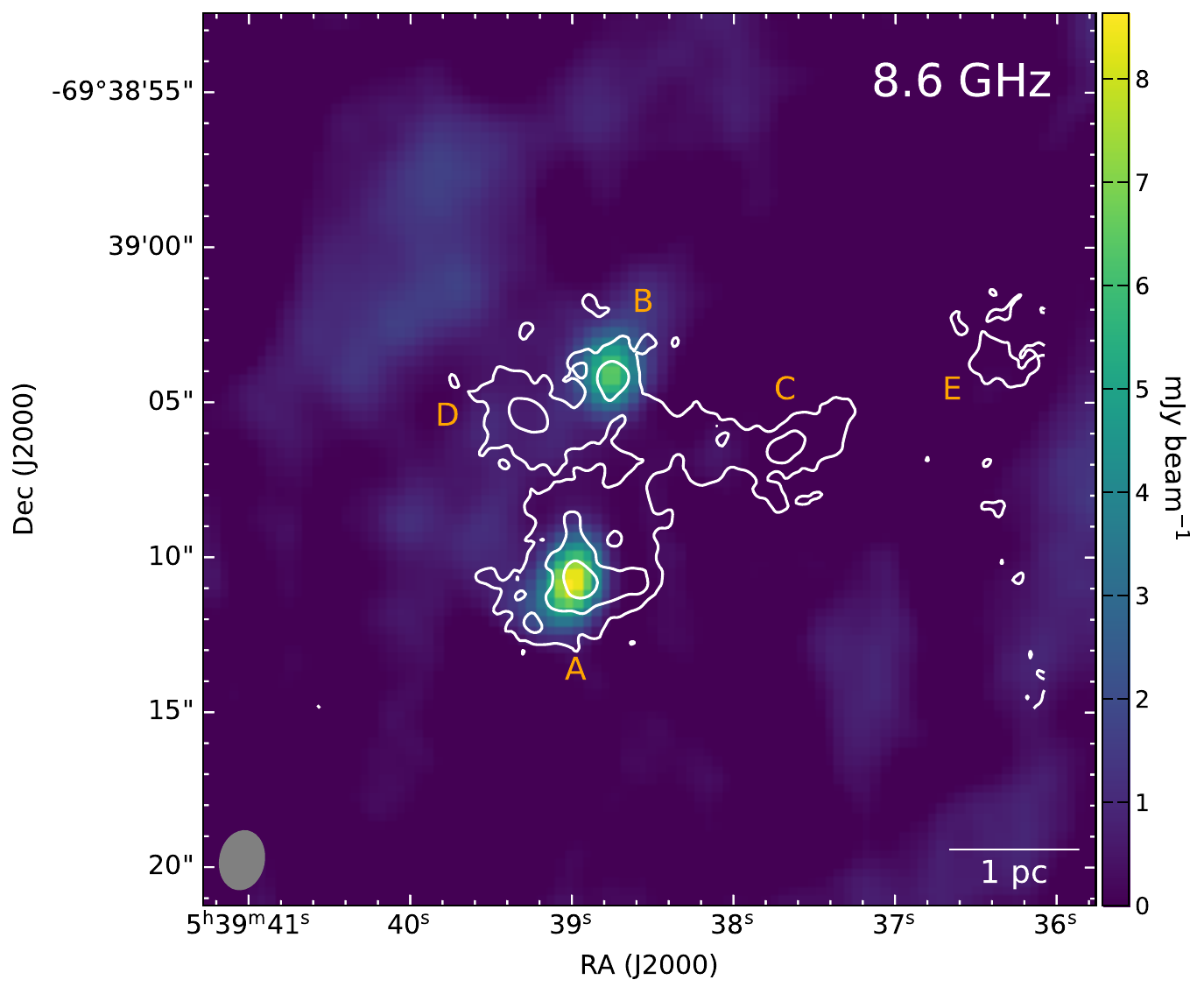}
    \caption{The 4.8 GHz and 8.6 GHz Australia Telescope Compact Array (ATCA) images of the ALMA field N\,160A--mm. The 1.2 mm continuum contours are overlaid on both images. The positions of two radio sources B0540-6940(5) and B0540-6940(4) are labeled in the left panel \citep{indebetouw2004}. The ALMA 1.2 mm continuum cores are labeled in the right panel. The beam size is $2\rlap.{''}077\times1\rlap.{''}689$ and $1\rlap.{''}897\times1\rlap.{''}413$ at 4.8 GHz and 8.6 GHz, respectively. The radio emission is associated with two of the 1.2 mm continuum sources in N\,160A--mm: A and B.}
    \label{fig:ATCA}
\end{figure*}

N\,160A--mm\,A is also associated with the radio emission (source B0540-6940(4) in \citealt{indebetouw2004}) as shown in the Australia Telescope Compact Array (ATCA) 4.8 GHz and 8.6 GHz images in Figure \ref{fig:ATCA}. The second radio source in our ALMA field, B0540-6940(5), coincides with N\,160A--mm\,B. 

\begin{figure*}
    \centering
    \includegraphics[width=0.75\textwidth]{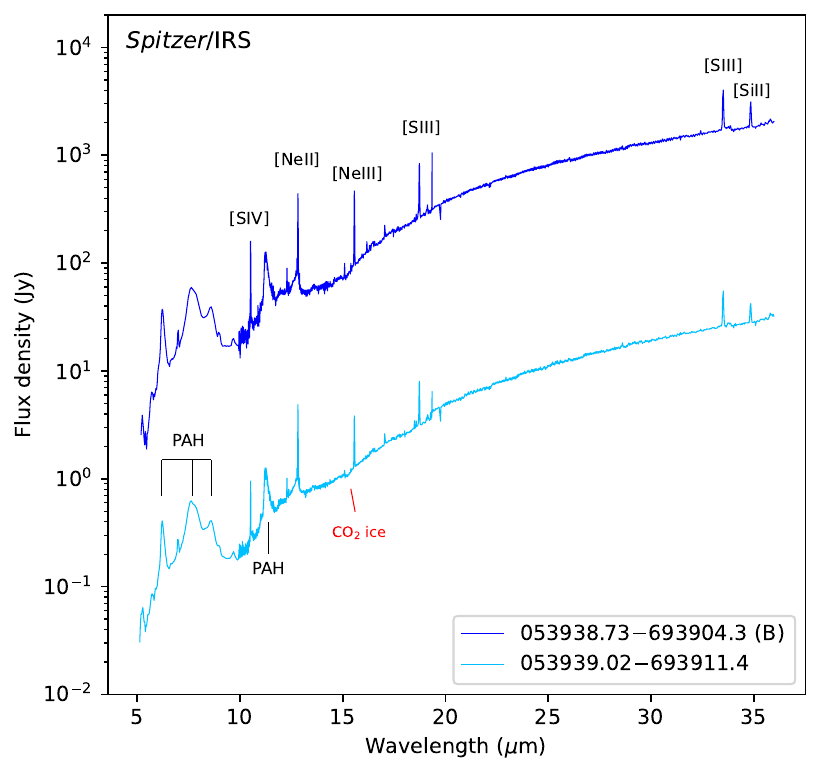}
    \caption{Spitzer/IRS spectra of the two YSOs in the ALMA field N\,160A--mm. One of the YSOs coincides with source N\,160A--mm\,B (053938.73-693904.3) and the other lies approximately 0$\rlap.{''}$75/0.12 pc to the southeast of N\,160A--mm\,A (053939.02-693911.4). The spectra are color-coded as indicated in the legend. For legibility, the top spectrum has been scaled by a factor of 100. The detailed analysis of the spectra is described in \citet{seale2009} and \citet{jones2017}. Note: the two spectra are classified as H\,{\sc ii}/YSO3 by \citet{jones2017}, but the faint CO$_2$ ice feature is present in both spectra -- they are 2$\sigma$ detections, thus no reliable analysis of these spectral features is possible. However, they provide a hint that the sources may be embedded YSOs (YSO1 classification in \citealt{jones2017}). Due to the large size of the Spitzer/IRS's slit, the YSO spectra are contaminated by the environment (H\,{\sc ii} region).}
    \label{fig:spectra}
\end{figure*}

\begin{figure*}
    \centering
    \includegraphics[width=0.49\textwidth]{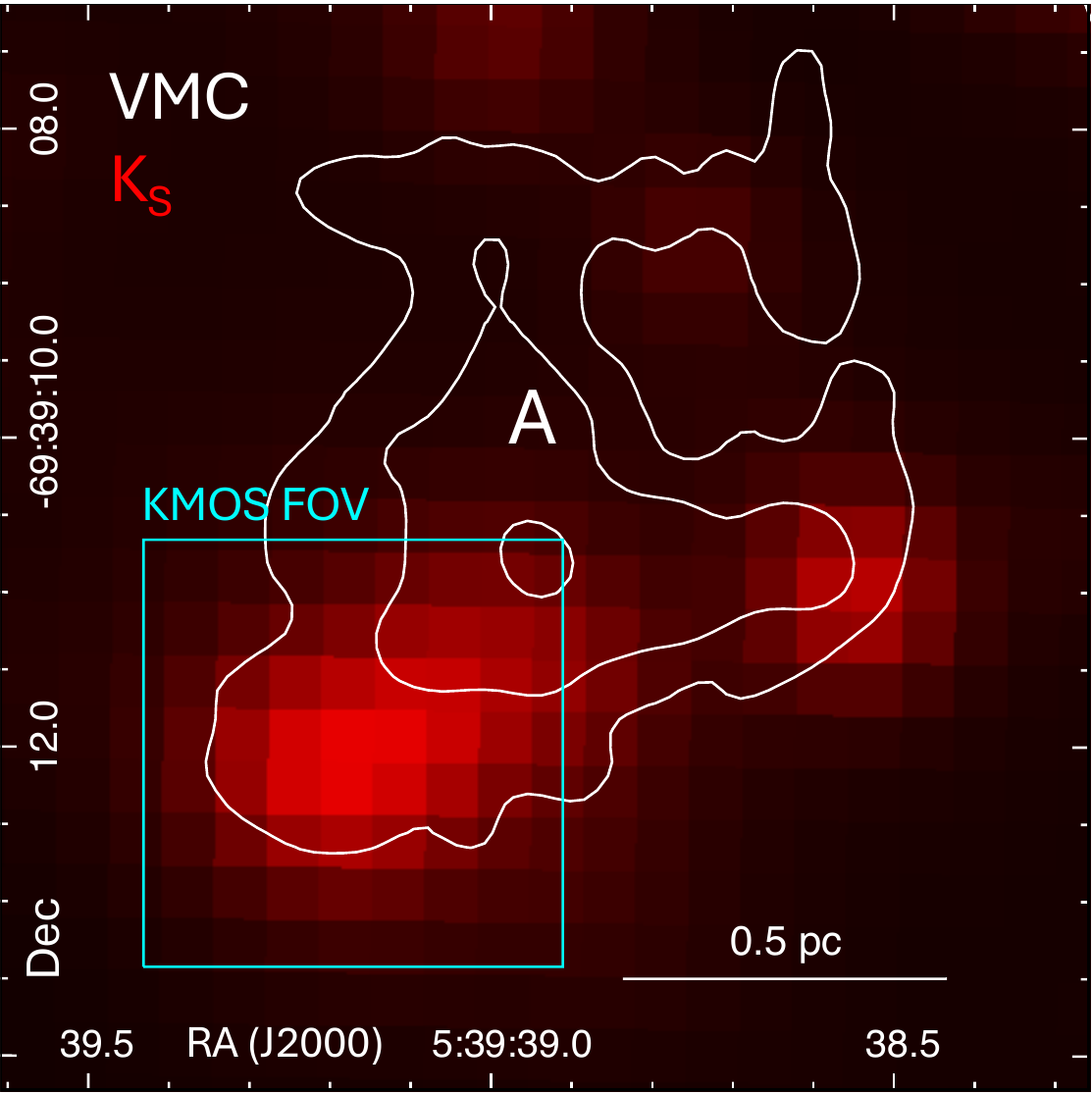} \includegraphics[width=0.49\textwidth]{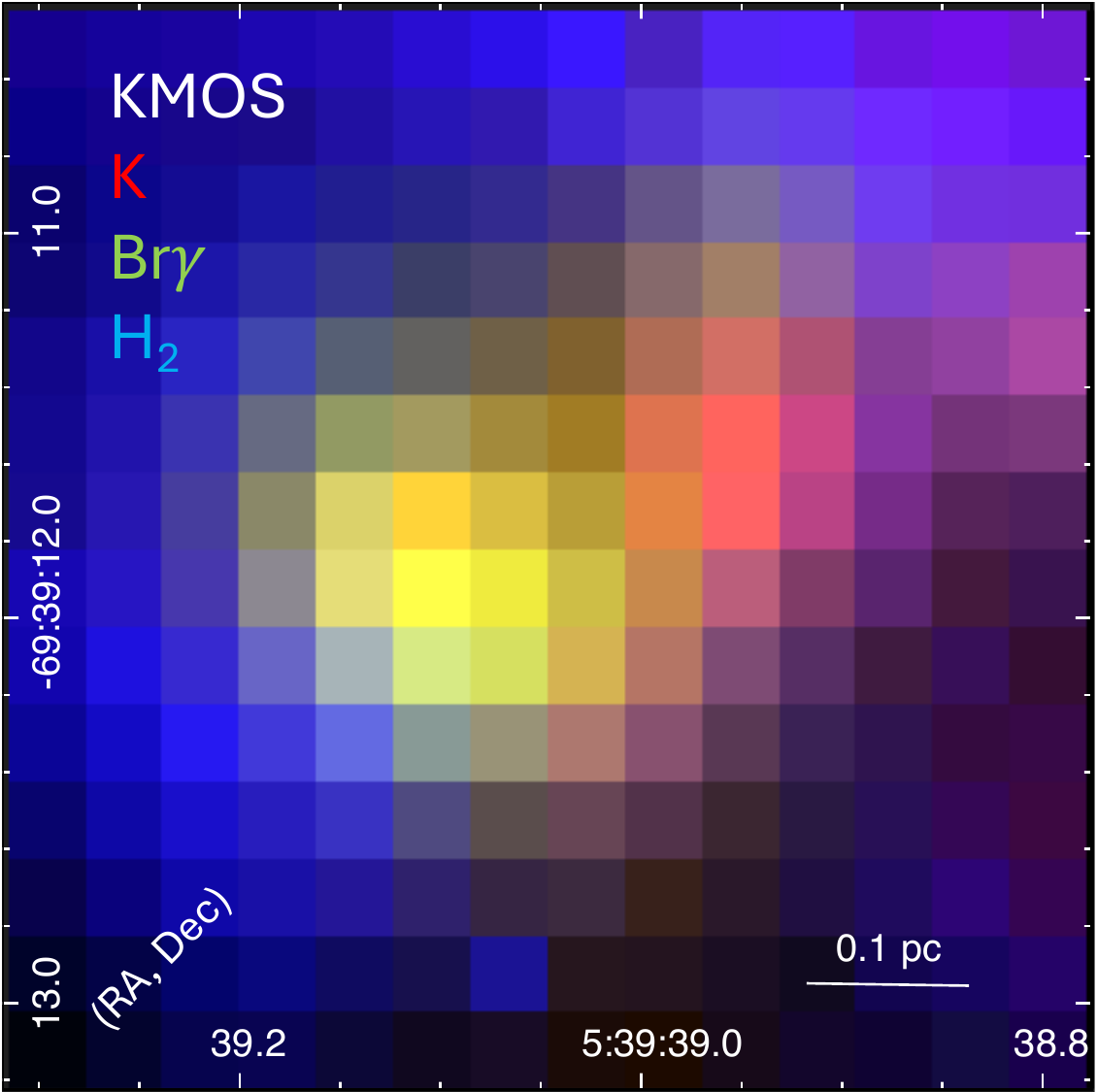}
    \caption{Left panel: The VMC K$_s$-band image (red) with the 1.2 mm continuum contours overlaid (white); the contour levels are 1\%, 5\%, and 20\% of the continuum peak. The cyan box indicates the approximate size and location of the KMOS field. Right: Three-color mosaic combining the KMOS $K$-band continuum (red), Br$\gamma$ (green), and H$_2$ (blue) images.}
    \label{fig:KMOS}
\end{figure*}

The Spitzer/IRS mid-IR spectra of the two YSOs located in the ALMA field N\,160A--mm are shown in Figure \ref{fig:spectra}. The near-IR view of the region around the hot core N\,160A--mm\,A is provided in Figure \ref{fig:KMOS} that compares the VMC $K_s$-band and KMOS $K$-band continuum, Br$\gamma$, and H$_2$ images. 

\begin{figure*}
    \centering
    \includegraphics[width=0.48\textwidth]{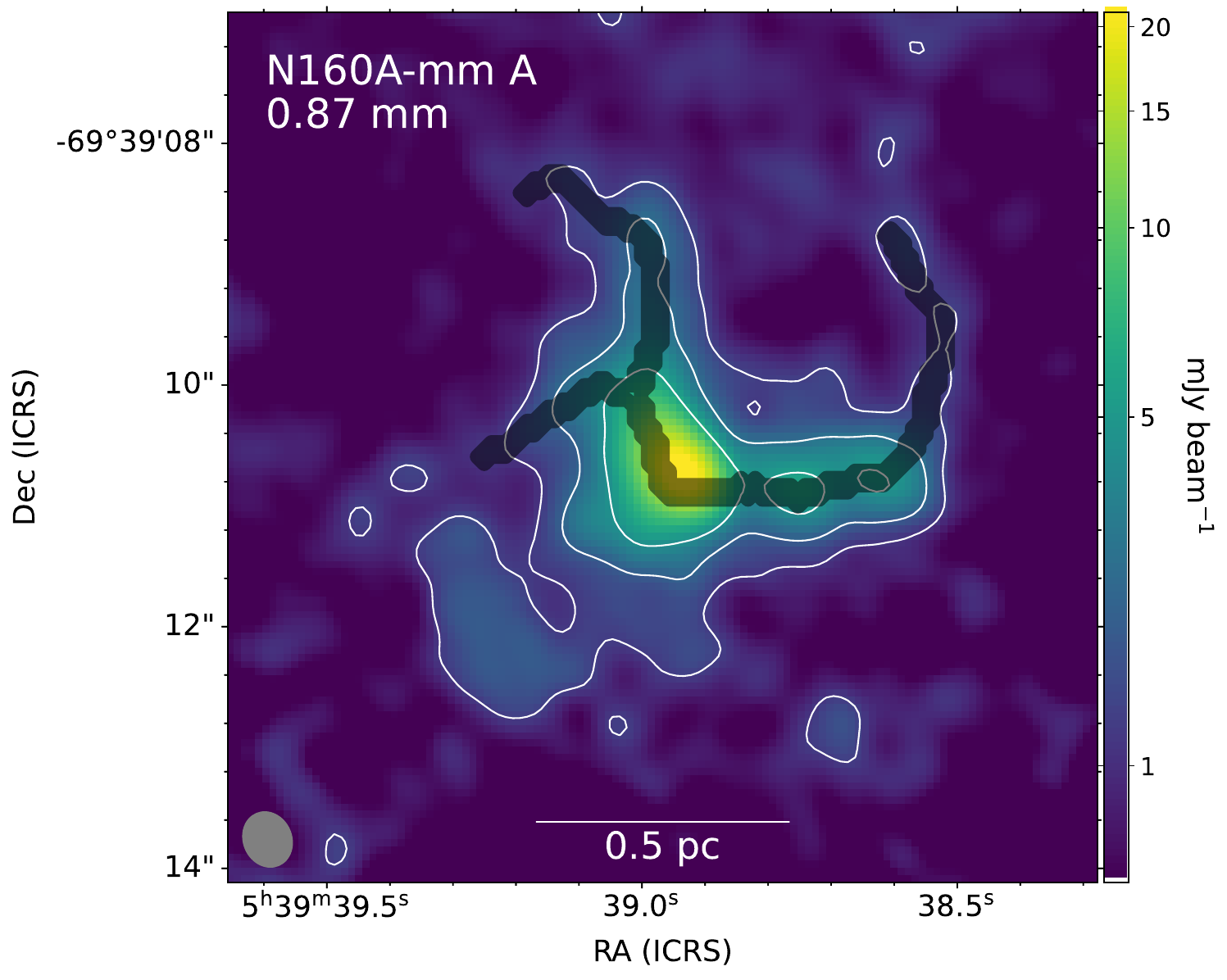} \includegraphics[width=0.48\textwidth]{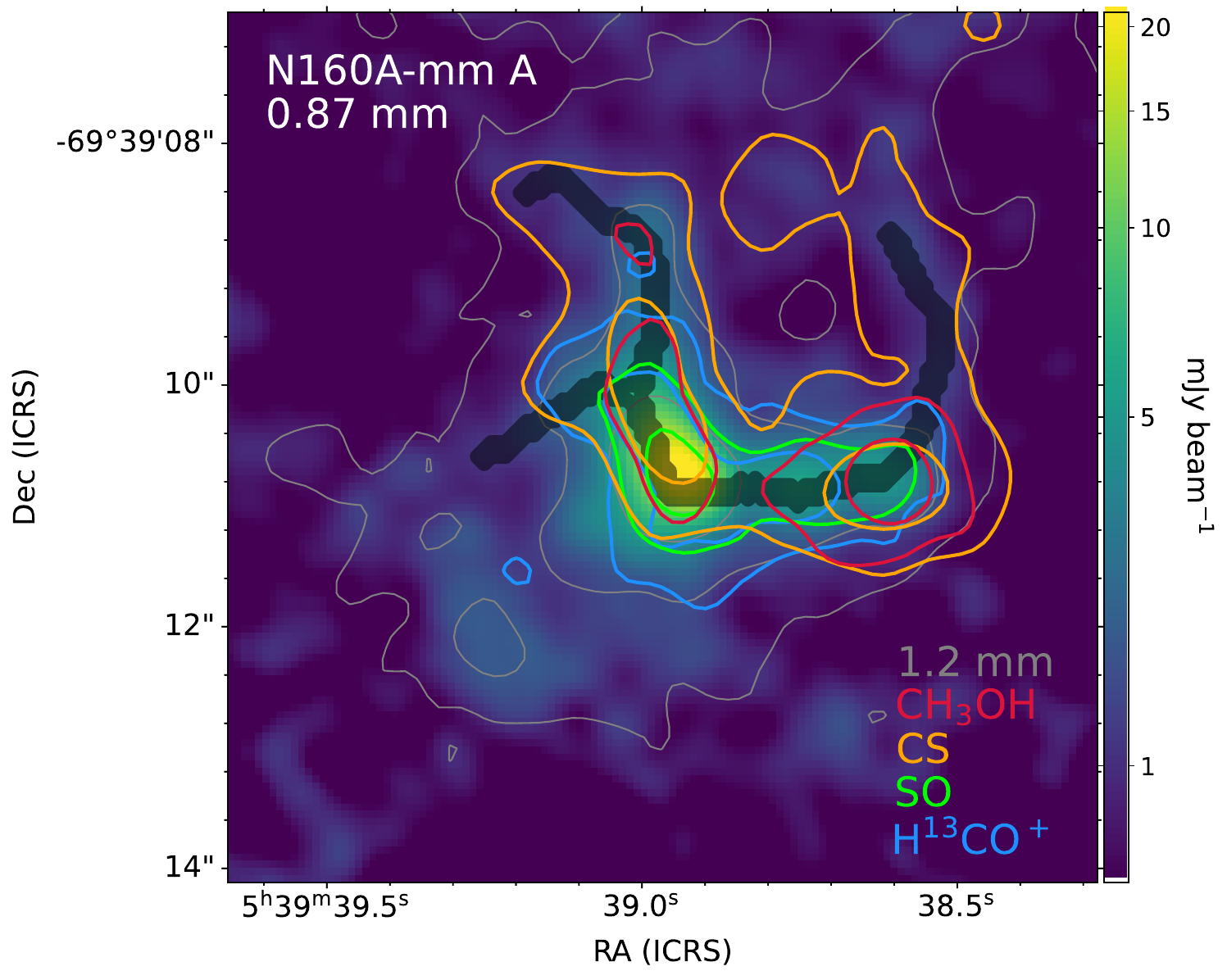}
    \caption{Left panel: The archival 0.87 mm continuum image of N\,160A--mm\,A with the beam size of $0\rlap.{''}464\times0\rlap.{''}395$ (indicated in the lower left corner). White contours correspond to the 0.87 mm continuum emission with contour levels of 5\%, 10\%, and 20\% of the continuum peak of 0.01531 Jy beam$^{-1}$. The black lines indicate the spines of the filaments identified toward N\,160A--mm\,A using FilFinder (for details, see Section~\ref{sss: hot core} and \citealt{Tokuda2023}). Right panel: The same image and filaments as in left panel with the 1.2 mm continuum contours overlaid in gray; the contour levels are the same as in Figure \ref{fig:Mom0Amaps}. Red, orange, green, and blue contours show 20\% and 60\% of the peak molecular emission for CH$_3$OH, CS, SO, and H$^{13}$CO$^+$, respectively.}
    \label{fig:Filaments}
\end{figure*}

In the ALMA Band 7 0.87 mm continuum image in Figure \ref{fig:Filaments}, we indicate the position of the spines of filaments identified in the vicinity of the hot core N\,160A--mm\,A with FilFinder. The filaments and their correlation with the molecular line emission are discussed in Section~\ref{sss: hot core}. 

\begin{figure*}
    \centering
    \includegraphics[width=0.95\textwidth]{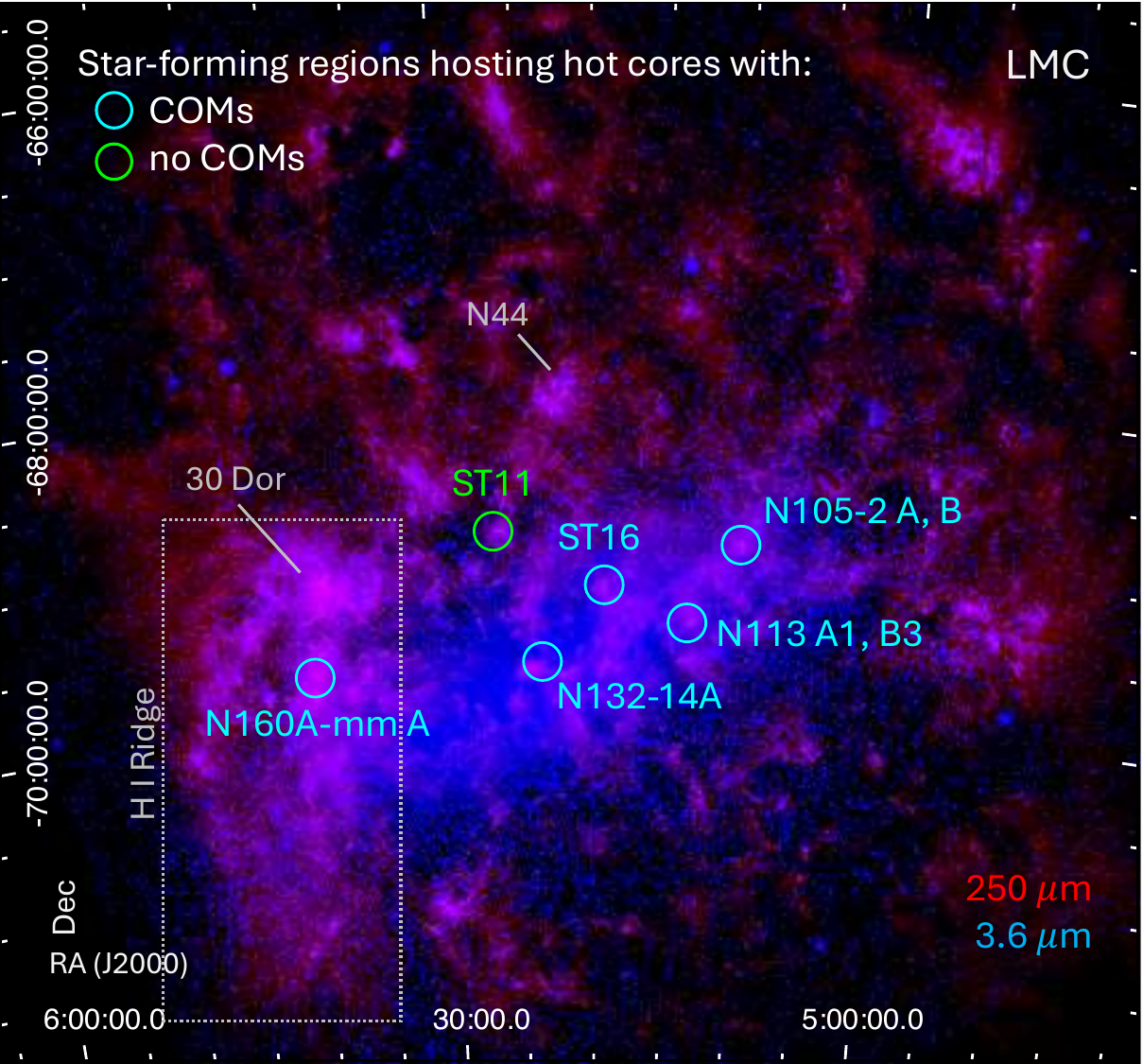} 
    \caption{The two-color mosaic of the LMC combining the Herschel/HERITAGE 250 $\mu$m (red) and Spitzer/SAGE 3.6 $\mu$m (blue) images. The 250 $\mu$m emission reveals cold dust in star-forming regions. The 3.6 $\mu$m emission traces the stellar population in the LMC, highlighting the LMC's stellar bar. The cyan and green circles indicate star-forming regions hosting hot cores with and without COM detections, respectively. The names of hot cores in each region are indicated. The newly detected hot core N\,160A--mm\,A lies outside the LMC's stellar bar.}
    \label{fig:bar}
\end{figure*}

The Herschel Inventory of the Agents of Galaxy Evolution (HERITAGE) Spectral and Photometric Imaging Receiver (SPIRE) 250 $\mu$m image \citep{meixner2013} and the SAGE 3.6 $\mu$m image \citep{meixner2006} are shown in Figure \ref{fig:bar}. The stellar bar of the LMC can be seen in the 3.6 $\mu$m with the star forming regions appearing in the 250 $\mu$m. Prior to the identification of N\,160A--mm\,A, all hot cores with COMs were found in the stellar bar.

\clearpage
\bibliographystyle{aasjournal}
\bibliography{refs_all.bib}

\end{document}